\documentclass[iop,apj,numberappendix,appendixfloats]{emulateapj}
\usepackage{graphicx}
\pdfoutput=1

\LongTables
\newcommand{\km}{${\rm km\,s}^{-1}$}
\newcommand{\nhi}{$N_{\rm H\,I}$}

\newcommand{\hi}{H$\;${\small\rm I}\relax}
\newcommand{\hii}{H$\;${\small\rm II}\relax}
\newcommand{\neviii}{Ne$\;${\small\rm VIII}\relax}
\newcommand{\mgx}{Mg$\;${\small\rm X}\relax}

\newcommand{\cii}{C$\;${\small\rm II}\relax}

\newcommand{\ciii}{C$\;${\small\rm III}\relax}
\newcommand{\civ}{C$\;${\small\rm IV}\relax}

\newcommand{\niii}{N$\;${\small\rm III}\relax}

\newcommand{\oi}{O$\;${\small\rm I}\relax}
\newcommand{\oii}{O$\;${\small\rm II}\relax}
\newcommand{\oiii}{O$\;${\small\rm III}\relax}
\newcommand{\oiv}{O$\;${\small\rm IV}\relax}

\newcommand{\ovi}{O$\;${\small\rm VI}\relax}

\newcommand{\mgii}{Mg$\;${\small\rm II}\relax}

\newcommand{\siii}{Si$\;${\small\rm II}\relax}
\newcommand{\sv}{S$\;${\small\rm V}\relax}
\newcommand{\siiii}{Si$\;${\small\rm III}\relax}
\newcommand{\sixii}{Si$\;${\small\rm XII}\relax}

\newcommand{\alii}{Al$\;${\small\rm II}\relax}
\newcommand{\Siii}{S$\;${\small\rm III}\relax}
\newcommand{\siiv}{Si$\;${\small\rm IV}\relax}
\newcommand{\siv}{S$\;${\small\rm IV}\relax}
\newcommand{\svi}{S$\;${\small\rm VI}\relax}
\newcommand{\feii}{Fe$\;${\small\rm II}\relax}

\newcommand{\hit}{H$\;${\scriptsize\rm I}\relax}

\newcommand{\ciit}{C$\;${\scriptsize\rm II}\relax}

\newcommand{\ciiit}{C$\;${\scriptsize\rm III}\relax}
\newcommand{\civt}{C$\;${\scriptsize\rm IV}\relax}

\newcommand{\niit}{N$\;${\scriptsize\rm II}\relax}
\newcommand{\niiit}{N$\;${\scriptsize\rm III}\relax}

\newcommand{\oiit}{O$\;${\scriptsize\rm II}\relax}
\newcommand{\oiiit}{O$\;${\scriptsize\rm III}\relax}
\newcommand{\oivt}{O$\;${\scriptsize\rm IV}\relax}

\newcommand{\ovit}{O$\;${\scriptsize\rm VI}\relax}

\newcommand{\mgiit}{Mg$\;${\scriptsize\rm II}\relax}

\newcommand{\siiit}{Si$\;${\scriptsize\rm II}\relax}
\newcommand{\svt}{S$\;${\scriptsize\rm V}\relax}
\newcommand{\siiiit}{Si$\;${\scriptsize\rm III}\relax}

\newcommand{\aliit}{Al$\;${\scriptsize\rm II}\relax}
\newcommand{\Siiit}{S$\;${\scriptsize\rm III}\relax}
\newcommand{\siivt}{Si$\;${\scriptsize\rm IV}\relax}
\newcommand{\sivt}{S$\;${\scriptsize\rm IV}\relax}
\newcommand{\svit}{S$\;${\scriptsize\rm VI}\relax}
\newcommand{\feiit}{Fe$\;${\scriptsize\rm II}\relax}
\newcommand{\zniit}{Zn$\;${\scriptsize\rm II}\relax}

\newcommand{\lya}{Ly\,$\alpha$\relax}
\newcommand{\lyb}{Ly\,$\beta$\relax}

\newcommand{\hst}{{\em HST}}

\submitted{Received by ApJ 2012 December 19; accepted by ApJ 2013 April 29}
\shortauthors{Lehner et al.}
\shorttitle{The Bimodal Metallicity Distribution of the CGM}

\begin{document}
\title{The Bimodal Metallicity Distribution of the Cool Circumgalactic Medium at $\lowercase{z}\la 1$\altaffilmark{1}}

\author{N.\ Lehner\altaffilmark{2},
	J.C. \ Howk\altaffilmark{2}, 
	T.M. \ Tripp\altaffilmark{3},
	J. Tumlinson\altaffilmark{4},
	J.X. \ Prochaska\altaffilmark{5},
	J.M. \ O'Meara\altaffilmark{6},
	C. Thom\altaffilmark{4},
	J.K. \ Werk\altaffilmark{5},
	A.J. Fox\altaffilmark{4},
	J. \ Ribaudo\altaffilmark{7} 
	}
\altaffiltext{1}{Based on observations made with the NASA/ESA Hubble Space Telescope,
obtained at the Space Telescope Science Institute, which is operated by the
Association of Universities for Research in Astronomy, Inc. under NASA
contract No. NAS5-26555.}
\altaffiltext{2}{Department of Physics, University of Notre Dame, 225 Nieuwland Science Hall, Notre Dame, IN 46556}
\altaffiltext{3}{Department of Astronomy, University of Massachusetts, Amherst, MA 01003, USA}
\altaffiltext{4}{Space Telescope Science Institute, Baltimore, MD 21218}
\altaffiltext{5}{UCO/Lick Observatory, University of California, Santa Cruz, CA }
\altaffiltext{6}{Department of Physics, Saint Michael's College, Vermont, One Winooski Park, Colchester, VT 05439}
\altaffiltext{7}{Department of Physics, Utica College, 1600 Burrstone Road, Utica, New York 13502}

\begin{abstract}
We assess the metal content of the cool ($\sim$$10^4$ K) circumgalactic medium (CGM) about galaxies at $z \la 1$ using an \hi-selected sample of 28 Lyman limit systems (LLS, defined here as absorbers with $16.2 \la \log N_{\rm H\,I} \la 18.5$) observed in absorption against background QSOs by the Cosmic Origins Spectrograph on-board the {\em Hubble Space Telescope}. The \nhi\ selection avoids metallicity biases inherent in many previous studies of the low-redshift CGM.  We compare the column densities of weakly ionized metal species (e.g., \oii, \siii, \mgii) to \nhi\ in the strongest \hi\ component of each absorber.  We find that the metallicity distribution of the LLS (and hence the cool CGM) is bimodal with metal-poor and metal-rich branches peaking at $[{\rm X/H}] \simeq -1.6$ and $-0.3$ (or about $2.5\%$ and $50\%$ solar metallicities).  The cool CGM probed by these LLS is predominantly ionized. The metal-rich branch of the population likely traces winds, recycled outflows, and tidally stripped gas; the metal-poor branch has properties consistent with cold accretion streams thought to be a major source of fresh gas for star forming galaxies.  Both branches have a nearly equal number of absorbers.  Our results thus demonstrate there is a significant mass of previously-undiscovered cold metal-poor gas and confirm the presence of metal enriched gas in the CGM of $z\la 1$ galaxies.
\end{abstract}

\section{Introduction}

One of the most pressing problems in galaxy formation is to understand how gas accretion and feedback influence the evolution of galaxies and the intergalactic medium (IGM). Modern cosmological simulations cannot explain the observed mass-metallicity relationship \citep{tremonti04} and color bimodalities observed in galaxies without invoking such large-scale flows \citep[e.g.,][]{keres05}. Stars cannot continue to form in galaxies over billions of years without a replenishment of gas from the IGM \citep[e.g.,][]{maller04,dekel06}, while feedback from star formation and AGN activity can fuel massive outflows from a gas-rich galaxy that may choke off star formation \citep[e.g.,][]{kacprzak08,oppenheimer10,oppenheimer12,fumagalli11b,faucher11}. The competition between these regulating processes is played out in the circumgalactic medium (CGM), the interface between galaxies and the IGM, where outflows may cool and stall, and where infalling gas may encounter winds or be shock-heated and halted.

Cosmological simulations have indicated the importance of feedback and accretion through the CGM for several decades \citep[e.g.,][]{white91,dekel06}. Modeling these phenomena with more realistic physics and modified methods to solve the hydrodynamical equations have become possible with improved computational power \citep[e.g.,][]{springel03,kobayashi04, oppenheimer10,springel10,smith11,keres12,sijacki12,vogelsberger12}. All these simulations show in particular that \hi\ absorbers with $N_{\rm H\,I} >10^{15}$ cm$^{-2}$ are excellent tracers of the CGM of galaxies within $\sim$300 kpc (about the virial radius, \citealt{ford12}). Several observational results from the analyses of the galaxy-absorber connection also show that the virialized  CGM is traced by the strong \lya\ absorbers \citep[e.g.,][]{morris91,bergeron94,lanzetta95,tripp98,penton02,bowen02,chen05,morris06,wakker09,prochaska11}. These observations show that there is significant clustering between galaxies and strong \hi\ ($N_{\rm H\,I}\ga 10^{15}$ cm$^{-2}$) absorbers, which points to a strong physical connection. In contrast, the weak \hi\ absorbers ($N_{\rm H\,I}\la 10^{14}$ cm$^{-2}$) are at best very loosely connected to galaxies, as evidenced by a weak or absent clustering signal in the galaxy-absorber two-point cross-correlation function.

The most sensitive way to study the large-scale CGM around galaxies  is therefore to investigate the properties of the strong \hi\ absorption (and associated metal lines associated with it) along QSO sightlines, as first proposed by \citet{bahcall69} and then observed by \citet{bergeron86}. Here, we focus on the Lyman limit systems (LLS) as tracers of the CGM \citep[e.g.,][]{tytler82,steidel90,bergeron94}, absorbers with large enough surface density of \hi\ atoms to produce a decrement in the UV flux at the Lyman limit. Traditionally, they have been separately categorized as partial Lyman systems ($16.2 \la \log N_{\rm  H\,I} < 17$) and LLS ($17 \le \log N_{\rm  H\,I} < 19$). In this paper, we will just define both categories as LLS for simplicity. Since all these absorbers trace CGM gas, in this context there is not a need to distinguish absorbers with $\tau_{\rm LL}< 1$ (partial-LLS) and $\tau_{\rm LL}>1$ (LLS), where $\tau_{\rm LL}$ is the optical depth at the Lyman limit. The importance of the LLS for characterizing the CGM and phenomena therein (large-scale flows) has been demonstrated in recent cosmological simulations \citep[e.g.,][]{faucher11b,fumagalli11b,stewart11a,voort12,shen13} as well as observationally as we now discuss in more detail.

Two recent results from our group have not only confirmed that LLS probe the CGM, but also demonstrated they can trace widely varying processes important for galaxy evolution. \citet{tripp11} show that a LLS ($\log N_{\rm H\,I} \simeq 17.0$) at $z=0.927$ traces a super-solar metallicity post-starburst galactic wind that extends at least to $>68$ kpc from its host.  On the other end of the metallicity spectrum, \citet{ribaudo11b} reported the discovery of a $z = 0.274$ LLS with the same \nhi\ as in the \citeauthor{tripp11} study, but  with a 2\% solar metallicity. They show that this absorber resides at impact parameter $\rho = 37$ kpc from a $\sim 0.3 L^*$ galaxy whose metallicity is a factor 30 higher than the LLS metallicity. The metallicity of this LLS is so low at this redshift that the majority of its gas is unlikely to have been recently incorporated into a galaxy.  Despite similar \nhi\ values, the metal content, amount of highly ionized gas, kinematics, galactic environments, and origins, are all very different between these two absorbers. The rest-frame equivalent width  of the strong \mgii\ $\lambda$2796 is $W_\lambda=0.1$ and 1 \AA\ for the metal-poor and metal-rich LLS, respectively; the low \mgii\ equivalent absorber would have been missed in most \mgii\ surveys. 

Prior to these results, only a handful of \hi-selected LLS or super-LLS (SLLS, $19 \le \log N_{\rm  H\,I}< 20.3$) with information on their metallicity were reported at $z \la~1$ \citep[e.g.,][]{chen00,zonak04,prochaska04,jenkins05,tripp05,cooksey08,lehner09}. In \citet{ribaudo11b}, we argued that the metal-poor LLS are the best-yet observational evidence for cold flow accretion in galaxies, while the higher metallicity absorbers are tracing a mixture of physical processes, such as outflows, recycled gas, and galaxy interactions. Knowing the metallicity of these absorbers is critical for placing constraints on their origins, but prior to the installation of the Cosmic Origins Spectrograph (COS) on the {\it Hubble Space Telescope}\ (\hst), the sample was too small to reliably determine the metallicity distribution and the range of possible metallicity values in the CGM at $z\la 1$.  Thanks to the installation of COS, the situation has dramatically improved in just a few years. 

In this paper, we present the first results of an analysis of the metallicity of the CGM at $z\la 1$ using LLS selected solely based on their \hi\ content. The new LLS described in this work were discovered blindly in our Cycle 17 and in GTO COS programs (see below). In these LLS, we can measure accurately the column densities of \hi\ and metal ions (in particular, the low ions such as \cii, \siii, \mgii) that allow us to model the ionization conditions and derive the metallicity following a similar methodology used in previous studies \citep[e.g.,][]{cooksey08,lehner09,ribaudo11b}. Our survey of \hi-selected LLS provides the first unbiased metallicity study of the LLS (and thus the cool CGM) metallicity distribution at $z\la 1$.

We compare our analysis of the \hi-selected with  metallicity estimates from the literature for SLLS and damped \lya\ absorbers (DLAs, $N_{\rm  H\,I}\ge 20.3$) over the same redshift range. However, we emphasize that SLLS and DLAs were mostly selected based on knowing {\it a priori}\ that they show strong \mgii\ absorption. Our sample of low-$z$ metallicities consists of 28 \hi-selected LLS, 29 SLLS, and 26 DLAs. Seventeen LLS have never been published previously, so we have increased the sample of LLS by a factor 2 and nearly a factor 5 compared to the pre-COS era. 

 Our paper is organized as follows. In \S\ref{s-newlls} we describe the new and existing LLS samples as well as the determination of their metallicities based on photoionization models (\S\ref{s-phot}) with additional technical details (including the description of each absorber) provided in the appendix for interested readers.  Our main results are presented in \S\ref{s-metres}, where in particular in \S\ref{s-metlls} we describe the bimodal distribution  function of the LLS metallicity. In this section, we also compare the results from the LLS and higher \hi\ column density absorber samples. In \S\ref{s-origins} we discuss implications of our new observational results for the physical processes and the properties of the CGM. Finally, in \S\ref{s-sum} we summarize our main results. 

\section{LLS Sample}\label{s-newlls}
\subsection{Sample Description}

We search in the UV spectra of QSOs for absorbers at $z\la 1$ based  on their \hi\ content,  i.e., they all satisfy the condition $16.2 \la \log N_{\rm H\,I} \la 18.5$ within the errors. Although the lower \nhi\ cut-off is somewhat arbitrary, it was selected because a break in the continuum at the Lyman limit could still be seen in high-signal to noise data. These absorbers were found blindly in the UV QSO spectra that were obtained for other scientific reasons, i.e., in particular, the absorbers were not pre-selected based on knowledge of their metal content. But for that reason, most of the absorbers have  $16.2 \la \log N_{\rm H\,I} \la 17.3$ where $\tau_{\rm LL}<1$, i.e., the UV flux is only partially attenuated. Prior to COS, QSO sightlines were chosen to avoid $\tau_{\rm LL} > 1$ LLS and very few QSOs were at high enough $z$ to produce a large number of LLS. 

We searched for the LLS in four high resolution COS programs:  11520 (``COS-GTO: QSO Absorbers, Galaxies and Large-scale Structures in the Local Universe", PI Green), 11598 (``How Galaxies Acquire their Gas: A Map of Multiphase Accretion and Feedback in Gaseous Galaxy Halos", a.k.a. the ``COS-Halos'' program, PI: Tumlinson), 11692 (``The LMC as a QSO Absorption Line System", PI: Howk), 11741 (``Probing Warm-Hot Intergalactic Gas at $0.5 < z < 1.3$ with a Blind Survey for \ovi, \neviii, \mgx, and \sixii\ Absorption Systems, PI: Tripp). Note that targeted systems in the COS-Halos program are not in our sample \citep[see a description of these absorbers in][]{tumlinson11a,werk12b} because the \nhi\ for the LLS is not well constrained in that sample.  At this time, we have not searched systematically the entire COS G130M/G160M QSO archive, although this will be underway soon. The redshifts of the 12 QSOs are in the range $0.33 \le z_{\rm QSO}\le 1.44 $. None of the LLS described here are proximate (i.e., with $z_{\rm abs}\sim z_{\rm QSO}$) systems. 

The choice of the present data sample is motivated by three reasons: 1) a large portion of these data was already reduced by our team to undertake the original scientific goals of these COS programs, which in particular helped with the line identification process; 2) several of these QSOs were observed with Keck HIRES, allowing us to measure \mgii\ $\lambda$$\lambda$2796, 2803 absorption, which is a key ion for determining the metallicity and to determine the velocity component structure of the absorbers; 3) most of the spectra  have relatively high signal-to-noise (S/N$\,\ga 15$--30 per resolution element).

\begin{deluxetable*}{lccccccc}
\tabcolsep=3pt
\tablecolumns{7}
\tablewidth{0pc}
\tablecaption{\label{t-sum}}
\tabletypesize{\footnotesize}
\tablehead{\colhead{Name} & \colhead{$z_{\rm abs}$}& \colhead{$\log N_{\rm HI}$} & \colhead{$[{\rm X/H}]$} & \colhead{$\log U$} & \colhead{$\rho$}  & \colhead{References}\\ %
\colhead{} &\colhead{} & \colhead{[cm$^{-2}$]} &\colhead{}  &\colhead{} & \colhead{(kpc\,h$^{-1}$)} &\colhead{}  %
 }
\startdata
\cutinhead{LLS -- $16.2 \le \log N_{\rm HI} < 19$ } %
          J0943+0531   & $   0.3542 $ & $   16.11 \pm     0.09 $ & $   <-1.30 		  $ & $\le-3.2$ & $95 $&                  Tho11$^a$ \\
          J1419+4207   & $   0.4256 $ & $   16.17 \pm     0.06 $ & $   -1.40 \pm 0.20	  $ & $ -2.9 $ & \nodata &                  This paper \\
          J1435+3604   & $   0.3878 $ & $   16.18 \pm     0.05 $ & $   < -1.40		  $ & $\ge -3.0$ & \nodata &                  This paper \\
          PG1116+215   & $   0.1385 $ & $   16.20 \pm     0.05 $ & $   -0.50 :            $ & $    -2.5$ & $   127 $ &                  Sem04,Tri98 \\
          PG1522+101   & $   0.5185 $ & $   16.22 \pm     0.02 $ & $   -0.40 \pm     0.05 $ & $    -3.6$ &   \nodata &                 This paper \\
         SBS1122+594   & $   0.5574 $ & $   16.24 \pm     0.03 $ & $   -1.05 \pm     0.05 $ & $    -3.1$ &   \nodata &                 This paper \\
         HE0439-5254   & $   0.6153 $ & $   16.28 \pm     0.04 $ & $   -0.30 \pm     0.05 $ & $    -2.7$ &   \nodata &                 This paper \\
	PG1338+416     & $   0.3488 $ & $   16.30 \pm     0.13 $ & $   -0.65 \pm     0.15 $ & $    -3.2$ &   \nodata &                 This paper \\
         J1419+4207    & $   0.5346 $ & $16.34 \,^{+0.23}_{-0.12}$ &$   -0.20 \pm 0.20    $ & $   -3.9 $ &   \nodata &  	       This paper \\        	 	 
	  PG1407+265   & $   0.6828 $ & $   16.38 \pm     0.02 $ & $	-1.80 \pm 0.30	  $ & $    -2.0$ &   \nodata &                 This paper \\
	PG1216+069     & $   0.2823 $ & $   16.40 \pm     0.05 $ & $   <-1.65	 $ 	    & $   <-2.6$ &   3238	&                 This paper,Pro11a \\      	
	J1419+4207     & $   0.2889 $ & $   16.40 \pm     0.07 $ & $   -0.65 \pm     0.10 $ & $    -3.1$ &   \nodata &                 This paper \\      
          PG1338+416   & $   0.6865 $ & $   16.45 \pm     0.05 $ & $   +0.10 \pm 0.10	  $ & $    -3.9$ &   \nodata &                 This paper \\
         PKS0405-123   & $   0.1672 $ & $   16.45 \pm     0.05 $ & $   +0.10 :            $ & $    -3.1$ & $   100 $ &                 Che00,Sav10 \\
         J1619+3342    & $   0.2694 $ & $   16.48 \pm     0.05 $ & $   -1.60 \pm      0.10$ & $    -2.9$ &   \nodata &                 This paper \\
         PKS0637-752   & $   0.4685 $ & $   16.48 \pm     0.04 $ & $    -0.50 \pm 0.1      $ & $   -3.8$ &   \nodata &                 This paper \\
	HE0153-4520    & $   0.2261 $ & $16.61\,^{+0.12}_{-0.17}$ & $   -0.80 \pm    0.25 $ & $    -2.8$ &   \nodata &		   Sav11 \\
         J1435+3604    & $   0.3730 $ & $   16.65 \pm     0.07 $ & $   -1.85  \pm    0.10 $ & $    -3.5$ &   \nodata &                 This paper \\ 
          PG1522+101   & $   0.7292 $ & $   16.66 \pm     0.05 $ & $  <-2.00		  $ & $   >-3.2$ &   \nodata &                 This paper \\
             PHL1377   & $   0.7392 $ & $   16.72 \pm     0.03 $ & $   -1.45 \pm     0.05 $ & $    -2.9$ &   \nodata &                 This paper \\
         PKS0552-640   & $   0.3451 $ & $   16.90 \pm     0.08 $ & $ < -1.50		  $ & $   >-3.8$ &   \nodata &                 This paper \\
          PG1630+377   & $   0.2740 $ & $   16.98 \pm     0.05 $ & $   -1.71 \pm     0.06 $ & $    -2.8$ & $    37 $ &                Rib11b \\
          PG1206+459   & $   0.9270 $ & $   17.00 \pm     0.10 $ & $   +0.30 :            $ & $   <-2.5$ & $    68 $ &                   Tri11 \\
         PKS1302-102   & $   0.0985 $ & $   17.00 \pm     0.20 $ & $ < -1.60		  $ & $    -2.9$ & $    65 $ &            Coo08 \\
          PG1634+706   & $   1.0400 $ & $   17.30 \pm     0.30 $ & $   -1.50 \pm     0.20 $ & $    -5.2$ &   \nodata &                  Zon04 \\
	     PHL1811   & $   0.0810 $ & $   17.98 \pm     0.05 $ & $   -0.19 \pm     0.08 $ & $    -4.0$ & $    34 $ &                Jenk05 \\
          PKS0312-77   & $   0.2026 $ & $   18.22 \pm     0.20 $ & $   -0.60 \pm     0.15 $ & $    -3.2$ & $    38 $ &                 Leh09 \\
           TON153      & $   0.6610 $ & $   18.30 \pm     0.30 $ & $   -1.69 \pm     0.37 $ & $    -3.2$ & $   104 $ &                      Kac12$^b$ \\
          J1009+0713   & $   0.3588 $ & $   18.40 \pm     0.20 $ & $   -0.40 \pm 0.20	  $ & $    -3.5$ &  $43$    &   		 Tum11b,Wer12b \\        
\cutinhead{SLLS -- $19 \le \log N_{\rm HI} < 20.3$ } %
          Q0826-2230   & $   0.9110 $ & $   19.04 \pm     0.04 $ & $   +0.68 \pm     0.08 $ & $   <-4.1$ &   \nodata &			Mei09 \\
          Q0005+0524   & $   0.8514 $ & $   19.08 \pm     0.04 $ & $  <-0.47 :		  $ & \nodata    &   \nodata &			Mei09 \\
          Q2352-0028   & $   0.8730 $ & $   19.18 \pm    0.09 $ & $   <-0.14 : 		  $ & \nodata    &   \nodata &                       Mei09 \\
          Q1054-0020   & $   0.9513 $ & $   19.28 \pm     0.02 $ & $  <-0.66 :		  $ & \nodata    &   \nodata &                       Des09 \\
	   Q2149+212   & $   1.0023 $ & $   19.30 \pm     0.05 $ & $  <+0.20 :		  $ & \nodata    &   \nodata &                      Nes08 \\
          J1001+5944   & $   0.1140 $ & $   19.32 \pm     0.10 $ & $   -0.37 \pm     0.10 $ & $    -3.4$ &   \nodata &                 Bat12 \\
          PG1216+069   & $   0.0063 $ & $   19.32 \pm     0.03 $ & $   -1.60 \pm     0.10 $ & $    -3.1$ & $    82 $ &                  Tri05 \\
          J0928+6025   & $   0.1538 $ & $   19.35 \pm     0.15 $ & $   +0.31 \pm     0.17 $ & $    -3.7$ & $    38 $ &                 Bat12 \\
          J0021+0043   & $   0.9424 $ & $   19.38 \pm     0.13 $ & $  <-0.41 : 		  $ & \nodata    &   \nodata &                       Des09 \\
          Q1330-2056   & $   0.8514 $ & $   19.40 \pm     0.02 $ & $  <-0.46 :		  $ & \nodata    &   \nodata &                       Des09 \\
          Q1228+1018   & $   0.9376 $ & $   19.41 \pm     0.02 $ & $  <-0.37 : 		  $ & \nodata    & $    38 $ &                  Mei09 \\
          Q1009-0026   & $   0.8866 $ & $   19.48 \pm     0.05 $ & $   +0.24 \pm     0.15 $ & $    -3.7$ & $    35 $ &                       Des09 \\
          J1553+3548   & $   0.0830 $ & $   19.55 \pm     0.15 $ & $   -1.10 \pm     0.16 $ & $    -3.2$ &   \nodata &                 Bat12 \\
          J0925+4004   & $   0.2477 $ & $   19.55 \pm     0.15 $ & $   -0.29 \pm     0.17 $ & $    -3.1$ & $    85 $ &                 Bat12 \\
          Q1631+1156   & $   0.9004 $ & $   19.70 \pm     0.04 $ & $  <-0.15 :	          $ & \nodata    &   \nodata &                 Mei09 \\
          Q0153+0009   & $   0.7714 $ & $   19.70 \pm     0.09 $ & $  <-0.34 :		  $ & \nodata    &   \nodata &                   Mei09\\
          Q2335+1501   & $   0.6798 $ & $   19.70 \pm     0.30 $ & $   +0.07 \pm     0.34:$ & \nodata    &   \nodata &                   Mei09\\
          J1435+3604   & $   0.2027 $ & $   19.80 \pm     0.10 $ & $   -0.41 \pm     0.16 $ & $    -3.3$ & $    38 $ &                 Bat12 \\
          Q2352-0028   & $   1.0318 $ & $   19.81 \pm     0.13 $ & $   <-0.51 	 	  $ & \nodata    &   \nodata &                      Mei09,Des09 \\
          Q0138-0005   & $   0.7821 $ & $   19.81 \pm     0.09 $ & $   +0.28 \pm     0.16 $ & \nodata    &   \nodata &                   Per08,Mei09 \\
          J0134+0051   & $   0.8420 $ & $   19.93 \pm     0.12 $ & $  <-0.36  		  $ & \nodata    &   \nodata &                   Per06 \\
          J1028-0100   & $   0.6321 $ & $   19.95 \pm     0.07 $ & $  <-0.20 		  $ & \nodata    &   \nodata &                       Des09 \\
          J1028-0100   & $   0.7089 $ & $   20.04 \pm     0.06 $ & $  <-0.18 		  $ & \nodata    &   \nodata &                       Des09 \\
          Q1436-0051   & $   0.7377 $ & $   20.08 \pm     0.11 $ & $   -0.05 \pm     0.12 $ & \nodata    & $   <33 $ &                   Mei09\\
          Q1455-0045   & $   1.0929 $ & $   20.08 \pm     0.06 $ & $   <-0.80		  $ & \nodata    &   \nodata &            	 Mei09 \\
          Q1009-0026   & $   0.8426 $ & $   20.20 \pm     0.06 $ & $   <-0.98		  $ & \nodata    & $    >15 $ &                       Mei09 \\
          Q1220-0040   & $   0.9746 $ & $   20.20 \pm     0.07 $ & $  <-1.14 		   $ & \nodata    &   \nodata &               Mei09 \\
          J1323-0021   & $   0.7160 $ & $   20.21 \pm     0.20 $ & $   +0.61 \pm     0.20 $ & \nodata    & $     9 $ &           Per06,Kha04 \\
          Q1107+0003   & $   0.9542 $ & $   20.26 \pm     0.14 $ & $ < -0.51		  $ & \nodata    &   \nodata &                      Nes08 \\
\cutinhead{DLA -- $\log N_{\rm HI} > 20.3$ }  %
          J2328+0022   & $   0.6520 $ & $   20.32 \pm     0.07 $ & $   -0.49 \pm     0.22 $ & \nodata    &   \nodata &                   Per06 \\
           Q0302-223   & $   1.0090 $ & $   20.36 \pm     0.11 $ & $   -0.73 \pm     0.12 $ & \nodata    & $    25 $ &  	Pet00,LeB97,Per06 \\
           Q1622+238   & $   0.6560 $ & $   20.36 \pm     0.10 $ & $   -0.87 \pm     0.25 $ & \nodata    &   \nodata &          Chu00,Rao00 \\
           Q1122-168   & $   0.6820 $ & $   20.45 \pm     0.05 $ & $   -1.00 \pm     0.15 $ & \nodata    &   \nodata &                Var00 \\
          Q1323-0021   & $   0.7156 $ & $   20.54 \pm     0.15 $ & $   +0.12 \pm     0.26 $ & \nodata    &   \nodata &                      Nes08 \\
          J1619+3342   & $   0.0963 $ & $   20.55 \pm     0.10 $ & $   -0.63 \pm     0.13 $ & \nodata    &   \nodata &                 Bat12 \\
          J1616+4154   & $   0.3213 $ & $   20.60 \pm     0.20 $ & $   -0.35 \pm     0.23 $ & \nodata    &   \nodata &                 Bat12 \\
          Q0139-0023   & $   0.6825 $ & $   20.60 \pm     0.12 $ & $  <-0.08 		  $ & \nodata    &   \nodata &                      Nes08 \\
          J1009+0713   & $   0.1140 $ & $   20.68 \pm     0.10 $ & $   -0.58 \pm     0.16 $ & \nodata    &   \nodata &                 Bat12 \\
           Q0454+039   & $   0.8600 $ & $   20.69 \pm     0.06 $ & $   -0.79 \pm     0.12 $ & \nodata    &   \nodata &                  Pet00 \\
          J0256+0110   & $   0.7250 $ & $   20.70 \pm     0.15 $ & $   -0.11 \pm     0.04 $ & \nodata    &   \nodata &                   Per06 \\
          Q0256+0110   & $   0.7252 $ & $   20.70 \pm     0.22 $ & $   -0.13 \pm     0.24 $ & \nodata    &   \nodata &                      Nes08 \\
          Q1733+5533   & $   0.9981 $ & $   20.70 \pm     0.04 $ & $   -0.44 \pm     0.07 $ & \nodata    &   \nodata &                      Nes08 \\
           Q2149+212   & $   0.9111 $ & $   20.70 \pm     0.10 $ & $  <-0.93		  $ & \nodata    &   \nodata &                      Nes08 \\
           Q1229-021   & $   0.3950 $ & $   20.75 \pm     0.07 $ & $   -0.47 \pm     0.15 $ & \nodata    & $     7 $ &  	Boi98,Leb97,Ste94 \\
          Q0253+0107   & $   0.6316 $ & $   20.78 \pm     0.12 $ & $  <-0.35 		  $ & \nodata    &   \nodata &                      Nes08 \\
          Q0449-1645   & $   1.0072 $ & $   20.98 \pm     0.07 $ & $   -0.96 \pm     0.08 $ & \nodata    &   \nodata &                   Per08 \\
          J1107+0048   & $   0.7400 $ & $   21.00 \pm     0.03 $ & $   -0.54 \pm     0.20 $ & \nodata    &   \nodata &           Per06,Kha04 \\
          Q1007+0042   & $   1.0373 $ & $   21.15 \pm     0.24 $ & $   -0.51 \pm     0.24 $ & \nodata    &   \nodata &                      Nes08 \\
          Q1727+5302   & $   0.9445 $ & $   21.16 \pm     0.05 $ & $   -0.56 \pm     0.06 $ & \nodata    &   \nodata &                      Nes08 \\
          J1431+3952   & $   0.6019 $ & $   21.20 \pm     0.10 $ & $   -0.80 \pm     0.20 $ & \nodata    &   \nodata &                  Ell12 \\
           Q1328+307   & $   0.6920 $ & $   21.25 \pm     0.06 $ & $   -1.20 \pm     0.09 $ & \nodata    & $     6 $ &  	Mey95,Leb97,Ste94 \\
          Q1225+0035   & $   0.7728 $ & $   21.38 \pm     0.12 $ & $   -0.78 \pm     0.14 $ & \nodata    &   \nodata &                      Nes08 \\
          Q1727+5302   & $   1.0306 $ & $   21.41 \pm     0.03 $ & $   -1.36 \pm     0.08 $ & \nodata    &   \nodata &                      Nes08 \\
           2353-0028   & $   0.6043 $ & $   21.54 \pm     0.15 $ & $   -0.92 \pm     0.32 $ & \nodata    &   \nodata &                      Nes08 \\
           Q0235+164   & $   0.5240 $ & $   21.65 \pm     0.15 $ & $   -0.58 \pm     0.15 $ & \nodata    & $     6 $ &      Tur04,Bur96 
\enddata
\tablecomments{We express the abundances with the usual logarithmic notation. $U$ is the ionization parameter derived from the Cloudy simulations and $\rho$ is the impact parameter to generally the closest galaxy (there might be evidence of other galaxies within 150 kpc, e.g., \citealt{jenkins05}, \citealt{kacprzak12}). For the LLS, the metallicity is estimated from $\alpha$ elements (e.g, O, Mg, Si). For the SLLS and DLA, the metallicities were estimated from Zn, except those estimated by Bat12 and Tri05, where X is an $\alpha$ element. For the SLLS with $\log N_{\rm HI}\la 19.7$ and no $U$ value, no ionization correction was estimated. For these absorbers, ionization corrections may be necessary, and we emphasize this uncertainty by adding a colon in the metallicity column and adopt the measured value as an upper limit. $a$: this absorber was analyzed by  \citealt{thom11}, but no error was given on \nhi\ and $1\sigma$ upper limits were used to estimate the metallicity. Using \hit\ $\lambda$$\lambda$926, 920, 919, 918, we re-estimated \nhi; we also re-estimated the metallicity using 3$\sigma$ upper limits. Our and Thom et al. results are consistent within 1$\sigma$. $b$: As this absorber was pre-selected on \mgiit\ absorption it is not included in our sample of \hit-selected LLS. }
\tablerefs{Bat12: \citealt{battisti12}; Boi98: \citealt{boisse98}; Bur96: \citealt{burbidge96}; Che00: \citealt{chen00}; Chu00: \citealt{churchill00}; Coo08: \citealt{cooksey08};  Des09: \citealt{des09}; Ell12: \citealt{ellison12}; Kac12: \citealt{kacprzak12}; Jen05: \citealt{jenkins05}; Kha04: \citealt{khare04}; Leb97: \citealt{lebrun97}; Leh09: \citealt{lehner09}; Mei09: \citealt{meiring09}; Mey95: \citealt{meyer95}; Nes08: \citealt{nestor08}; Per06: \citealt{peroux06}; Per08: \citealt{peroux08}; Pet00: \citealt{pettini00}; Pro11a: \citealt{prochaska11a}; Rao00: \citealt{rao00}; Rib11a: \citealt{ribaudo11b}; Sav10: \citealt{savage10}; Sav11:  \citealt{savage11}; Sem04:  \citealt{sembach04}; Ste94:  \citealt{steidel94}; Tho11:  \citealt{thom11}; Tri98:  \citealt{tripp98}; Tri05: \citealt{tripp05}; Tri11: \citealt{tripp11}; Tum11b: \citealt{tumlinson11b}; Tur04: \citealt{turnshek04}; Zon04: \citealt{zonak04}; Wer12b: \citealt{werk12b}}
\end{deluxetable*}

To complement our sample of unpublished LLS, we systematically surveyed the literature for \hi-selected LLS at $z\la 1$ for which the metallicity has been estimated. Our entire LLS sample is summarized in Table~\ref{t-sum}, where the redshift of the absorbers, the \hi\ column density, the metallicity, ionization parameter (see below), and the impact parameter ($\rho$) are listed. We assume the absorber is associated with the galaxy near the QSO with a redshift consistent with that of the absorber. In some cases, multiple galaxies may be within about $100$ kpc \citep[e.g.,][]{jenkins05,cooksey08,tumlinson11b}. In cases where more than one galaxy is observed, the galaxy with the smallest impact parameter $\rho$ and nearest redshift was selected. We emphasize that the current galaxy redshift surveys are quite heterogeneous in terms of depth and completeness, and it remains possible, e.g., that there are very low luminosity galaxies closer to the sightline or associated luminous galaxies. 

 To the best of our knowledge all LLS but one summarized in Table~\ref{t-sum} were discovered serendipitously in the \hst\ UV spectra, i.e., they were not pre-selected owing to existing optical observations of \mgii\ absorption. For most of the QSOs observed prior the COS era, a member of our group was involved in the original scientific program that led to the acquisition of the QSO. For the other two QSOs (PG1634+704 and TON153), we searched the MAST archive for the abstract of the original program to make sure that the program was not pre-selected based on the strength of a metal line. The only exception is the metal-poor LLS observed toward TON153 \citep{kacprzak12}, which was originally targeted for its known weak ($W \sim 0.3$ \AA) \mgii\ absorber at $z=0.6610$. The number of \hi-selected LLS is 28, while our total sample of LLS at $z<1$ consists of 29 LLS.

\subsection{New LLS Sample}
\subsubsection{Data Reduction and Presentation}

The 17 new absorbers were all observed with COS. Information on the design and performance of COS can be found in \citet{green12}. The data reduction and co-addition procedures of the COS data are described in \citet{thom11} and \citet{meiring11b}, and we refer the reader to these papers for more details on COS data processing.  In short, the individual exposures were co-added in photon counts and the shifts between the individual exposures were determined using the Galactic interstellar absorption lines or well-detected lines from multiple-line extragalactic systems.  In either case, alignment was based on comparison of transitions of comparable strength. The exposures were shifted to the common heliocentric frame and co-added.  The resolution of the COS G130M and G160M gratings is $R\approx 17,000$. The redshift was determined from the peak optical depth of the strongest \hi\ component using the weak \hi\ transitions. When they are detected, we find that \cii, \oii, and \mgii\ are at the same redshifts as the strongest component of \hi. We normalized the QSO continua with low order Legendre polynomials near the absorption under consideration. Typically polynomials with orders $m \le 3$ were fitted. 

In  the Figs.~\ref{f-pg1522a} to \ref{f-pg1216} of the Appendix, we show for each LLS the normalized profiles of \hi\ (generally a weak and a strong transition) and all or most of the observed metal lines covered by COS as a function of the rest-frame velocity. We also show the normalized profiles  of \mgii\ when available. For all the QSOs, except for two sightlines near the LMC  (PKS0552-640, PKS0637-752) and the two GTO QSOs (HE0439-5254, SBS1122+594), high-resolution (about 6 \km\ FWHM) optical spectra were obtained with HIRES on the Keck\,I telescope. For the LMC sightlines (program 11692), optical spectra were obtained with the MagE spectrograph on the Magellan Clay telescope at Las Campanas Observatory. The MagE resolution is much cruder than Keck HIRES resolution, about 55 \km, but sufficient to constrain the total equivalent width and column density of \mgii. Information about the reduction and normalization of the Keck and Magellan data can be found in \citet{thom11} and \citet{werk12}, respectively. 

Because the \mgii\ $\lambda$$\lambda$2796, 2803 doublet is so strong, we will demonstrate below that it is an excellent metallicity tracer for the LLS when coupled with \nhi\ derived from the UV data and within the studied \nhi\ interval. The high resolution \mgii\ spectra also show that its component structure is relatively simple for most of the absorbers in our new LLS sample. In general one component largely dominates the column density profile. The exception in the new sample is the absorber $z=0.2889$ toward J1419+4207 where there are at least 3 components, two of them dominating the column density profile.

\subsubsection{Column Densities of the Metal Ions}
\begin{deluxetable}{lcc} 
\tabcolsep=3pt
\tablecolumns{3}
\tablewidth{0pc}
\tablecaption{Adopted Column Densities$^a$\label{t-data}}
\tabletypesize{\footnotesize}
\tablehead{\colhead{Species} & \colhead{$\log N$} & \colhead{$v_1,v_2$} \\ %
\colhead{} & \colhead{[cm$^{-2}$]} & \colhead{(\km)}  %
 }
\startdata
\cutinhead{PHL1377 -- $z_{\rm abs}= 0.7392$$^b$}  
\hit\     & $ 16.72 \pm 0.03	     $      & $-82,100$  	\\
\ciit\    & $13.35 \pm 0.08	     $      & $-42,45$   \\
\ciiit\   & $>14.11 		     $      & $-82,100$   \\
\civt\    & $ 14.17 \pm 0.05	     $      & $-82,100$   \\
\oiit\    & $ \le 14.30 \pm 0.04	$   & $-82,100$    \\
\oiiit\   & $\ge 14.83 \pm 0.02		$   & $-82,100$   \\
\oivt\    & $\ge 14.90 \pm 0.03		$   & $-82,100$   \\
\ovit\    & $ 14.22 \pm 0.11	     $      & $-82,100$   \\
\mgiit\   & $ 11.98 \pm 0.07	     $ 	    & $-42,45$    \\
\siivt\   & $ 13.25 \pm 0.09	     $      & $-82,100$   \\
\Siiit\   & $ 13.17 \pm 0.05	     $      & $-82,100$   \\
\sivt\    & $ <13.30			$   & $-82,100$   \\
\svt\     & $ 12.68 \pm 0.10	     $      & $-82,100$   \\
\svit\    & $ <13.06			 $  & $-82,100$    \\
\aliit\   & $ <12.10		     $      & $-42,45$   \\
\cutinhead{PG1338+416 -- $z_{\rm abs}= 0.3488$}  
\hit\     & $ 16.30 \pm 0.13	     $      & $-40,75$   \\
\ciit\    & $13.90 \pm 0.05	     $      & $-40,75$   \\
\ciiit\   & $\ge 14.12 \pm 0.07	     $      & $-90,75$   \\
\ovit\    & $ 14.56 \pm 0.04	     $      & $-90,75$   \\
\mgiit\   & $12.49 \pm 0.02		$   & $-40,75$   \\
\siiit\   & $12.65 \pm 0.15		$   & $-40,75$    \\
\siiiit\  & $\ge 13.21 \pm 0.06	     $      & $-40,75$   \\
\cutinhead{PG1338+416 -- $z_{\rm abs}= 0.6865$}  
\hit\     & $ 16.41 \pm 0.02	     $ & $-75,102$ \\
\ciit\    & $ 14.21 \pm 0.05	     $ & $-45,102$ \\
\ciiit\   & $>14.11 		     $ & $-57,125$ \\
\oiit\    & $ 14.30 \pm 0.04	     $ & $-45,102$ \\
\oiiit\   & $>14.92 		     $ & $-75,125$ \\
\oivt\    & $>15.10 		     $ & $-75,125$ \\
\ovit\    & $ 14.75 \pm 0.05	     $ & $-75,135$ \\
\mgiit\   & $ 13.12 \pm 0.04	     $ & $-35,75$ \\
\Siiit\   & $ 13.52 \pm 0.08	     $ & $-35,75$ \\
\sivt\    & $ 13.58 \pm 0.04	     $ & $-75,110$ \\
\svt\     & $ 12.94 \pm 0.06	     $ & $-75,110$ \\
\svit\    & $ 13.22 \pm 0.06	     $ & $0,100$ \\
\cutinhead{PG1407+265 -- $z_{\rm abs}= 0.6828$}  
\hit\     & $16.38 \pm 0.02	     $ & $-75,80$\\
\ciit\    & $<12.75		     $ & $-75,80$\\
\ciiit\   & $\ge 13.97 \pm 0.04	     $ & $-75,80$\\
\oiit\    & $<13.18		     $ & $-75,80$\\
\oiiit\   & $\ge 14.65 \pm 0.01	     $ & $-75,80$\\
\oivt\    & $\ge 14.87 \pm 0.03	     $ & $-75,80$\\
\ovit\    & $14.07 \pm 0.03	     $ & $-75,80$\\
\mgiit\   & $<11.47		     $ & $-75,80$\\
\Siiit\   & $13.11 \pm 0.04	     $ & $-75,80$\\
\sivt\    & $13.17 \,^{+0.15}_{-0.24}$ & $-75,80$\\
\svt\     & $12.80 \pm 0.05	     $ & $-75,80$\\
\svit\    & $12.73 \,^{+0.10}_{-0.13}$ & $-75,80$\\
\cutinhead{PG1522+101 -- $z_{\rm abs}= 0.5185$}  
\hit\     & $ 16.22 \pm 0.02	     $ & $-60,50$\\
\ciit\    & $ 13.37 \pm 0.05	     $ & $-20,50$\\
\ciiit\   & $ 13.23 \pm 0.04	     $ & $-30,70$\\
\oiit\    & $ 14.08 \pm 0.04	     $ & $-30,30$\\
\oiiit\   & $ 13.73 \pm 0.06	     $ & $-20,60$\\
\oivt\    & $<13.49 		     $ & $-30,30$\\
\ovit\    & $<13.34 		     $ & $-30,30$\\
\mgiit\   & $ 12.28 \pm 0.01	     $ & $-25,25$\\
\cutinhead{PG1522+101 -- $z_{\rm abs}= 0.7289$} 
\hit\     & $ 16.66 \pm 0.05	     $ & $-60,60$\\
\ciit\    & $ <12.80		     $ & $-60,60$\\
\ciiit\   & $ 13.40 \pm 0.15	     $ & $-30,60$\\
\oiit\    & $<13.42 		     $ & $-60,60$\\
\oiiit\   & $ 13.88 \pm 0.05	     $ & $-50,60$\\
\ovit\    & $ 13.92 \pm 0.07	     $ & $-30,40$\\
\mgiit\   & $<11.49 		     $ & $-60,60$\\
\Siiit\   & $<12.44		     $ & $-60,60$\\
\sivt\    & $<12.70		     $ & $-60,60$\\
\svt\     & $<12.28		     $ & $-60,60$\\
\cutinhead{PKS0552-640 -- $z_{\rm abs}= 0.3415$} 
\hit\     & $ 16.90 \pm 0.08	     $ & $-80,80$\\
\ciiit\   & $ 12.92 \pm 0.07	     $ & $-30,60$\\
\ovit\    & $<13.24 		     $ & $-60,60$\\
\mgiit\   & $<12.22		     $ & $-60,60$\\
\siiit\   & $<12.27 		     $ & $-60,60$\\
\siiiit\  & $ 12.41 \pm 0.09	     $ & $-60,60$\\
\cutinhead{PKS0637-752 -- $z_{\rm abs}= 0.4685$} 
\hit\     & $ 16.48 \pm 0.04	     $ & $-60,60$\\
\ciit\    & $ 13.72 \pm 0.03	     $ & $-50,50$\\
\ciiit\   & $\ge 13.82 \pm 0.03	     $ & $-90,60$\\
\oiit\    & $ 14.03 \pm 0.04	     $ & $-50,50$\\
\oiiit\   & $\ge 14.45 \pm 0.04:     $ & $-90,60$\\
\oivt\    & $ 14.32 \pm 0.05	     $ & $-90,60$\\
\ovit\    & $ 13.87 \pm 0.05	     $ & $-80,60$\\
\mgiit\   & $ 12.82 \pm 0.04	     $ & $-60,60$\\
\sivt\    & $<13.40		     $ & $-90,60$\\
\svt\     & $ 12.61 \pm 0.08	     $ & $-90,60$\\
\cutinhead{HE0439-5254 -- $z_{\rm abs}= 0.6153$}  
\hit\     & $ 16.28 \pm 0.04	     $ & $-45,70$ \\
\ciit\    & $ 13.93 \pm 0.03	     $ & $-45,50$ \\
\ciiit\   & $>14.49 		     $ & $-81,89$ \\
\oiit\    & $ 14.19 \pm 0.03	     $ & $-45,50$ \\
\oiiit\   & $>14.95 :		     $ & $-112,50$ \\
\oivt\    & $>15.27 :		     $ & $-112,70$ \\
\ovit\    & $ 14.84 \pm 0.03	     $ & $-150,210$ \\
\Siiit\   & $ 13.68 \pm 0.04	     $ & $-45,60$ \\
\sivt\    & $ 13.69 \pm 0.06	     $ & $-45,60$ \\
\svt\     & $ 13.24 \pm 0.04	     $ & $-45,60$ \\
\svit\    & $ 13.13 \pm 0.10	     $ & $-45,60$ \\
\cutinhead{SBS1122+594 -- $z_{\rm abs}= 0.5574$}  
\hit\     & $ 16.24 \pm 0.03	     	$ & $-60,40$ \\ 
\ciit\    & $< 13.60			$ & $-40,40$ \\
\ciiit\   & $ 12.99 \pm 0.08	     	$ & $-40,40$ \\
\oiit\    & $ 13.61 \,^{+0.10}_{-0.15}	$ & $-40,40$ \\
\oiiit\   & $ 13.90 \pm 0.08		$ & $-40,40$ \\
\oivt\    & $<13.43			$ & $-40,50$ \\
\ovit\    & $13.99 \pm 0.12		$ & $-40,40$ \\
\sivt\    & $<13.49			$ & $-40,40$ \\
\cutinhead{J1419+4207 -- $z_{\rm abs}= 0.2889$}  
\hit\     & $ 16.40 \pm 0.07		$ & $-80,90$  \\
\ciit\    & $13.92 \pm 0.13		$ & $-80,90$  \\
\ciiit\   & $\ge 14.00 \pm 0.08  	$ & $-80,90$  \\
\ovit\    & $14.54 \pm 0.04		$ & $-80,90$  \\
\mgiit\   & $12.58 \pm 0.02		$ & $-30,40$  \\
\siiit\   & $<12.96			$ & $-80,90$  \\
\siiiit\  & $13.33 \pm 0.08		$ & $-80,90$  \\
\cutinhead{J1419+4207 -- $z_{\rm abs}= 0.4256$}  
\hit\     & $ 16.17 \pm 0.06	     	$ & $-70,70$  \\
\ciit\    & $<13.32			$ & $-50,60$  \\
\ciiit\   & $13.33 \pm 0.10		$ & $-50,60$  \\
\ovit\    & $13.76 \pm 0.11	     	$ & $-80,20$  \\
\mgiit\   & $<11.78			$ & $-50,60$  \\
\siiiit\  & $12.85 \pm 0.13 	     	$ & $-50,60$  \\
\cutinhead{J1419+4207 -- $z_{\rm abs}= 0.5346$}  
\hit\     & $ 16.34 \,^{+0.23}_{-0.12}		$ & $-40,50$\\
\ciit\    & $ \le 13.64 \,^{+0.10}_{-0.12}	$ & $-40,50$\\
\ciiit\   & $13.56 \pm 0.09			$ & $-45,50$\\
\oiit\    & $13.97 \pm 0.11 	     		$ & $-40,50$\\
\oiiit\   & $<14.04				$ & $-40,50$\\
\ovit\    & $<13.90				$ & $-40,50$\\
\mgiit\   & $ 12.64 \pm 0.02			$ & $-34,30$\\
\cutinhead{J1435+3604 -- $z_{\rm abs}= 0.3730$}  
\hit\     & $ 16.65 \pm 0.07	    	$ & $-70,60$\\
\ciiit\   & $13.15 \pm 0.09		$ & $-70,60$\\
\oiit\    & $<14.02			$ & $-70,60$\\
\oiiit\   & $<14.31			$ & $-70,60$\\
\ovit\    & $<13.51			$ & $-70,60$\\
\mgiit\   & $ 11.56 \pm 0.12		$ & $-20,20$\\
\siiit\   & $<12.54			$ & $-70,60$\\
\siiiit\  & $<12.58 	     		$ & $-70,60$\\
\cutinhead{J1435+3604 -- $z_{\rm abs}= 0.3878$}  
\hit\     & $ 16.18 \pm 0.05		$& $-70,60$ \\
\ciit\    & $<13.64			$& $-70,60$ \\
\ciiit\   & $\ge 14.01 \pm 0.07		$& $-130,100$ \\
\oiit\    & $<13.99			$& $-70,60$ \\
\oiiit\   & $\ge14.45	\pm 0.13	$& $-110,30$ \\
\ovit\    & $14.24 \pm 0.09		$& $-110,30$ \\
\mgiit\   & $<11.66			$& $-70,60$	 \\
\siiit\   & $<12.69			$& $-70,60$ \\
\siiiit\  & $<12.71 	     		$& $-70,60$ \\   
\cutinhead{J1619+3342 -- $z_{\rm abs}= 0.2694$}  
\hit\     & $ 16.48 \pm 0.05		$& $-60,70$ \\
\ciit\    & $<13.37		 	$& $-60,60$ \\
\ciiit\   & $13.65 \pm 0.05  		$& $-60,60$ \\
\ovit\    & $<13.93		     	$& $-60,60$ \\
\mgiit\   & $11.75 \pm 0.12		$& $-40,40$\\ 
\siiit\   & $<12.54			$& $-60,60$ \\
\siiiit\  & $13.02 \pm 0.03		$& $-60,60$ \\
\siivt\   & $<12.90 	     		$& $-60,60$ \\
\cutinhead{PG1216+069 -- $z_{\rm abs}= 0.2823$$^c$}  
\hit\     & $ 16.40 \pm 0.07		$ & $-60,70$  \\
\ciit\    & $<13.15 			$ & $-40,40$  \\
\ciiit\   & $>13.63	 		$ & $-35,40$  \\
\niit\    & $<13.03 			$ & $-25,25$  \\
\niiit\   & $<13.03 			$ & $-25,25$  \\
\ovit\    & $13.60 \pm 0.05		$ & $-25,50$  \\
\siiit\   & $<11.85			$ & $-25,25$  \\
\siiiit\  & $\ge 13.10 \pm 0.09		$ & $-25,25$  
\enddata
\tablecomments{$^a$The reader should refer to the Appendix for a description of each of these absorbers (in particular for the complete information about the velocity structure; the velocity interval used for the integration of the $N_a(v)$ profile for each species is listed in the last column. $^b$\civ, \siiv, and \alii\ are from STIS E230M data. $^c$The analysis of PG1216+069 is based on both COS and STIS E140M data; see Appendix for details. The symbol ``$>$" indicates that the absorption profile is saturated. The symbol ``$\ge$" indicates that the absorption profile might be saturated.  The symbol ``$<$" indicates that there is no detection at the $3\sigma$ level. The symbol ``$\le $" indicates that the profile might be contaminated by an unrelated absorber. A value followed by ``:'' indicates that it is uncertain (e.g., a line is saturated, but could also be contaminated).}
\end{deluxetable}

We employed the apparent optical depth (AOD) method described by \citet{savage91} to estimate the column density of the metal ions. The absorption profiles are converted into apparent column densities per unit velocity, $N_a(v) = 3.768\times 10^{14} \ln[F_c(v)/F_{\rm obs}(v)]/(f\lambda)$ cm$^{-2}$\,(\km)$^{-1}$, where $F_c(v)$ and $F_{\rm obs}(v)$ are the modeled continuum and observed fluxes as a function of velocity, respectively, $f$ is the oscillator strength of the transition and $\lambda$ is the wavelength in \AA. The atomic parameters are for the UV transitions from \citet{morton03} and for the EUV transitions from \citet{verner94}.  The total column density was obtained by integrating over the absorption profile  $N_a = \int_{v_1}^{v_2} N_a(v)dv$. When no detection was observed, we estimated a 3$\sigma$ upper limit following the method described by \citet{lehner08}. 

Figs.~\ref{f-pg1522a} to \ref{f-pg1216} in the Appendix show that often several transitions of the same ion are available to estimate the column density (e.g., for \oii, \oiii, \ovi, \Siii, respectively). This allows us to assess if the lines are saturated and/or contaminated. For the \mgii\ doublet, the only evidence of saturation is for the $z=0.6865$ absorber toward PG1338+416, but even in this case the saturation is mild since the difference in the $N_a$ values between the strong and weak transitions is small \citep[implying a correction of only 0.10 dex following the method described in][]{savage91}. Other doublets (e.g., \ovi) or lines with multiple transitions (e.g., \oii) generally give similar integrated column densities within $1\sigma$ error, except when there is evidence for contamination. For assessing the line saturation in species with a single transition (e.g., \siiii, \ciii), we use the information from the profiles of other species with multiple transitions that trace similar type of gas (i.e., where the temperature and turbulence should be similar). In Table~\ref{t-data}, we summarize our adopted column densities and the velocity interval for the integration of the profiles. When more than one transition is available (see Figs.~\ref{f-pg1522a} to \ref{f-pg1216}), the adopted column density represents the average value between the different estimates when there is no evidence of saturation or contamination (i.e., the $N_a$ values agree within 1$\sigma$ error and there is not a systematic increase of the $N_a$ with decreasing $f\lambda$).

\subsubsection{\hi\ Column Densities}
We estimate the \hi\ column density from the weak transitions of the Lyman series down to $\lambda$916 (typically, the Lyman series lines become too blended and/or the continuum placement too uncertain at $\lambda<916$ \AA\ to estimate reliable \nhi\ or $W_\lambda$) and from the flux decrement at the Lyman limit.  In the latter method, by estimating the optical depth $\tau_{\rm LL}$ at the Lyman limit ($\lambda_{\rm LL} = 912$ \AA), we can directly deduce \nhi\ since $N_{\rm H\,I} = \tau_{\rm LL}/\sigma_{\rm LL}$ where $\sigma_{\rm LL} = 6.3\times 10^{-18}$ cm$^2 (\lambda/912)^3$ is the absorption cross section of the hydrogen atom at the Lyman limit \citep{spitzer78}. To define the continua of the QSO before and after the break, we adopted the method described by \citet{ribaudo11a} by fitting a composite QSO spectrum developed by \citet{zheng97}. Depending on the complexity of the QSO continua and interstellar spectra (in particular near the Milky Way \lya\ absorption), this method could be reliably used to determine \nhi\ for absorbers with  $\log N_{\rm H\,I} \ga 16.3$ or $ \tau_{\rm LL}\ga 0.15$ for the typical S/N levels in our sample.

To estimate the \hi\ column density from the Lyman series, we employed the AOD method described above. We use all the available weak Lyman transitions that are ot contaminated and for which the continuum can be reliably modeled.  As for the metal lines, we use several \hi\ transitions to assess the saturation level, if any.  As we are solely interested in the components with  $\log N_{\rm H\,I} \ga 16.2$, we only integrate the $N_a(v)$ profiles over the velocity interval that defines these components. For example, for the $z=0.6153$ absorber toward HE0439-5254 (Fig.~\ref{f-he0439}), \lyb\ has several components at $v > 75$ \km\ and $v<-75$ \km. The strongest \hi\ component for which  $\log N_{\rm H\,I} \ga 16.2$ is observed over $-75 <v < 75$ \km, and its column can be estimated using only the weaker transitions (e.g., \hi\ $\lambda$$\lambda$923, 919).  This ensures the metallicity estimates are accurate, since the low ions are generally seen in the strongest \hi\ component only.

We also use a third method where we measure the equivalent width ($W_\lambda$) for all the Lyman transitions that do not contain any evidence of widely displaced components to estimate \nhi\ from a single-component curve-of-growth (COG).  For example, for the absorber toward PHL1377, we use both strong and weak lines for the COG, while for the absorber toward HE0439-5254, we only use the weak transitions as there are several components in the stronger ones both blueshifted and redshifted with respect to the main principal components. The COG method used a $\chi^2$ minimization and $\chi^2$ error derivation approach outlined by \citet{savage90}. The program solves for $\log N$ and $b$ independently in estimating the errors. 

In the Appendix, we summarize for each absorber which methods were used to determine \nhi. In general, there is a good agreement between these methods when we were able to use two or three of them. The three methods explore different parameters and continuum models, allowing us to reliably estimate  \nhi\ and its error. When more than one method was used, we averaged the values and propagated the errors accordingly.

\subsection{Determination of the Metallicity of the LLS}\label{s-phot}

Large ionization corrections are required to determine the metallicity from the comparison of \hi\ with singly or doubly ionized metal species, except where \oi\ is observed, which only happened for the three strongest LLS in the sample (with $\log N_{\rm H\,I}>18$). As displayed in Figs.~\ref{f-pg1522a} to \ref{f-pg1216} (see also references given in Table~\ref{t-sum}), the ionization structure can be quite complicated in some cases, with the detection of species in different ionizing stages and sometimes with  velocity shifts between the different ions. However, in other cases, it is much simpler with very few ions detected, and, as we will see below, these generally correspond to low metallicity absorbers. While it is evident that several ionization processes are at play for some of the absorbers (see below), our main aim here is not to understand the detailed origins of the physical processes that give rise to the observed ionization and velocity structures (we will explore this further in the future). Instead our goal is to determine the metallicity of the weakly ionized gas that is observed over the same velocity interval as the strongest \hi\ component with  $\log N_{\rm H\,I} \ga 16.2$. This requires we concentrate on modeling the ionization of the singly and doubly ionized species.

\begin{figure*}[tbp]
\epsscale{1} 
\plottwo{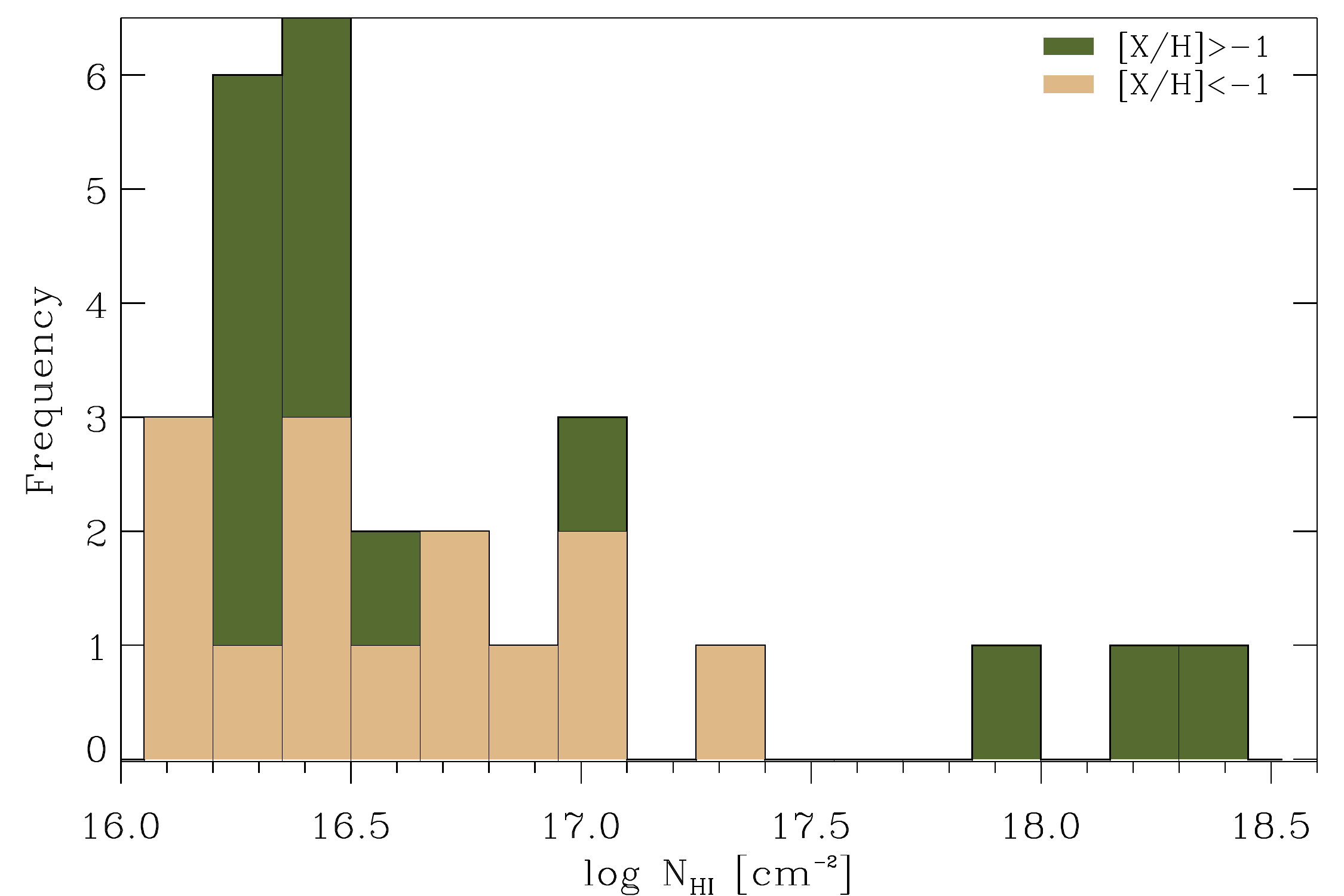}{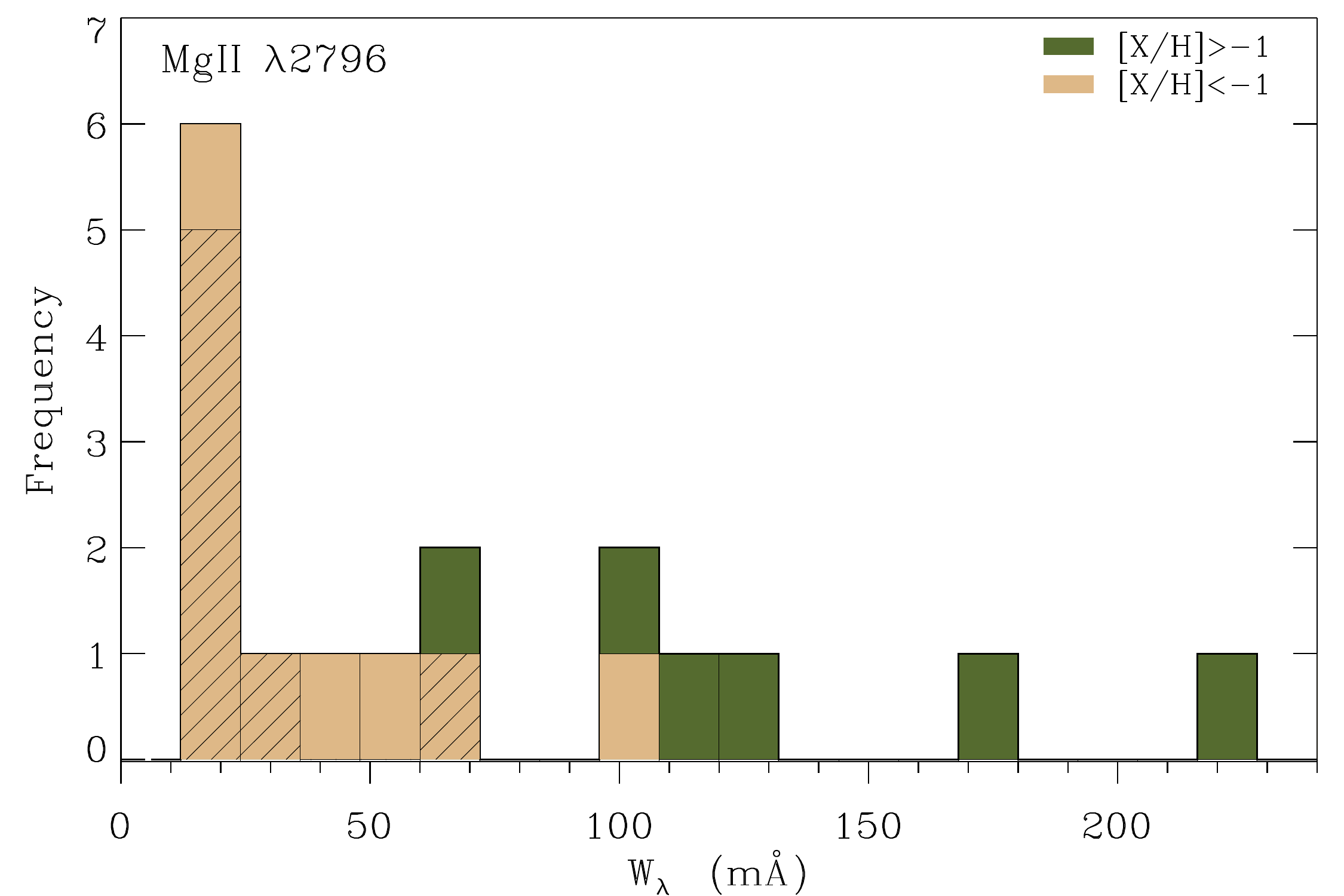}  
  \caption{Distributions of \nhi\ ({\it left}) and the \mgiit\ $\lambda$2796 rest-frame equivalent width ({\it right}) for the metal-poor and more metal-rich absorbers. Only absorbers from the \hit-selected LLS sample are considered. The hashed histograms highlight values that are upper limits, i.e., the strong transition of the \mgiit\ doublet is not detected. Visually and statistically, the differences between the metal-poor and more metal-rich samples are not significantly different for the \nhi\ distribution, but they are for the $W_\lambda(\mbox{\mgiit})$ distribution.
\label{f-disttest}}
\end{figure*}

\begin{figure}[tbp]
\epsscale{1.1} 
\plotone{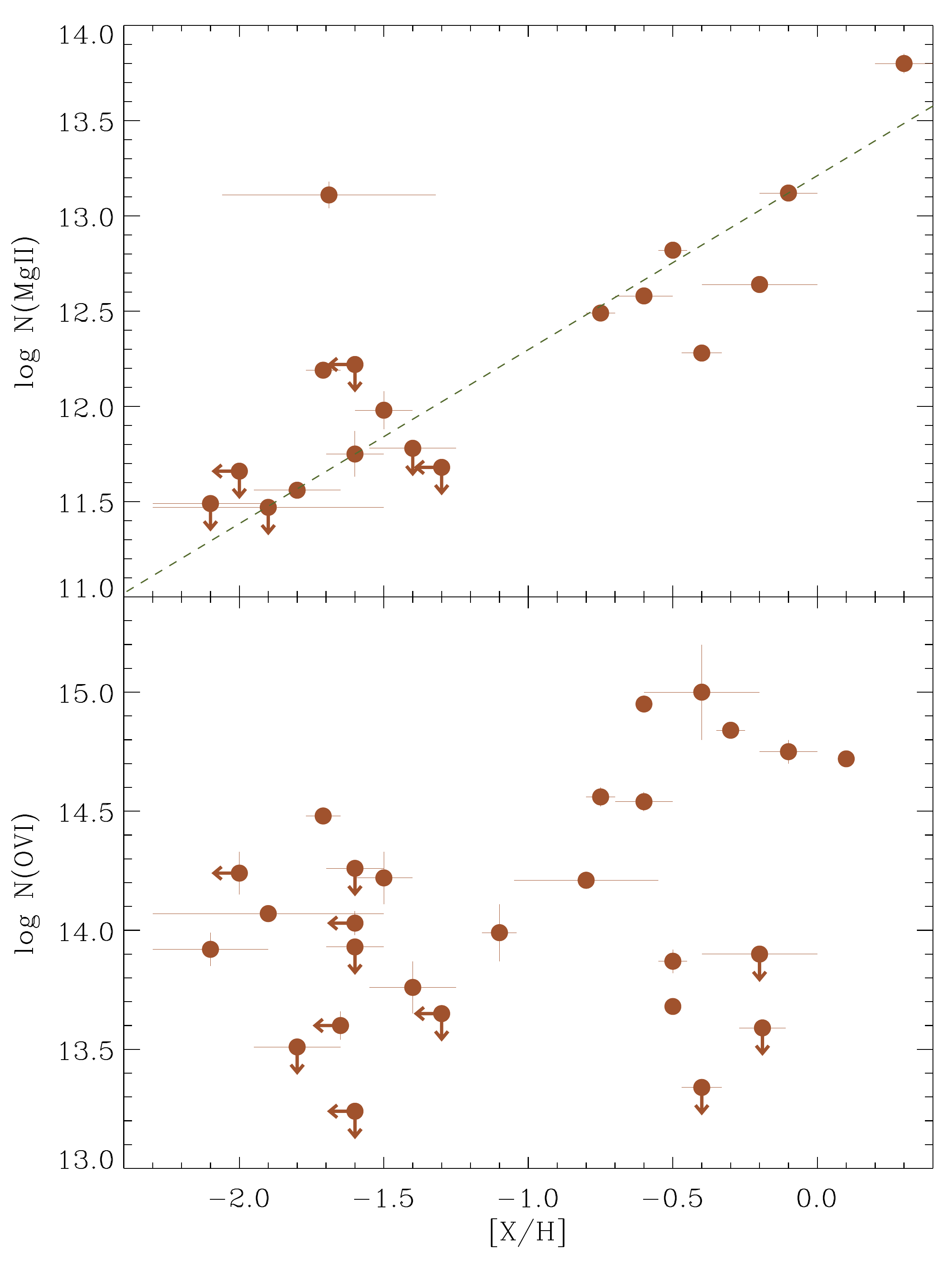}  
  \caption{The column densities of \mgiit\ and \ovit\ versus the metallicity of the cool, photoionized gas for the LLS in our sample.  The $N_{\rm O\,VI}$ versus ${[\rm X/H]}$ is essentially a scatter diagram, but there is a strong correlation between $N_{\rm Mg\,II}$ and the metallicity. The dotted line show a linear fit to the data with error bars using a ``robust" least absolute deviation method ($\log N_{\rm Mg\,II}= 0.9\, {[\rm X/H]} + 13.2$). The departing point at low metallicity is the \mgiit-selected LLS with a large \nhi\ (18.3 dex) compared to the other LLS (16.1--17.1 dex). 
 \label{f-mg2met}}
\end{figure}

\subsubsection{Photoionization Models: Methodology}\label{s-photmet}

To determine the metallicity and physical conditions of the strong \hi\ component in each absorber, we model its ionization using Cloudy \citep[version c10, last described by][]{ferland98}.  We assume the gas is photoionized, modeling it as a uniform slab in thermal and ionization equilibrium. We illuminate the slab with the  Haardt-Madau  background radiation field from quasars and galaxies (HM05, as implemented within Cloudy). We describe in detail two examples of Cloudy photoionization models in the Appendix. In summary, for each absorber, we vary the ionization parameter, $U =n_\gamma/n_{\rm H}=$\,H ionizing photon density/total hydrogen number density (neutral + ionized), and the metallicity, $[{\rm X/H}] \equiv \log N_{X}/N_{\rm H} - \log ({\rm X/H})_\sun $, to search for models that are consistent with the constraints set by the column densities determined from the observations. We assume solar relative heavy element abundances from \citet{asplund09}. We do not include  a priori the effects of dust or nucleosynthesis on the relative abundances, although we consider these possibilities a posteriori. The preferred species to constrain the ionization parameter are those where their velocity structures follow the \hi\ velocity profile the best. The velocity structure of the singly ionized species (e.g., \oii, \cii, \siii, and \mgii) match well the \hi\ velocity structure, and, in some cases, doubly ionized species can also be used reliably based on their similar velocity profiles compared to the low ions and \hi. The excellent correspondence between the peak optical depths of the \hi\ and the strongest low-ion lines implies the \hi\ and low ions must be co-spatial. This gives us confidence that the metallicity distribution resulting from this modeling is not a simple consequence of unrelated absorption. However, that is why we do not blindly compare the total column densities: various velocity components may have different metallicities, and While the total \hit\ column density may not change by more than $\sim$0.2--0.3 dex compared to the \nhi\ value in the strongest component, that may not be the case for metal ions such \ciiit\ and \oiiit, where most of the column density may in fact be associated with the low \nhi\ component (see, e.g., Figs.~\ref{f-pg1338b} and \ref{f-sbs1122}). When singly and doubly ionized species can be used, ionic ratios of the same element can be used to determine $U$ and [X/H], alleviating uncertainties from possible non-solar relative abundances. When singly ionized species are not detected, the $3\sigma$ upper limit on $N_{\rm Mg\,II}$ provides a stringent constraint on the metallicity (see next), and we also use doubly ionized species (\ciii, \oiii, \siiii) if their velocity profiles follow well the profile of the strongest \hi\ component. The metallicities of the LLS were determined using $\alpha$ elements (O, Si, Mg, S).

For the LLS in our sample, low ionization species like \mgii\ are powerful metallicity indicators, as we illustrate in the right panel of  Fig.~\ref{f-disttest} and the top panel of Fig.~\ref{f-mg2met}. In  Fig.~\ref{f-disttest}, we show the distribution of the rest-frame equivalent width, $W_{\lambda}$, of  \mgii\ $\lambda$2796 color coded according to the metallicity of the absorbers: metal-poor absorbers with $[{\rm X/H}<-1]$ have a relatively weak \mgii\ $\lambda$2796 absorption ($W_\lambda < 100$ m\AA, with $W_\lambda \la 30$ m\AA\ for more than half of the weak \mgii\ absorbers), while more metal-rich absorbers have typically strong  \mgii\ absorption ($W_\lambda \ga 100$ m\AA, this figure does not include the extremely strong \mgii\ absorber -- $W_\lambda \sim 1$ \AA\ -- with supersolar metallicity toward PG1206+459; see Table~\ref{t-sum} and \citealt{tripp11}). Fig.~\ref{f-mg2met} demonstrates further a strong correlation between [X/H] and  $N_{\rm Mg\,II}$. This strong metallicity dependency is a result both of the limited range of \nhi\ (mostly $16.2 \la \log N_{\rm H\,I} \la 17$) and  narrow  $\log U$ interval (see Appendix) sampled by our absorbers.  The column densities of the doubly ionized species (like \ciii\ or \siiii) can, however, vary by a large factor over the constrained  $U$ interval.

The LLS that are already in the literature were modeled following a similar method in the original publications (see Table~\ref{t-sum} for the references). For a very few absorbers (especially pre-2005), the HM96 (instead of the HM05) UV background \citep{haardt96} was adopted,  and sometimes the UV background from QSOs only was favored. This may introduce some differences in the metallicity \citep[e.g.,][]{howk09}, but no more than a factor about 2--3, and, specifically for the strong \hi\ absorbers, J. Werk et al. 2013 (in preparation) show that the impact of changing the UV background is quite small on the derived metallicities. Further, as this concerns only four systems, it would  not affect our conclusions anyway. Because most studies also use principally Si and Mg to constrain the metallicities of the photoionized gas, the change in the solar abundances over the last 8--10 years has little effect  since the changes are minor for these elements. 

In the Appendix, we discuss the results of our photoionization modeling for each absorber (and see the references given in Table~\ref{t-sum} for the absorbers that are already in the literature). The adopted metallicities are summarized in Table~\ref{t-sum}. The errors on the metallicity and $U$ reflect the range of values allowed by the $1\sigma$ uncertainties on the observed column densities. They do not include errors from the limitation of the models used to estimate the ionization corrections.

\subsubsection{Photoionization Models: Systematic Uncertainties}\label{s-moreion}

For all the LLS in our sample, the fraction of H that is ionized exceeds 90\% (see \S\ref{s-llsprop}), and the metallicity estimates are therefore photoionization model dependent. We now assess whether we can identify subtle ionization effects that may produce spurious metallicity systematics and whether trends can be observed between the observed absorber properties  and the metallicity that might suggest unidentified systematics. As we discuss in \S\ref{s-metlls}, we find that the metallicity distribution of the LLS is bimodal, and it is important to demonstrate that this result is robust and not an artifact from the large ionization corrections required to estimate the metallicity. 

Since the absorbers straddle the boundary of the optically thin and thick regimes, there might be some dependence of the results with \nhi. In Fig.~\ref{f-disttest} (left panel), we show the \nhi\ distribution for two metallicity intervals, larger or smaller than 10\% solar (see \S\ref{s-metres} for the justification of these two metallicity intervals). Both visually and statistically, the differences between the metal-poor and more metal-rich samples are not significant. There is no dependence of our derived metallicities on \nhi\ (see also the Appendix).  

In order to reduce potential systematics from different ionization processes, we (and others, see references in Table~\ref{t-sum}) typically do not rely on higher ionization stages (e.g., comparing \oii\ and \oiv) to constrain the ionization models. High ions  can be produced both by photoionization and collisional ionization processes and disentangling these processes can be quite complicated and can lead to erroneous metallicity estimates. For absorbers with $16.2 \la \log N_{\rm H\,I} \la 18.5$, the photoionization models that match the properties of the low ions systematically underestimate the column densities of the high ions (see Appendix and references in Table~\ref{t-sum}). The interpretation of this discrepancy is that the low and high ions are not cospatial and that other physical processes are at play. The fact that we do not rely on the high ions to constrain our models also has the benefit of reducing the uncertainties from leaking photons from galaxies near the absorbers: for the Milky Way, \citet{fox05} showed that \cii/\ciii\ (\siii/\siiii) for gas at $\sim 50$ kpc from the Milky Way  is almost negligibly affected if  the gas is immersed in the extragalactic UV background alone or in the UV background plus the Galactic radiation field. Larger effects are typically observed for \cii/\civ\ (or \siii/\siiv) between the UV background alone and the UV background  plus the inclusion of the local radiation field because the Galactic field does not include many photons above the ionization edge.

\begin{figure*}[t]
\epsscale{0.9} 
\plotone{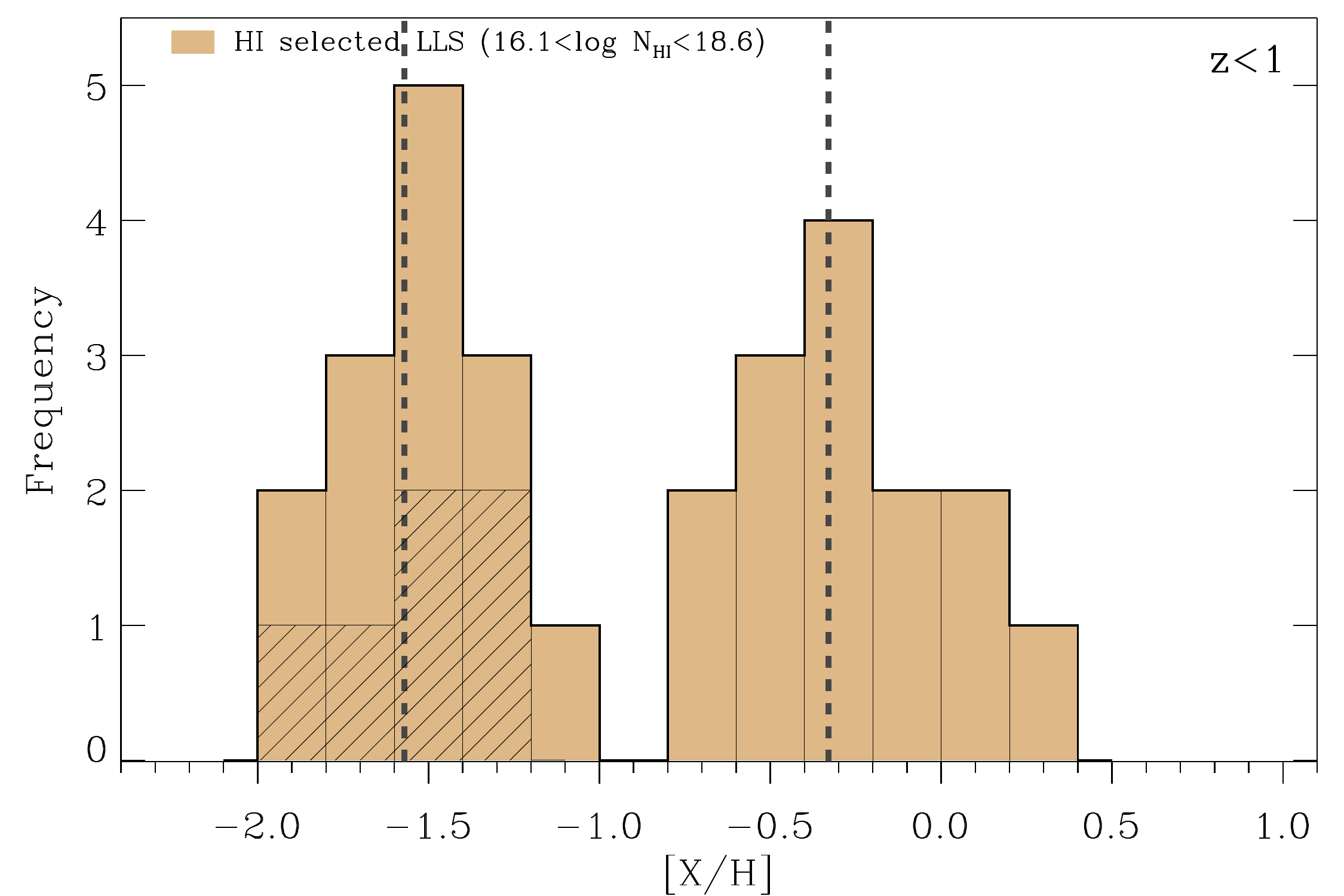}  
  \caption{Metallicity distribution function of the LLS at $z\la 1$. The hashed histograms highlight values that are upper limits. The metallicity distribution is bimodal where the peak values are shown by the vertical dotted lines. For the LLS with $[{\rm X/H}]\le -1$, the mean metallicity is an upper limit as 6/14 of the [X/H] estimates are upper limits. 
 \label{f-zdistm}}
\end{figure*}

We minimize dust or nucleosynthesis effects by comparing when possible single elements in different ionization stages (e.g., \cii/\ciii, \oii/\oiii). We also reduce the nucleosynthesis effects by considering when possible ratios of $\alpha$ elements (O, Mg, Si, S), as their relative abundances should remain close to solar values (except in the presence of strong dust depletion,which should be negligible for the LLS in our sample as discussed below). On the other hand, a large fraction of C is produced in intermediate-mass stars, while, e.g., O is thought to be synthesized almost entirely in massive stars. This leads to a time lag in the production of these elements and hence  $[{\rm O/C}]\equiv \log N_{\rm O}/N_{\rm C} - \log ({\rm O/C})_\sun$ may not be zero. Instead, in galaxies,  $[$C/O$]\sim -0.4$ to $-0.2$  when $[\rm{O/H}]<-0.5$ \citep[with a large scatter, e.g.,][and references therein]{garnett99,henry00}.  In the diffuse environment that the LLS probe, dust depletion is expected to be small or absent \citep[but see][]{menard12} for the species considered in this work (in particular, C, O, Mg, Si, S) and many of the species (C, O, Mg, Si) are depleted by a similar  amount in ``halo'' condition based on the results from gas depletion studies in the Milky Way halo \citep{savage96,jenkins09}. Furthermore, in the Galactic halo, C, O, Mg, Si are all depleted by the same amount. Our Cloudy photoionization models do not suggest strong departures from non-solar relative abundances for O, Mg, Si, S. Nevertheless it is possible that there could be a factor $\sim$2 departure from solar relative abundances, which could then affect by a similar factor our metallicity estimates. However, while this effect may shift the mean metallicity in both branches of the metallicity distribution, the effect would be too small to alter the bimodality nature of the metallicity distribution described in \S\ref{s-metlls}. For [C/$\alpha$], we refer the reader to \S\ref{s-relabu} where this ratio is discussed in more detail.

We also checked whether collisional ionization models could explain the observations as well as the photoionization models discussed in \S\ref{s-phot}. We used  the collisional ionization equilibrium and non-equilibrium models of \citet{gnat07} and shock ionization models of \citet{gnat09} to compare the predicted and observed ionic ratios of \cii/\ciii, \siii/\mgii, and \oii/\mgii. For the majority of the absorbers, the observed ionic ratios are well outside the predicted curves by these models. For 2 absorbers, only solar or super-solar metallicity shock ionization models may be adequate (for these, only limits on the ionic ratios are available). Hence the low ions are best explained by photoionization models.

We therefore conclude that there does not seem to be any hidden artifact from the ionization corrections on the metallicity distribution of the LLS discussed in \S\ref{s-metlls}. We did not find any relation between the metallicity and \nhi\ (see also appendix), which spans a factor $>200$ and 160, respectively, between the lowest and highest values. On the other hand, the photoionization models naturally explain the strength of the observed absorption profiles for the singly ionized metal species. 
  
\section{Properties of the LLS} \label{s-metres}
\subsection{Metallicity Distribution Function}\label{s-metlls}

Fig.~\ref{f-zdistm} shows the metallicity distribution function for the 28 \hi-selected LLS at $z\la 1$  summarized in Table~\ref{t-sum}. Visually the distribution is strikingly bimodal. The upper half of the sample (14/28) has a mean and dispersion $\langle {\rm [X/H]}\rangle = -0.33 \pm 0.33$ (median $-0.40$). The lower half has $\langle {\rm [X/H]}\rangle < -1.57 \pm 0.24$ (median $<-1.60$); the mean and median are strictly an upper limit because only an upper limit could be derived for the metallicity of 6/14 absorbers. The above is based on a visual inspection of the histogram distributions of the LLS metallicity. Unsurprisingly, a K-S test shows that the LLS metallicity distribution is not consistent with a normal distribution. 

To further quantify whether the LLS metallicity distribution is unimodal or bimodal, we use a gaussian mixture modeling (GMM) algorithm developed by \citet{muratov10} as well as an independent test of the bimodality based on the dip statistic. These statistical tests were used to quantify and test the bimodality of the metallicity distribution of the globular clusters in the Milky Way, which also shows two peaks in the metallicity. The dip test does not make any assumption on the distribution of the data and is a true test of modality.  The dip probability of the LLS sample being bimodal is 88\% (91\% if limits are removed from the sample); hence the distribution is very unlikely to be a skewed unimodal distribution.  We now apply the GMM algorithm to the LLS sample assuming that limits are actual values. For the description of the GMM method, its advantages and limitations, we refer the reader to the appendix in \citet{muratov10}.  The GMM algorithm rejects the unimodal distribution at a confidence level of better than 99.7\%.  A heteroscedastic split gives $\mu_1 =  -1.59 \pm 0.07$, $\sigma_1 =0.20 \pm 0.05$ and $\mu_2 =  -0.38 \pm 0.12$, $\sigma_2 =0.35 \pm 0.09$ ($\mu_i, \sigma_i$ being the mean and dispersion of the metallicity, respectively, in each branch of the bimodal distribution). A homoscedastic split gives $\mu_1 =  -1.55 \pm 0.08$ and $\mu_2 =  -0.33 \pm 0.10$, $\sigma_1 = \sigma_2 =0.28 \pm 0.05$. The separation of the means relative to their widths, $D = |\mu_1 - \mu_2|/[(\sigma^2_1+\sigma^2_2)/2]^{0.5}$, is $D = 4.5 \pm 0.8$, which is significantly larger than 2, the minimum value for a clean separation between the modes \citep{ashman94}. The  two peaks of the distribution are therefore markedly separated. In both cases, the sample is evenly split between the metal-poor and metal-rich groups. If limits are removed from the sample, the conclusions would be essentially the same, with $D= 4.5 \pm 0.9$, except that the metal-poor group would consist of about 36\% of the entire sample. Hence the GMM method and the dip test demonstrate with a high level of confidence that the (unbinned) LLS metallicity distribution is bimodal. Note that, as above, $\mu_1$ should be considered an upper limit.

\begin{figure}[tbp]
\epsscale{1.1} 
\plotone{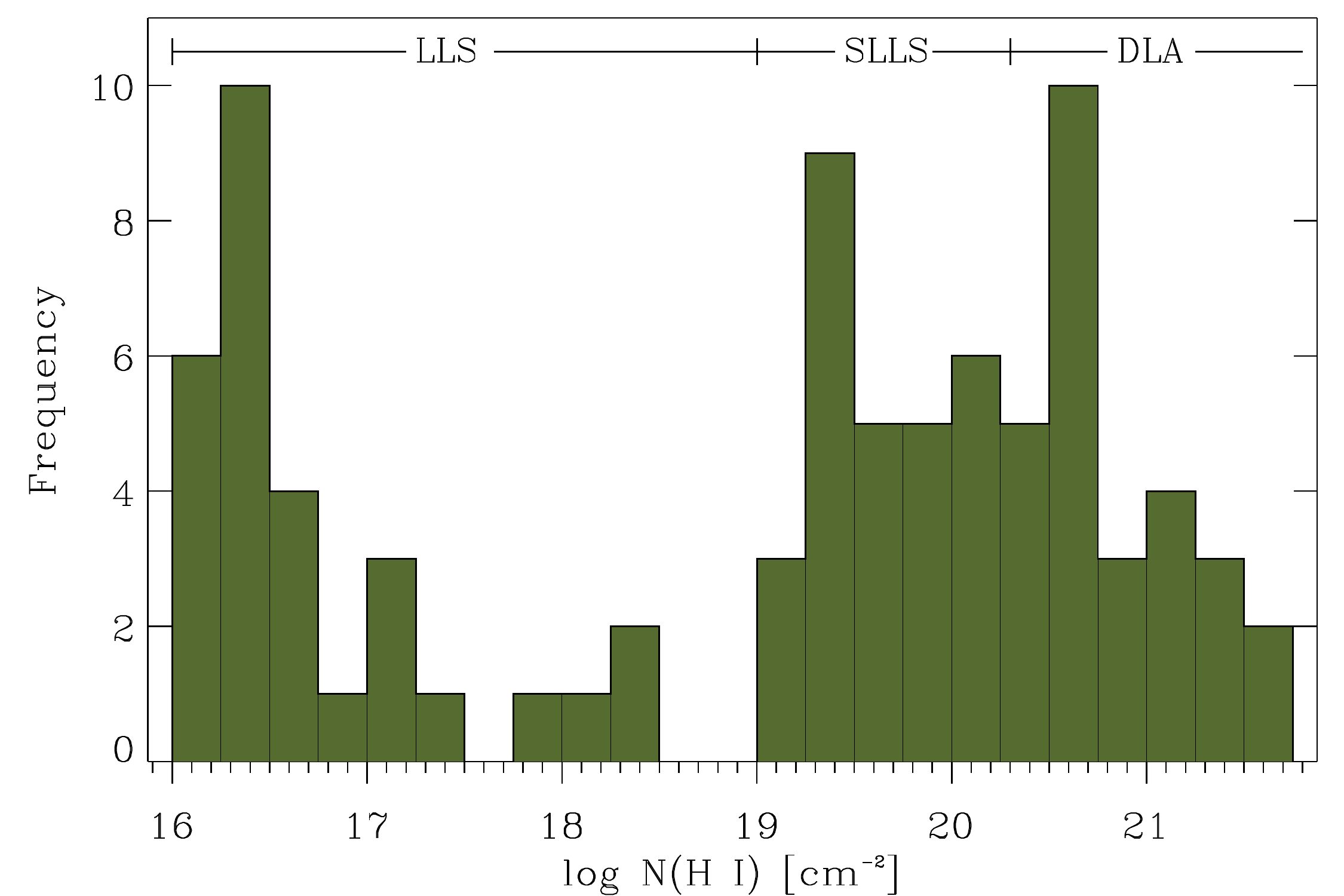}  
  \caption{Distribution of the \hit\ column density for the entire sample (LLS, SLLS, DLAs). 
 \label{f-ndist}}
\end{figure} 

\begin{figure}[tbp]
\epsscale{1.1} 
\plotone{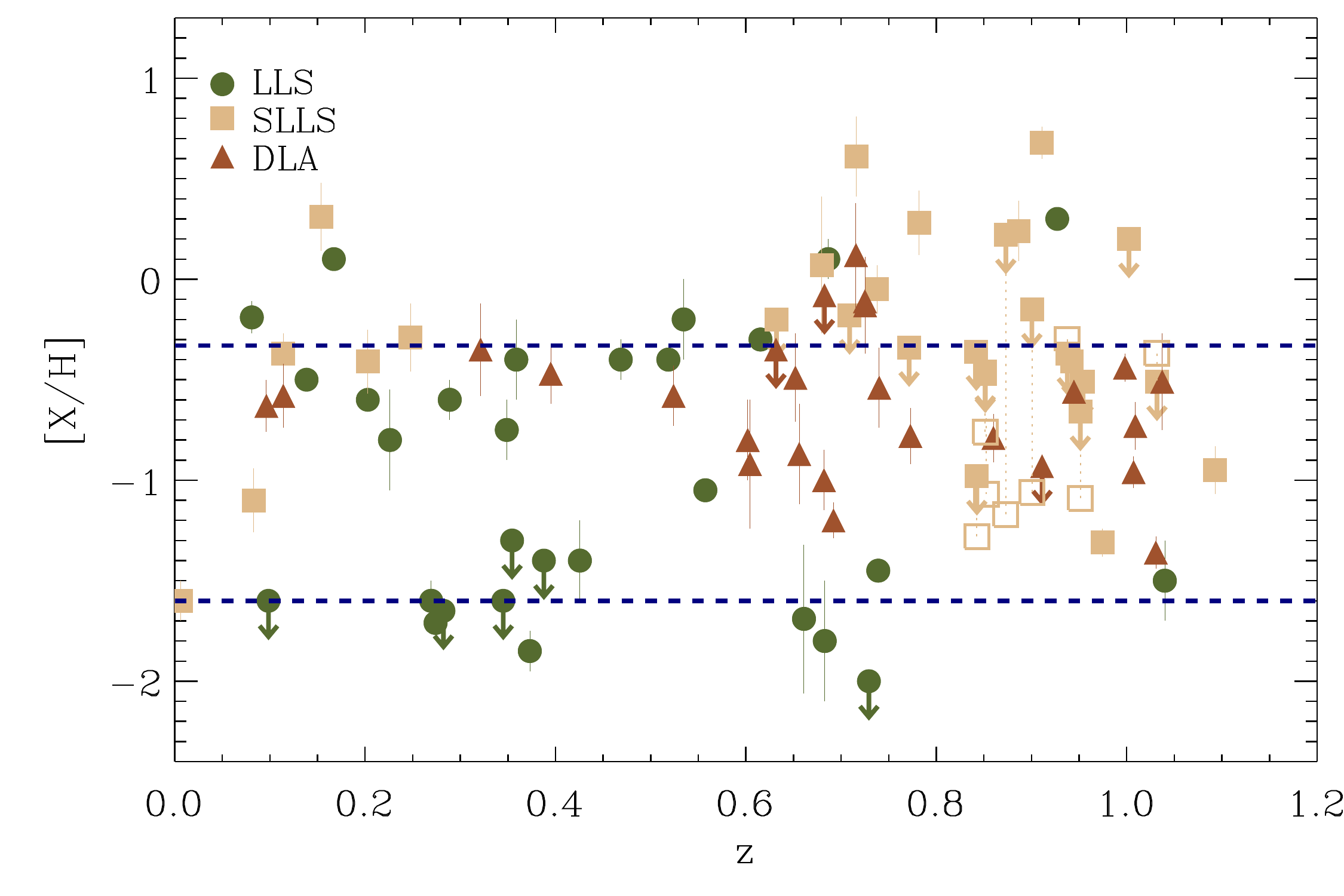}  
  \caption{The metallicity of the LLS, SLLS, and DLAs as a function of redshift. For the LLS, X is an $\alpha$ element. For the SLLS and DLAs, most of the metallicities were estimated using \zniit. The open squares show the [\feiit/\hit] values (error bars are typically around $\pm 0.1$ dex) for the absorbers where \zniit\ was not detected and $N_{\rm Fe\,II}$ was estimated. The dotted and dashed horizontal lines show the average metallicity of the metal-rich and metal-poor branches, respectively. Note that metal-poor and metal-rich LLS are found at any $z$ at $z\la 1$.
 \label{f-zmetal}}
\end{figure}

In order to compare the LLS metallicity distribution with that of the optically thick absorbers (SLLS and DLAs),  we systematically searched the literature for SLLS and DLAs at $z\la 1$ for which the metallicity has been estimated. The entire sample is summarized in Table~\ref{t-sum} where the absorbers are ordered by increasing \nhi\ values. SLLS and DLAs were largely pre-selected based on the observations of strong \mgii\ $\lambda$$\lambda$2796, 2802 absorption, with the exception of the SLLS and DLAs from \citet{tripp05} and \citet{battisti12}. The typical selection for the DLA sample used a criterion of rest equivalent width of \mgii\ $\lambda$2796 $W_\lambda\ge 0.6$ \AA\ \citep[e.g.,][]{rao06}. For this cut-off \citet{rao06} argued that the DLA sample is essentially complete and therefore free of metallicity bias. On the other hand, this might not be the case for the SLLS. We will therefore treat the metal-selected SLLS sample with that caveat in mind.  

In Fig.~\ref{f-ndist}, we show the \hi\ column density distribution for the entire sample summarized in Table~\ref{t-sum}. There are 29 LLS (28 \hi-selected, 1 \mgii-selected), 29 SLLS, and 26 DLAs. For the LLS, most of the absorbers have $16.2 \la \log N_{\rm H\,I} \la 17$, while 5/29 have higher $N_{\rm H\,I} $.  The lack of LLS with $\log N_{\rm H\,I} \ga 17.5$ ($\tau_{\rm LL}> 1$) is primarily an observational bias owing to the UV selection of the QSOs, and, secondarily due to the difficulty to reliably estimate \nhi\ at $17.5 \la \log N_{\rm H\,I} \la 18.7$.  This observational selection effect therefore affects the \hi\ distribution, but there is not an obvious bias imparted to the metallicity distribution. 

In Fig.~\ref{f-zmetal}, we show  the metallicity of the LLS, SLLS, and DLAs\footnote{The metallicity reported for the DLAs and SLLS is generally estimated using \zniit\ and \feiit\ (excluding the results from \citealt{battisti12} and \citealt{tripp05} where $\alpha$ elements and Fe were used). As Fe can be depleted significantly onto dust grains and Zn is not, we always reported the metallicities derived from Zn. Several SLLS and DLAs have, however, only upper limits quoted for Zn based on its non-detection. For these, the abundance determined from \feiit\ does not suggest extremely low metallicities as observed in the LLS, i.e., typically [\feiit/\hit$]\ga -1.3$ \citep[see][]{meiring09}. The studies of [S/Zn] as a function of the metallicity in stars (non-LTE analysis) and DLAs show no trend with the metallicity \citep{nissen07,rafelski12}. There might be a scatter of about $\pm 0.2$ dex, although within $1\sigma$ error, [S/Zn$]\simeq 0$, i.e., we can directly compare the metallicity distributions derived from Zn and $\alpha$ elements. } as a function of redshift over $0<z\la 1$  (i.e., over the last 8 Gyr of the cosmic time). There is no sign of evolution within this timeframe (see Fig.~\ref{f-zmetal}). Previous authors have shown a general increase of the metallicity of the DLAs from $z>5$ to $z\sim 0$, but the metallicity appears to plateau from $z\sim 1$ to $z\sim 0$ as observed in Fig.~\ref{f-zmetal}  \citep[e.g.,][]{prochaska03,battisti12,rafelski12}. Low and high metallicity LLS and SLLS are found at any $z$. The metallicity distribution of the DLAs is narrower, with a central value near the present-day SMC metallicity (see next). This suggests the DLAs and at least part of the absorbers with $\log N_{\rm H\,I}\la 20$ have different origins. For the SLLS, the metallicity distribution is broader than for the DLAs, with most of the values in the range $-0.6 \le {\rm [X/H]_{SLLS}} \le +0.4$.  However, the distribution might be biased toward the more metal-rich absorbers owing to the selection methodology for these systems.

\begin{figure}[tbp]
\epsscale{1.} 
\plotone{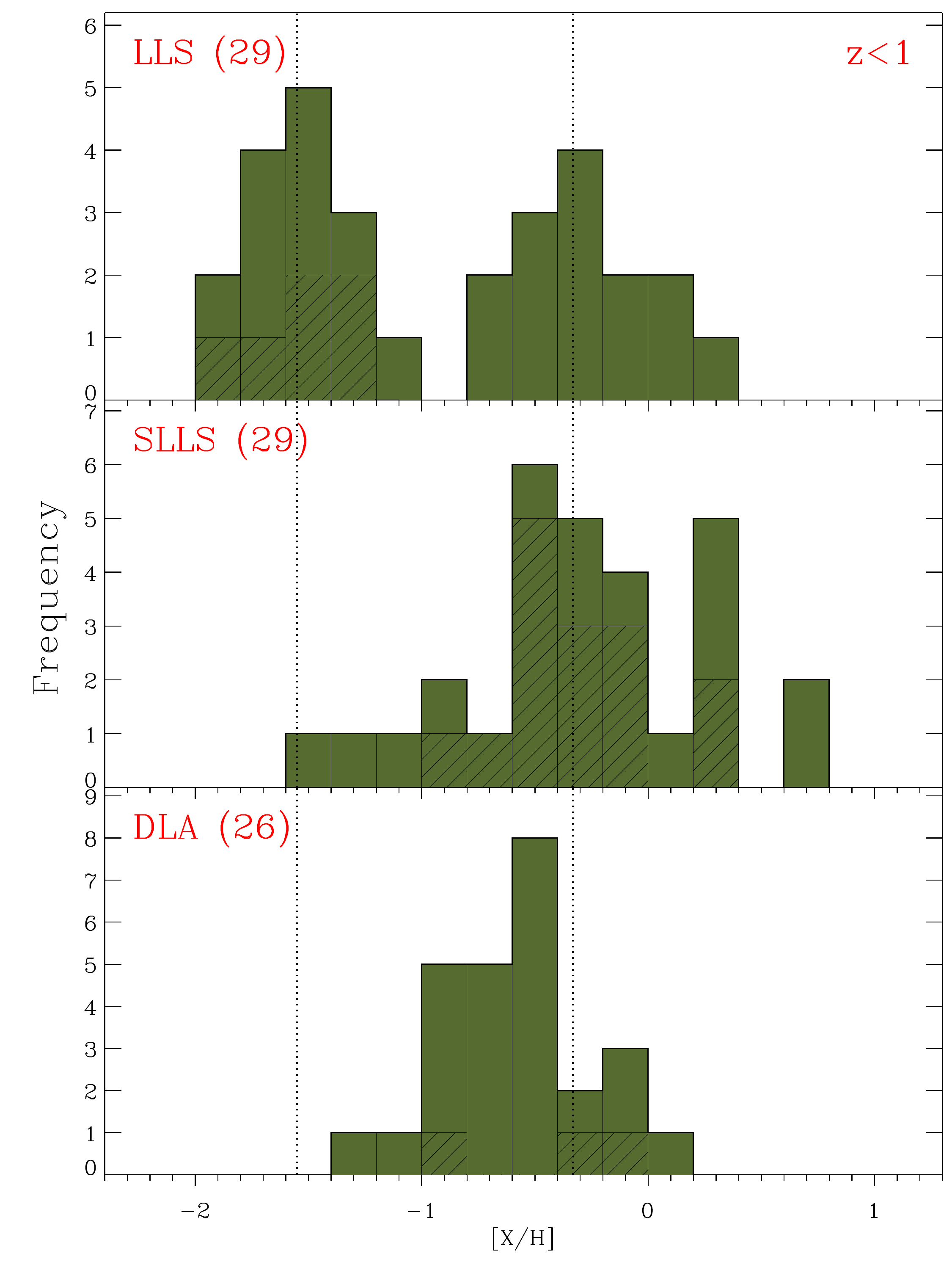}  
  \caption{Metallicity distribution function of the LLS, SLLS, and DLAs at $z\la 1$. The hashed histograms highlight values that are upper limits. The SLLS sample was mostly pre-selected based on the presence of strong \mgiit\ absorption and therefore might be biased toward the higher metallicity. Most of the DLAs were also pre-selected based on strong \mgiit\ absorption, but this should not bias as much the metallicity distribution function  owing to the much larger overall column density. The mean of each peak of the LLS metallicity distribution is shown by the vertical dotted lines. For the LLS with $[{\rm X/H}]\le -1$, the mean metallicity is strictly an upper limit as 6/14 of the [X/H] estimates are upper limits. 
 \label{f-zdist1}}
\end{figure}

\begin{figure}[tbp]
\epsscale{1.2} 
\plotone{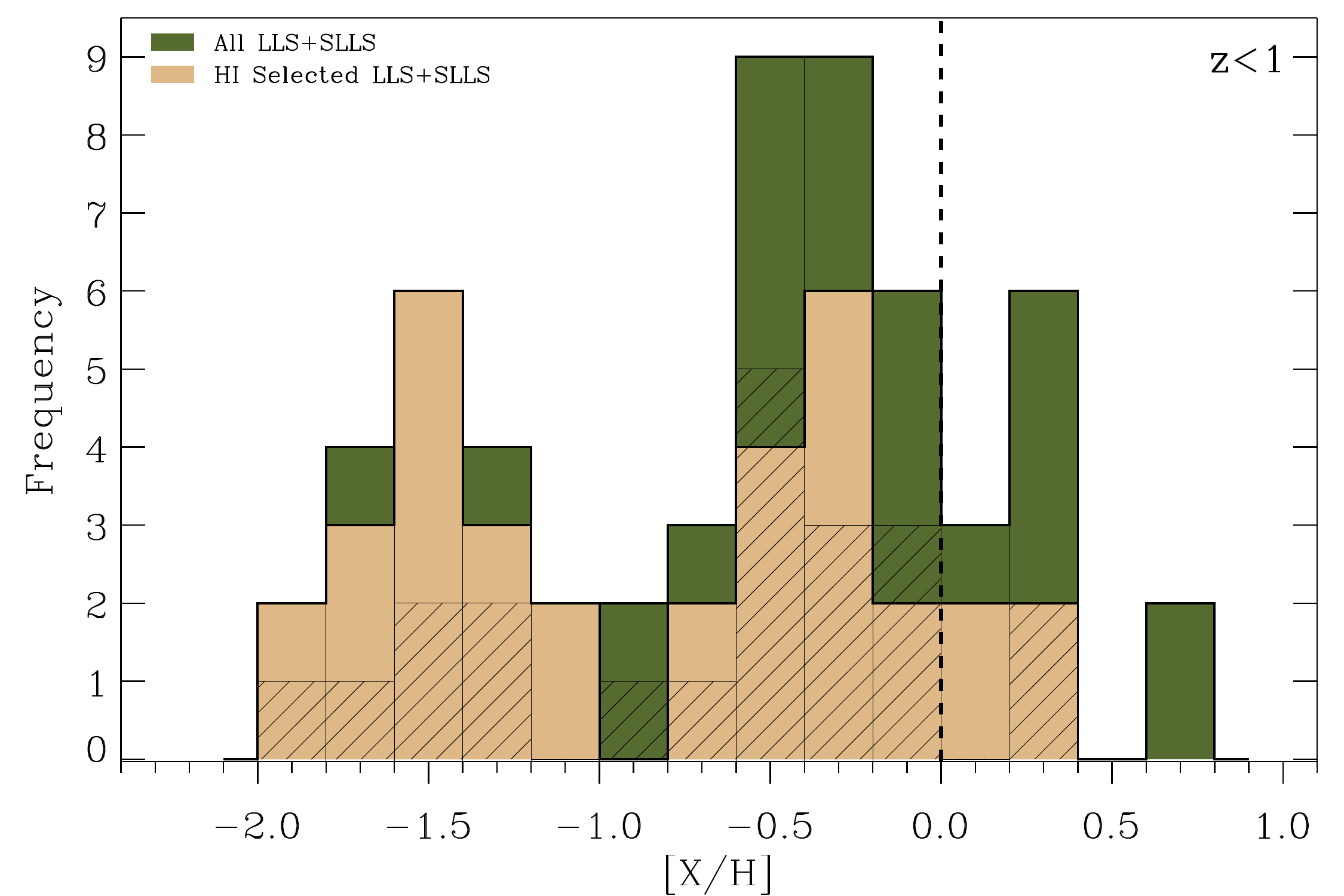}  
  \caption{Distribution of the metallicity for the LLS and SLLS. In this figure, the LLS and SLLS have been combined in one single population. The hashed histograms highlight values that are upper limits for the LLS+SLLS. This shows that the metal-selected absorbers generally miss the metal-poor (${\rm [X/H]}<-1$) branch. The two \mgiit-selected absorbers in the metal-poor branch have a large \hit\ column density (\nhi\,$=10^{18.30}, 10^{19.55}$ cm$^{-2}$).   
 \label{f-zdist}}
\end{figure}

In Fig.~\ref{f-zdist1}, we compare the distribution of metallicities for the LLS, SLLS, and DLAs. The DLA metallicity distribution is visually consistent with an unimodal distribution with $\langle {\rm [X/H]_{DLA}}\rangle = -0.62 \pm 0.35$ (1 $\sigma$ dispersion) and the median $-0.58$. A K-S test rejects the null hypothesis of no difference between the DLA and LLS datasets at the 99.1\% level. The K-S test also informs us that the DLA distribution is consistent with a normal distribution. In Fig.~\ref{f-zdist}, we have combined the LLS and SLLS and separately highlight the absorbers initially selected based on their \hi\ content. Metal-selected absorbers are unmistakably biased against the low metallicity population; the only detection of metal-poor absorbers based on their metal content are strong \hi\ absorbers ($\log N_{\rm H\,I}=18.3$ and 20.2).  If we consider the \hi-selected LLS+SLLS, the GMM method would yield essentially the same results as listed above; in that case 46\% of the absorbers are metal-poor. The dip probability of the LLS samples being bimodal is even larger, 99\%. On the other hand, if the entire LLS+SLLS sample is considered, the dip test remains inconclusive.

\subsection{Relative Abundances of C/$\alpha$}\label{s-relabu}

\begin{figure}[tbp]
\epsscale{1.1} 
\plotone{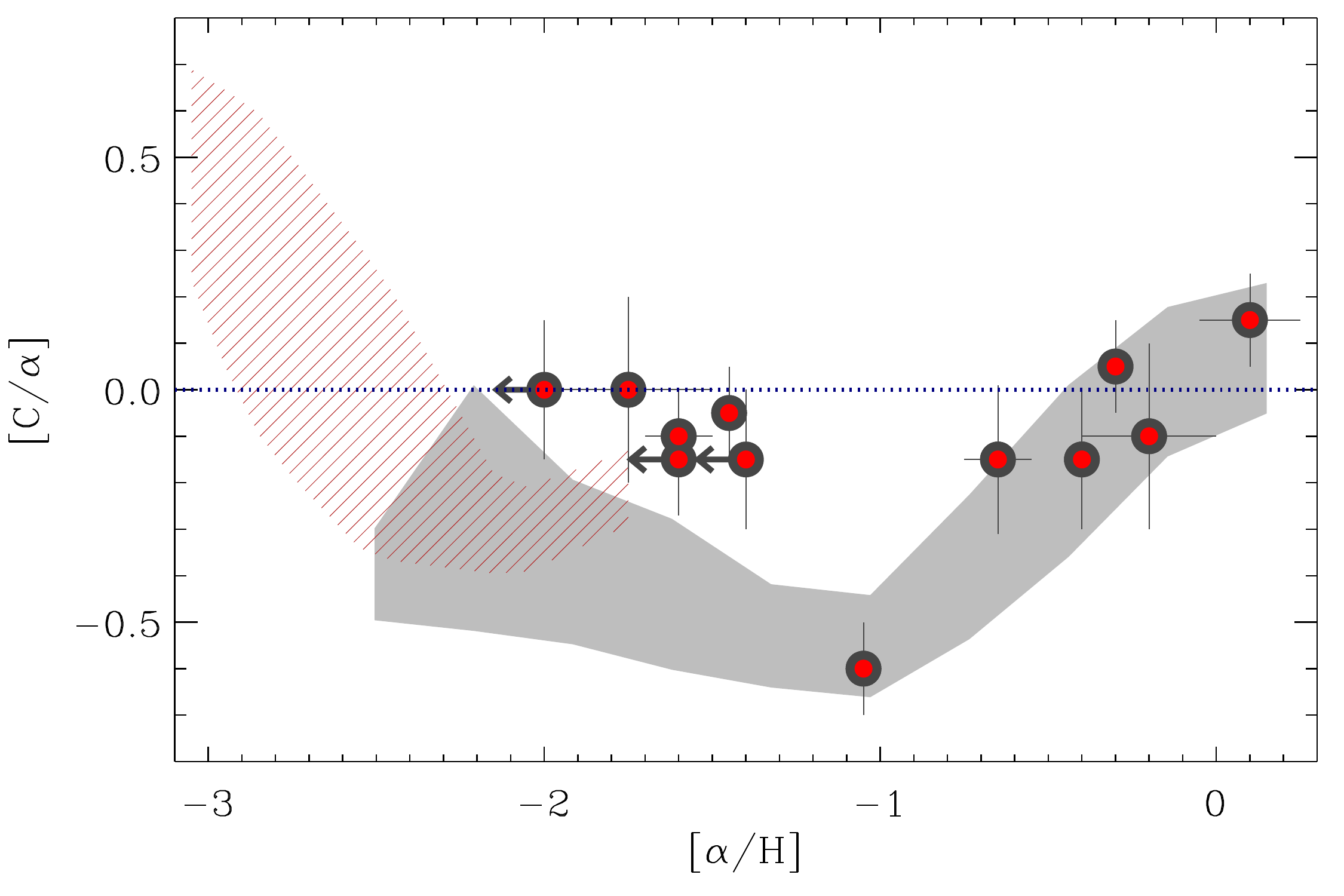}  
  \caption{The evolution of  [C/$\alpha$] as a function of the metallicity of the LLS at $z\la 1$ is shown with the circle symbols. The solid area shows the approximate trend and spread of [C/$\alpha$] vs. [$\alpha$/H]  for the Milky Way stars  \citep[from][]{akerman04,fabbian09}, while the hatched area shows the trend for $z>2$ DLA and SLLS \citep[from][]{pettini08,penprase10}. The width of the areas correspond to about the 1$\sigma$ spread at a given metallicity.   
 \label{f-calpha}}
\end{figure}

We noted in \S\ref{s-photmet} that the relative abundances of O, Mg, Si did not depart from solar values within the errors, and hence, since these elements have a similar nucleosynthesis history, there is no evidence for significant dust depletion ($\la 0.2$ dex) in the LLS. Similar to these elements, carbon is not expected to be strongly depleted into dust grains at these densities, but  the C/$\alpha$ ratio is sensitive to nucleosynthesis effects with a time-lag between the production of $\alpha$ elements and carbon \citep[e.g.,][]{cescutti09,mattsson10}. In Fig.~\ref{f-calpha}, we show [C/$\alpha$]-[$\alpha$/H] dependency determined from stellar  \citep{akerman04,fabbian09} and $z>2$ SLLS and DLAs \citep{pettini08,penprase10} observations. Chemical evolution models can reproduce that trend for $[{\rm \alpha/H}]> -1$, but have had more problem for the  upturn of [C/$\alpha$] at  $[{\rm \alpha/H}]\la -1$ \citep[e.g.,][]{cescutti09,mattsson10}.  Recently, \citet{kobayashi11} show that the observed high [C/$\alpha$] values in very metal-poor environments are consistent with expected nucleosynthesis yields in core-collapse type-III supernova of primordial 20--50 M$_\sun$ stars.  

In Fig.~\ref{f-calpha}, we also show the results for the LLS at $z<1$ where we were able to reliably estimate $[{\rm C}/\alpha]$. For the LLS with $-1.1\la [{\rm \alpha/H}]\la +0.1$, $[{\rm C}/\alpha]$ follows a similar pattern as the stars in our Milky Way or \hii\ regions in nearby galaxies \citep[e.g.,][]{henry00}. However, for $[{\rm \alpha/H}]\la -1.4$, we find $ -0.2\la  [{\rm C}/\alpha] \la 0$ for LLS, higher than seen in stars or high redshift DLAs. If there is no subtle ionization-dust depletion effect, the solar values of [C/$\alpha$] at metallicities $ [{\rm \alpha/H}] \la -1.4$ dex could be indicative of a different chemical evolution history for the metal-poor LLS as it scatters around a solar value rather than $-0.4$ to $-0.2$ dex solar (although we note that all the various samples overlap within the observed scatter and several values of $[\alpha/{\rm H}]$ for the LLS are upper limits at $[\alpha/{\rm H}]<-1$). The metallicities of the $z\la 1$ LLS  are so low for these redshifts that most of their metals are unlikely to have been incorporated recently into a galaxy, although the metals must have been produced and subsequently ejected from some galaxies at a much earlier epoch.  The solar C/$\alpha$ at low metallicity suggests that the gas has only been enriched by massive and metal-poor stars at an early epoch (see above).

\subsection{Physical Properties} \label{s-llsprop}

We  now review the physical properties of the LLS and, in particular, assess whether the physical properties are similar (or not) between the metal-poor and metal-rich LLS samples. Based on the Cloudy models (see Appendix and Table~\ref{t-cloudy}), typically the cool gas of the LLS is predominantly ionized (${f_{\rm H\,II} = N_{\rm H\,II}/N_{\rm H}}>90\%$, and on average $f_{\rm H\,II} = 0.98 \pm 0.02$). This contrasts markedly from the DLAs that are predominantly neutral \citep{vladilo01}. From the amount of total $N_{\rm H}$ (estimated from the Cloudy simulations, see Appendix), we can make a rough estimate of the mass of the photoionized (``phot") CGM gas probed by the strong \hi\ absorbers with $16.2\la \log N_{H\,I}\la 18.5$. Assuming a spherical halo of radius $R$, its mass is $M^{\rm phot}_{\rm CGM} \approx \pi R^2 f_c \mu m_{\rm H} N_{\rm H},$ where $f_c$ is the covering factor, $\mu =1.3$ corrects for the presence of He, $m_H$ is the hydrogen mass, and $N_{\rm H} = N_{\rm H\,I} + N_{\rm H\,II} (\approx N_{\rm H\,II}$ for the LLS). We can then write:
$$
M^{\rm phot}_{\rm CGM} \sim 2\times 10^{9} {\rm M}_\sun  \;\frac{f_c}{0.3} \;\left(\frac{R}{\rm 150\, kpc}\right)^2 \; \frac{N_{\rm H}}{\rm 10^{19} \,cm^{-2}}.
$$
From the recent COS-Halos survey of absorbers around $0.2 <L/L^*<3$ galaxies within $\rho < 150$ kpc \citep{werk12b}, we can estimate the covering factor of LLS  with $\log N_{\rm H\,I}>16.2$: $f_c \simeq 30\%$ for $R \le 150$ kpc (see also \S\ref{s-origins}). 
The mean impact parameter to galaxies for these absorbers is $\langle \rho \rangle = 60 \pm 36$ kpc.  Based on the results presented in the Appendix, for the metal-poor sample the $\log N_{\rm H}$ values average around $19.0 \pm 0.5$, while for the more metal-rich sample it is $18.6 \pm 0.7$, implying that the metal-rich gas CGM on average may be somewhat less massive by a factor $\sim$2 if the sizes and covering factors are similar. Hence the gas that is contained in the photoionized phase traced principally by the strong \hi\ absorbers is of $10^8$--$ 10^{10}$ M$_\sun$. We emphasize that this does not include the mass included in the gas-phase traced by the observed higher ionization stages or by the lower \nhi\ components at higher absolute velocities relative to the strongest \nhi (see below), which is likely to be at least as large if not much larger \citep{tripp11,tumlinson11a}.

From the Cloudy models, we can also derive rough estimates of the linear scale ($l \equiv N_{\rm H}/n_{\rm H}$) of these absorbers (see Appendix and references in Table~\ref{t-sum}). They generally range from a few pc to a few kpc, with one that might be larger than 10 kpc. They are therefore not hundreds of kpc or Mpc-scale structures as some of the \lya\ forest absorbers \citep[i.e., with lower \nhi, see, e.g.,][]{prochaska04,lehner06}. There is no trend between $l$ and the metallicity (see also Appendix).

The Cloudy model-derived temperatures are typical of photoionized gas temperatures, varying from $0.6\times 10^4$ K to $3\times 10^4$ K (see Appendix and references in Table~\ref{t-sum}). As the \mgii\ was observed at high resolution, we can derive reliable $b$-values and hence sensitively constrain the temperatures. We find $3.0<b<6.3$ \km\ with the average $\langle b \rangle = 4 \pm 1$ \km\ (based on 8 measurements), implying $T<(0.6$--$3) \times 10^4$ K. The model-derived temperatures  are therefore consistent with the observations.

As we can see from Figs.~\ref{f-pg1522a} to \ref{f-pg1216}, hot gas (traced in particular by \ovi, but also by \sv, \svi, etc.)\footnote{We have not modeled the highly ionized gas, and therefore it might seem unjustified to associate the highly ionized gas with hot gas. However, based on other works \citep[e.g.,][]{lehner09,savage10,savage11,tripp11,tumlinson11b} and on the systematic failure of our Cloudy photoionization models to reproduce the observed column densities of the highly ionized traced by, e.g., \ovit, \sivt, \svt, \svit, we conclude that the highly ionized gas found at similar velocities as the LLS must trace collisionally ionized gas, likely at $T\ga 10^5$ K.} is often detected at velocities overlapping with low-ion absorption in these LLS.  We emphasize that although the profiles of the low and high ions overlap in velocities, this is not sufficient to conclude that the hot and cool ionized plasmas have the same origin or metallicity \citep[see, e.g.,][]{tripp11,meiring12}. To determine the metallicity, \nhi\ is required but is very difficult to determine for the hot gas. An interesting (but rare) example where the properties of the broad \lya\ absorption associated with the \ovi\ absorption could be estimated in a LLS was described in \citet{savage11}. That is the absorber at $ z = 0.2261$ toward HE0153-4520 (see Table~\ref{t-sum}). They show that the hot ($\sim 10^6$ K) gas traces a hot metal-enriched plasma, while the cooler photoionized gas traced by the LLS is much more metal poor, implying different origins (e.g., cool inflow of metal-poor gas through a hot metal-enriched halo gas).

We also note that the presence of hot gas is also not ubiquitous along our sightlines. In the bottom panel of Fig.~\ref{f-mg2met}, we show the total column density of \ovi\ as a function of the metallicity of the photoionized gas. Non-detections of \ovi\ (down to a level of $\la 10^{13.5}$ cm$^{-2}$) are found at any metallicity (25\% LLS have no \ovi\ detection; 11\% LLS have no \ovi\ with limits $\la 10^{13.5}$ cm$^{-2}$). The scatter seen in this figure may be expected as the amount of \ovi\ does not only depend on the abundance of O, but of course also on the ionization conditions and total hot gas column.  For the LLS with  $N_{\rm O\,VI}\la 10^{14.5}$ cm$^{-2}$, Fig.~\ref{f-mg2met} shows a scatter plot. However, while the sample is small, this figure also indicates that LLS with  $N_{\rm O\,VI}\ga 10^{14.5}$ cm$^{-2}$ are only found at high metallicity. Such high $N_{\rm O\,VI}$ values have been associated with galaxies with high star formation rates and may trace the coronal halo of galaxies \citep{heckman02,chen00,lehner09,tumlinson11a,fox11}. Other than this dependence, there does not seem  to be a clear dependence between the high and low ions for the metal-poor and metal-rich LLS samples. A future detailed analysis of the absorption profiles of the high and low ions as well as a comparison with the output of collisionally ionized gas models will be necessary to undertake to assess the relationship  between the hot and cool gas (or lack of connection).

\subsection{The Connection to Galaxies}\label{s-llsgal}
\begin{figure}[tbp]
\epsscale{1.2} 
\plotone{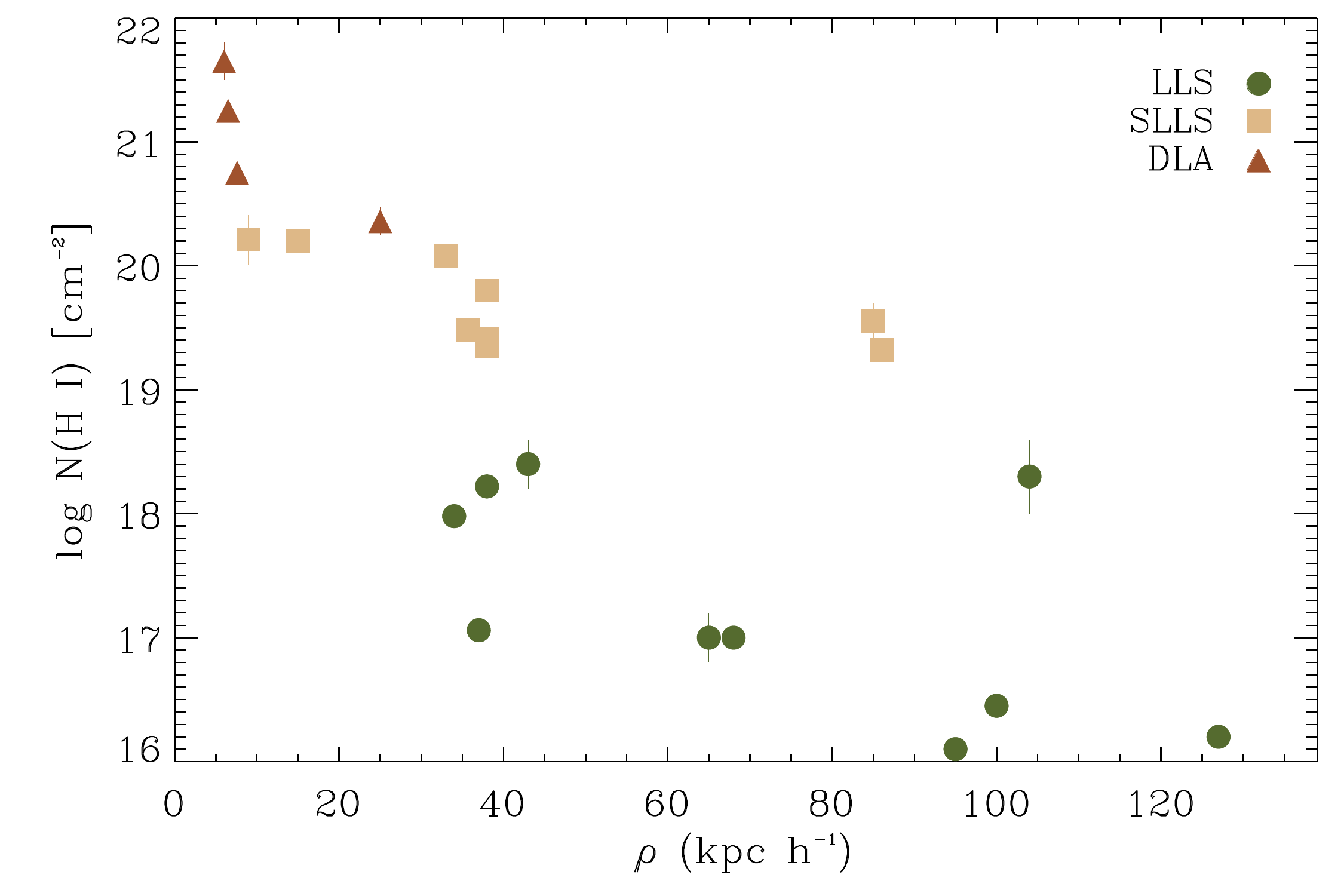}  
  \caption{The \hit\ column density versus the impact parameter. Only the ``closest'' galaxy (with a spectroscopic redshift) to a given absorber was considered \citep[see Table~\ref{t-sum} for the references and][]{meiring11a}.
\label{f-rhonh}}
\end{figure}

For some of the LLS in the literature, galaxy searches have identified the probable absorber-galaxy association. In Table~\ref{t-lls}, we tabulate in order of increasing LLS metallicity the properties of the LLS-galaxy association for 9 systems. We only list one galaxy, but there is evidence toward several sightlines of galaxy interactions or at least other galaxies \citep[e.g.,][]{jenkins05,lehner09,tumlinson11b} or a possible group of galaxies \citep[e.g.,][]{chen00,cooksey08} near the sightlines. In  Fig.~\ref{f-rhonh}, we show \nhi\ as a function of the impact parameter with respect to the associated galaxy for all classes of absorbers, confirming that LLS are typically within 100 kpc of a galaxy and that stronger \nhi\ absorbers probe the closer environments of galaxies. In that figure, we did not include  the $z= 0.2823$ LLS toward PG1216+069. This is the only absorber for which no galaxy was found within 100 kpc. In fact the closest galaxy is a $1.85 L*$ at $\rho \simeq 3$ Mpc based on the galaxy survey by \citet{prochaska11a}. This galaxy survey is magnitude limited  ($R = 19.5$), corresponding to about $>0.3 L*$ at $z\sim 0.3$. Hence it is quite possible that a less luminous galaxy is closer to PG1216+069. This highlights a major drawback of the current sample of galaxy surveys listed in Tables~\ref{t-sum} and \ref{t-lls}: they are quite heterogeneous in terms of both depth and completeness.

Understanding the galaxy environment is beyond the scope of this paper and the issue of galaxy redshift depth and completeness will be studied in detail in the future. Nevertheless, despite the current heterogeneous and small sample of galaxy surveys, combining our results with previous ones, we can conclude that the LLS typically probe the CGM of galaxies. The $\rho$ or $\delta v$ parameters do not appear to be a function of the metallicity, except possibly that there is more spread in $\delta v$ in the metal-poor branch (assuming the absorbers are not associated with fainter galaxies).  Both sub-$L^*$ and super-$L^*$ galaxies are associated with the LLS in both branches of the metallicity distribution, but, again, a better understanding of the completeness will need to be addressed to be able to draw stronger conclusions. 
\begin{deluxetable}{lccccc}
\tabcolsep=0pt
\tablecolumns{6}
\tablewidth{0pc}
\tablecaption{LLS and Possible Associated Galaxies$^a$\label{t-lls}}
\tabletypesize{\footnotesize}
\tablehead{\colhead{Name} & \colhead{$z_{\rm abs}$} & \colhead{$[{\rm X/H}]$} & \colhead{$\rho$} & \colhead{$\delta v$}  & \colhead{$L$}\\ %
\colhead{} &\colhead{} &\colhead{}   & \colhead{(kpc\,h$^{-1}$)} &\colhead{(\km)} &\colhead{($L*$)} }
\startdata
          J0943+0531   & $   0.3542 $ &  $   <-1.30 		$  &   $95   $ & $+285$  & 0.4 \\
          PG1630+377   & $   0.2740 $ &  $   -1.71 \pm     0.06 $  & $    37 $ & $-26$  & 0.3 \\
           TON153      & $   0.6610 $ &  $   -1.69 \pm     0.37 $  & $   104 $ & $0$	& 1.3 \\
           PG1216+069$^b$& $   0.2823 $ &  $  < -1.65		$  & $    3238 $ & $-152$ & 1.9 \\
         PKS1302-102   & $   0.0985 $ &  $ < -1.60  		$  & $    65 $ & $-354$ & 0.2 \\
\hline		 
          PKS0312-77   & $   0.2026 $ &  $   -0.60 \pm     0.15 $  & $    38 $ & $+16$  & 0.7 \\
          PG1116+215   & $   0.1385 $ &  $   -0.50 :			$  & $   127 $ & $+50$  & 2.8 \\
          J1009+0713   & $   0.3588 $ &  $   -0.40 \pm 0.20 	$  & $    43 $ & $+25$  & 0.3 \\
	     PHL1811   & $   0.0810 $ & $    -0.19 \pm     0.08 $  & $    34 $ & $+36$  & 0.5 \\
        PKS0405-123    & $   0.1672 $ &  $   +0.10 :			$  & $   100 $ & $-15$  & 3.4 \\
          PG1206+459   & $   0.9270 $ &  $   +0.30 :			$  & $    68 $ & $0$	& 1.8 
\enddata
\tablecomments{$^a$We emphasize that the galaxy redshift surveys used to find the associated galaxies are heterogeneous in terms of both depth and completeness (references for the galaxy surveys can be found in Table~\ref{t-sum}). The reported associated galaxies are generally the closest to the absorbers in physical space (smallest impact parameter $\rho$ and in redshift spaces (smallest $\delta v = c (z_{\rm abs} - z_{\rm gal})/(1+z_{\rm abs})$). To estimate $\delta v$, the strongest \hit\ component is considered. In some cases, several absorption components can spread over generally over about 100--200 \km\ (and in one case over 1000 \km\ for the absorber toward PG1206+459, see Tripp et al. 2011).  We express the abundances with the usual logarithmic notation. The references for all the sightlines can be found in Table~\ref{t-sum}. When no information was provided on the galaxy luminosity in the published literature, we used the information provided to give a rough estimate of $L$ using a K-correction. $^b$Based on a survey complete to about $>0.3 L*$ at $z\sim 0.3$ (see text for more details).  
}
\end{deluxetable}

\section{The Circumgalactic Medium at $\lowercase{z}\la 1$}\label{s-origins}

For the first time, we are able to probe sensitively the metallicity distribution function of the cool ($\sim 10^4$ K), photoionized  ($N_{\rm H\,I}/N_{\rm H\,II}\ll 0.1$) CGM about $z\la 1$ galaxies (for $16.2 \la \log N_{\rm H\,I} \la 18.5$). In our analysis we only consider the gas associated with the strongest \hi\ components, and therefore we have not yet characterized in a systematic manner the connection between the strongest and weaker \hi\ components. The weaker \hi\ components generally correspond to more highly ionized gas seen at higher absolute velocities in the rest frame of the LLS (e.g., $\ga 75$\% LLS have \ovi\ detected at similar or within $<100$ \km).  Analyses of the velocity components associated with the LLS show that a large fraction of baryonic mass can be present in the lower \hi\ column density components where the gas is more ionized and often highly ionized \citep[e.g.,][]{jenkins05,cooksey08,lehner09,thom11,tripp11,tumlinson11b,kacprzak12,werk12b}.

We find that very low metallicity LLS are not rare at $z\la 1$. There is a statistically significant bimodal distribution of the metallicity of the LLS with a peak around $[{\rm X/H}] <-1.57$ dex (hereafter the ``metal-poor branch'') and another peak at $-0.33$ dex (hereafter the ``metal-rich branch''). Half of the LLS population is found in each metallicity branch. However, we point out that metal-poor gas found in the same CGM as metal-rich gas along the sightline can be effectively hidden if they have similar line-of-sight velocities; this effect is illustrated in synthetic spectra depicting outflow and inflow in cosmological simulations \citep{shen13}.  In contrast to the LLS, the metallicities of the DLAs are normally distributed, peaking around $[{\rm X/H}]=-0.6$. Not only are the metallicity distribution functions of the LLS and DLAs very different, but so is the full range of metallicity values:  for the DLAs, $[{\rm X/H}]$ varies from $-1.4$ to about solar (a factor $\sim 25$ between the lowest and highest values), while for the LLS, it ranges from $[{\rm X/H}]<-2$  to about +0.3 dex solar (a factor $>200$ between the lowest and highest values). The large spread in the LLS metallicities is observed over the entire redshift range $0<z\la 1$ (see Fig.~\ref{f-zdist}). The fact that the metallicity distributions of the DLAs and LLS are so different implies that they must probe different environments or phenomena. In view of their larger neutral fraction and smaller impact parameter to galaxies, DLAs must probe the environments very near galaxies, if not the interstellar medium of galaxies themselves. 

The bimodal metallicity distribution of the LLS implies that the gas of the metal-poor and metal-rich LLS samples may not appreciably mix in the CGM or, when they do, one of the branches must dominate. Indeed if some metal-poor gas of mass $M_p$ and metallicity $Z_p$ is mixed with a metal rich gas of mass $M_r$ and metallicity $Z_r$, the metallicity of the mixed gas can be expressed as $Z_m = (Z_p M_p + Z_r M_r)/(M_p + M_r)$. If $M_r \gg M_p$ (or $M_r \ll M_p$), then $Z_m \sim Z_r$ (or $Z_m \sim Z_p$). If, however, $M_r \sim M_p$, then the metallicity of the mixed gas would be somewhat intermediate between the metal-poor and metal-rich branches, which is not observed (at least at a significant level in the present LLS sample). As we discuss below, the metal-poor and metal-rich groups may be principally signatures of inflows and outflows, respectively. In that case, the gas traced by the metal-poor and metal-rich branches may not be expected to interact much in the CGM. Indeed, in cosmological simulations, metal-poor cold flows not only feed the galaxies but also supply their angular momentum, and hence they naturally rotate with the galaxy disks \citep[e.g.,][]{dekel06,stewart11b,rubin12,shen13}, while on the other hand, stellar feedback in galaxies tends to be bi-conical, near perpendicular to the galaxy major axis \citep[e.g.,][]{strickland04,bouche12,brook11}.

It is difficult to assess a priori which population of galaxies our LLS observations probe. The rate of incidence of the LLS along the sightline is proportional to the product of the comoving number density of galaxies giving rise to LLS, $n_{\rm LLS}$, and the average physical cross-section of the galaxies, $\sigma_{\rm LLS}$, i.e., $d\mathcal{N}_{\rm LLS}/dz \propto n_{\rm LLS} \sigma_{\rm LLS}$ \citep[e.g.,][]{ribaudo11a}. Hence another consequence of the bimodal metallicity distribution is that the product $n_{\rm LLS} \sigma_{\rm LLS}$ must be similar in both metallicity branches as about the same number of absorbers are observed in each category. While it is well-known that the number density of galaxies is related to the galaxy luminosity (through the luminosity function), the cross section for LLS absorption may also depend on the luminosity \citep[e.g., through a Holmberg scaling, see, e.g.,][]{chen10,ribaudo11a}. While the number density increases with decreasing luminosity, the cross-section may decrease. It is a priori unclear  which of these quantities matters the most for understanding the bimodal metallicity distribution, i.e., whether there could be a difference in galaxy type probed by the two branches of the LLS metallicity distribution function. 

However, for 10 out of 29 LLS, we know that at least one galaxy (sometimes more than one) is found within $\la 100$ kpc with a redshift near the LLS (see Tables~\ref{t-sum} and \ref{t-lls} and Fig.~\ref{f-rhonh}). As we argue in \S\ref{s-llsgal} based on previous studies, it is likely that all the LLS probe the CGM of galaxies. The metal-rich LLS branch is therefore evidence that metal enriched gas can be found at several tens of kpc from galaxies. The [C/$\alpha$] ratio follows a pattern of chemical enrichment that is similar to that observed in our Milky Way or other nearby galaxies (see \S\ref{s-relabu}). The metal-rich LLS gas may therefore trace outflows, recycled material, or tidally stripped gas. Detailed examples of absorbers probing outflows can be found in \citet{tripp11} or \citet{prochaska04}, while examples of absorbers tracing galaxy interactions are discussed in \citet{jenkins05} and \citet{lehner09}.  

The metal-poor LLS population is evidence that near-pristine gas can also commonly be found within a few tens of kpc from galaxies. The \nhi\ values of the LLS are also too small to trace directly galaxies, and the bulk of the metal-poor sample is at $[{\rm X/H}]<-1.4$ (only one LLS has $[{\rm X/H}]\simeq -1.1$), i.e., at metallicities too low to trace even very low-metallicity galaxies at $z<1$ \citep[e.g.,][]{kobulnicky04,perez13}, with only exceptions being too uncommon to find with regularity in absorption line studies \citep[see also][]{ribaudo11b}. The finding of low metallicity LLS with near solar C/$\alpha$ (see \S\ref{s-relabu}) is also consistent with the metal-poor CGM gas being unlikely to have been recently incorporated into galaxies. Thus they may instead trace cold streams of ionized, metal-poor gas falling onto galaxies  with properties consistent with those predicted in cosmological simulations \citep[e.g.,][see below]{fumagalli11b,shen13}.

\citet{werk12b} have reported some properties (but not the metallicity) of the absorbers within 150 kpc of 42 galaxies from the COS-Halos project, probing a range of luminosity $0.2\la L/L^* < 3$ (about 60\% of galaxies being $<0.7 L^*$). The COS-Halos sample was not selected based on the \nhi\ content, but on the galaxy properties and impact parameter. The \nhi\ distribution of this sample follows more closely the differential \hi\ column density distribution \citep[e.g.,][]{lehner07,ribaudo11a}, and about 70\% (including the lower limits) of their absorbers have  $14.5\la \log N_{\rm H\,I}<16$. About 40\% of those have no detection of \mgii, but in this \nhi\ interval, they are not sensitive to metallicities  $[{\rm X/H}]\la -1$.  About 30\% of the absorbers have $\log N_{\rm H\,I}>16$, but only lower limits on \nhi\ have been determined
. The only exception is one absorber, which is in our sample, the metal-poor LLS toward J0943+0531 (see Table~\ref{t-sum}, \citealt{thom11}). It is therefore too early to determine if there is any difference between the COS-Halos and our samples regarding the metallicity distribution of the LLS. Furthermore since that there are only about 10 COS-Halos absorbers that are in the LLS \nhi\ interval,  a statistically robust comparison between the two samples may actually be impossible to undertake. 

Other observational techniques have also been used to empirically characterize the flows of matter between the galaxies and the CGM. One, in particular, has been extremely useful for characterizing the outflows of galaxies, whereby ``down the barrel'' spectra of galaxies are searched for blueshifted and redshifted absorption troughs relative to the galaxy systemic velocity as signatures of outflow and inflow, respectively \citep[e.g.,][]{heckman00,martin05,tremonti07,weiner09,steidel10,rubin11}. This method reveals that outflows are quite ubiquitous in star-forming galaxies. However, with this technique and over the redshift interval $0.4\le z\le 1.4$ that overlaps with our study, \citet{rubin12} and \citet{martin12} found  a very small fraction (only a few percent) of galaxies showing significant redshifted  (infalling) \feii\ or \mgii\ absorption relative to the host galaxy.\footnote{Although \citeauthor{rubin12} also noted that the presence of infall onto as many as 40\% of the galaxies could not be ruled out.} As these authors argue, the gas that is seen is metal enriched, and hence more likely tracing the recycling of gas via, e.g., a galactic fountain. Identifying infalling metal-poor gas has proven difficult using metal-lines (especially \mgii\ and \feii). Our results demonstrate there is a significant amount of this metal-poor gas, but that it would produce \feii\ and \mgii\ absorption too weak to be identified in complicated galaxy spectra where outflows are very often present. While the QSO absorption line technique does not provide the direction of the gas motion relative to the associated galaxy, and requires additional observations to characterize the galaxy properties, selecting absorbers drawn from an \hi-selected population is unique in providing insights into the interaction of galaxies with their surroundings and may be the only way to directly study the very low metallicity CGM in absorption (see, however, \citealt{rauch11} for \lya\ emission imaging techniques).

The present work and other recent results \citep[e.g.,][]{stocke10,stocke13,thom11,tripp11,tumlinson11a,kacprzak12,churchill12,werk12b} signal a shift in our ability to characterize the properties of the CGM and have already provided new stringent empirical results to test cosmological hydrodynamical simulations. Regarding the metallicity of the CGM, there are now several theoretical works that determine the metallicity of the inflowing and outflowing gas at $z\sim 2$--3, although there is not yet a theoretical prediction of the metallicity distribution function of the CGM in the LLS regime at $z<1$ for comparison with the present results. The recent simulations by \citet{fumagalli11b} include mild stellar feedback and show that cold streams are traced mostly by LLS within 1 or 2 virial radii of galaxies, where the gas has only been weakly enriched  ($[{\rm X/H}]\simeq -1.8 \pm 0.5$). The gas is also predominantly ionized as it is observed.  The simulations by \citet{shen13} include stronger galactic outflows, but still show that cold flows are metal-poor with a median value $-1.2$ dex. These simulations and those by \citet{voort12}  show that major outflows do not suppress significantly cold flow accretion. \citet{voort12} similarly show that cold mode accretion is generally metal-poor ($[{\rm X/H}]\la -1.5$) for any halo mass. Our results for the metal-rich LLS at $z<1$ show metallicities higher than predicted in their models, but their simulations are at $z=2$. The metallicities of the cold flow gas from these theoretical results compare quite well with our empirical results, and it is unlikely that the metallicity of the cold filaments is much dependent on the redshift. However, it would be important to confirm this in the simulations as well as to understand the type of galaxies and their properties associated with these absorbers within these simulations. While this comparison between the simulations and the present observational results is encouraging, it will also be important to determine if these recent theoretical results are confirmed with the AREPO simulations where different hydrodynamic schemes are employed to solve the equations \citep{springel10,vogelsberger12}. These simulations show in particular some critical differences in the covering factors of LLS and DLA and how galaxies get their gas \citep{keres12,bird13,nelson13}. Notwithstanding, and independently of the theoretical work,  our observational results show that large amounts of both metal-poor and metal-enriched cool, photoionized gas traced by the LLS are found around galaxies at $z<1$. 

We finally point out that the often adopted $10\%$ solar metallicity as a representative value for the IGM in the low redshift universe has no support from our observations. In fact, the \ovi\ absorbers often used to characterize the warm-hot ionized medium (WHIM) are quite likely to trace the CGM rather than the IGM based on their overall high metallicity \citep[e.g.,][]{sembach04,lehner06}, consistent with the metal-rich branch of the LLS. The majority of the \ovi\ absorbers observed at $z\la 0.5$ are also associated with sub-dwarf galaxies, not the diffuse WHIM predicted by cosmological simulations (\citealt{prochaska11}; see also \citealt{wakker09,stocke06}). Another consequence of this work is therefore that the IGM far from the virial radii of  galaxies could have a characteristic metallicity well below $2\%$ solar at $z<1$. The very low metallicity gas has been essentially missed in studies of the \lya\ forest because  absorbers with $\log N($\hi$)\la 14$ have a column density too small to detect metal lines at significant levels to constrain the metallicity of the gas below $10\%$ solar.

\section{Summary}\label{s-sum}
We have assembled the first sizable sample of LLS  at $z\la 1$ selected solely on their \hi\ content ($16.2\la \log N_{\rm H\,I} \la 18.5$), i.e., where the main criterion for assembling our sample of CGM absorbers is independent of metallicity. The \nhi\ of the LLS is too low to directly probe the galaxies, but is large enough to trace their CGM. The metallicity is determined by estimating the column densities of metal ions and \hi\ in the strongest \hi\ component (not over the entire velocity profile where metal-line absorption may be observed), and correcting for ionization effects using Cloudy photoionization models. There are 28 LLS in our sample, 16 of them being newly analyzed in this work. Our main results are as follows: 

\begin{enumerate}
\item  We empirically establish the  metallicity distribution function of the LLS  (and hence the cool CGM) at $z\la 1$: the distribution is bimodal with a peak around $[{\rm X/H}]\le -1.57 \pm 0.24 $ (the metal-poor branch) and another peak at $[{\rm X/H}] = -0.33 \pm 0.33$ (the metal-rich branch) with about equal number in each branch. The metallicity in the metal-poor branch is so low at $z\la 1$ that the majority of the gas in in these LLS is unlikely to have been recently incorporated into galaxies.
\item  In contrast, the DLA metallicities are normally distributed, peaking around $[{\rm X/H}] = -0.6$. The range of metallicity between the DLA and LLS samples is also quite different, varying from  $[{\rm X/H}] = -1.4$ to about 0 (a factor $\sim 25$ between the lowest and highest values) for the DLAs, while for the LLS, it varies from $<-2$ dex to about $+0.3$ dex (a factor $>200$ between the lowest and highest values).  
\item The gas associated with the strongest \hi\ component of the LLS is predominantly photoionized ($f_{\rm H\,II}> 90\%$), with a temperature around $10^4$ K. The mass of the photoionized component of a typical galaxy's CGM selected by these LLS is $\sim 10^8$--$ 10^{10}$ M$_\sun$. There is also often (but not always) evidence of more highly ionized gas as revealed by the presence of, e.g., \ovi, \svi, and other highly ionized species as well as other higher absolute velocity components with lower \nhi, which may contain as much mass. 
\item We find that C/$\alpha$ varies with metallicity for the $z\la 1$ LLS. For  $[{\rm \alpha/H}]\ga -1$, C/$\alpha$ follows the general pattern seen in stars and \hii\ regions with similar metallicities, suggesting a similar chemical enrichment. For  $[{\rm \alpha/H}]\la -1$, C/$\alpha$ is, however, near solar, different from the trend seen in low metallicity stars and high redshift DLAs. The solar C/$\alpha$ at low metallicity suggests that the gas has only been enriched by massive and metal-poor stars at an early epoch. 
\item The bimodal metallicity distribution of the LLS implies that the metal-poor and metal-rich samples  have similar $n_{\rm LLS} \sigma_{\rm LLS}$ and that the metal mixing between the two metallicity branches is either inefficient or dominated by one branch. 
\item Combining these results, we conclude that the metal-rich gas likely traces winds and recycled gas from outflows and galaxy interactions. The metal-poor LLS are very likely tracing cold accretion onto galaxies, with properties in very good agreement with those seen in cold flow accretion models and simulations. Independent from the simulations, our empirical results show there is not only a large mass of metal-rich gas around galaxies  at $z\la 1$, but also a significant mass of metal-poor gas that may become available for star formation. 
\item Finally, based on these results, the  metallicity of the diffuse IGM at $z\la 1$ may be well below $2\%$ solar.
\end{enumerate}

\section*{Acknowledgments}

It is a pleasure to thank Hsiao-Wen Chen for pertinent discussions during the earlier stage of this study and Oleg Gnedin for sharing his Gaussian Mixture Modeling (GMM) code. Support for this research was provided by NASA through grants  HST-GO-11741, HST-GO-11598,  and HST-AR-12854 from the Space Telescope Science Institute, which is operated by the Association of Universities for Research in Astronomy, Incorporated, under NASA contract NAS5-26555. This material is also based upon work supported by the National Science Foundation under Grant No. AST-1212012. All of the data presented in this paper were obtained from the Mikulski Archive for Space Telescopes (MAST). STScI is operated by the Association of Universities for Research in Astronomy, Inc., under NASA contract NAS5-26555. Some of the data presented herein were obtained at the W.M. Keck Observatory, which is operated as a scientific partnership among the California Institute of Technology, the University of California and the National Aeronautics and Space Administration. The Observatory was made possible by the generous financial support of the W.M. Keck Foundation. The authors wish to recognize and acknowledge the very significant cultural role and reverence that the summit of Mauna Kea has always had within the indigenous Hawaiian community.  We are most fortunate to have the opportunity to conduct observations from this mountain. This research has made use of the NASA's Astrophysics Data System Abstract Service and the SIMBAD database, operated at CDS, Strasbourg, France.

\begin{appendix}
\makeatletter 
\renewcommand{\thefigure}{A\@arabic\c@figure} 

\renewcommand{\thetable}{A\@arabic\c@table}

In this Appendix, we provide more details for the new sample of LLS, in particular about their velocity profiles and column densities, and the Cloudy photoionization models that were used to determine the metallicity of these absorbers. The redshift of each absorber is defined based on the strongest \hi\ component. In  the figures that follow, we show the normalized profiles of \hi\ (generally a weak and a strong transition) and  most of the detectable metal lines observed by COS as a function of the rest-frame velocity. We also systematically show the \mgii\ transitions when optical observations were acquired. 

\begin{figure*}[tbp]
\epsscale{0.9} 
\plottwo{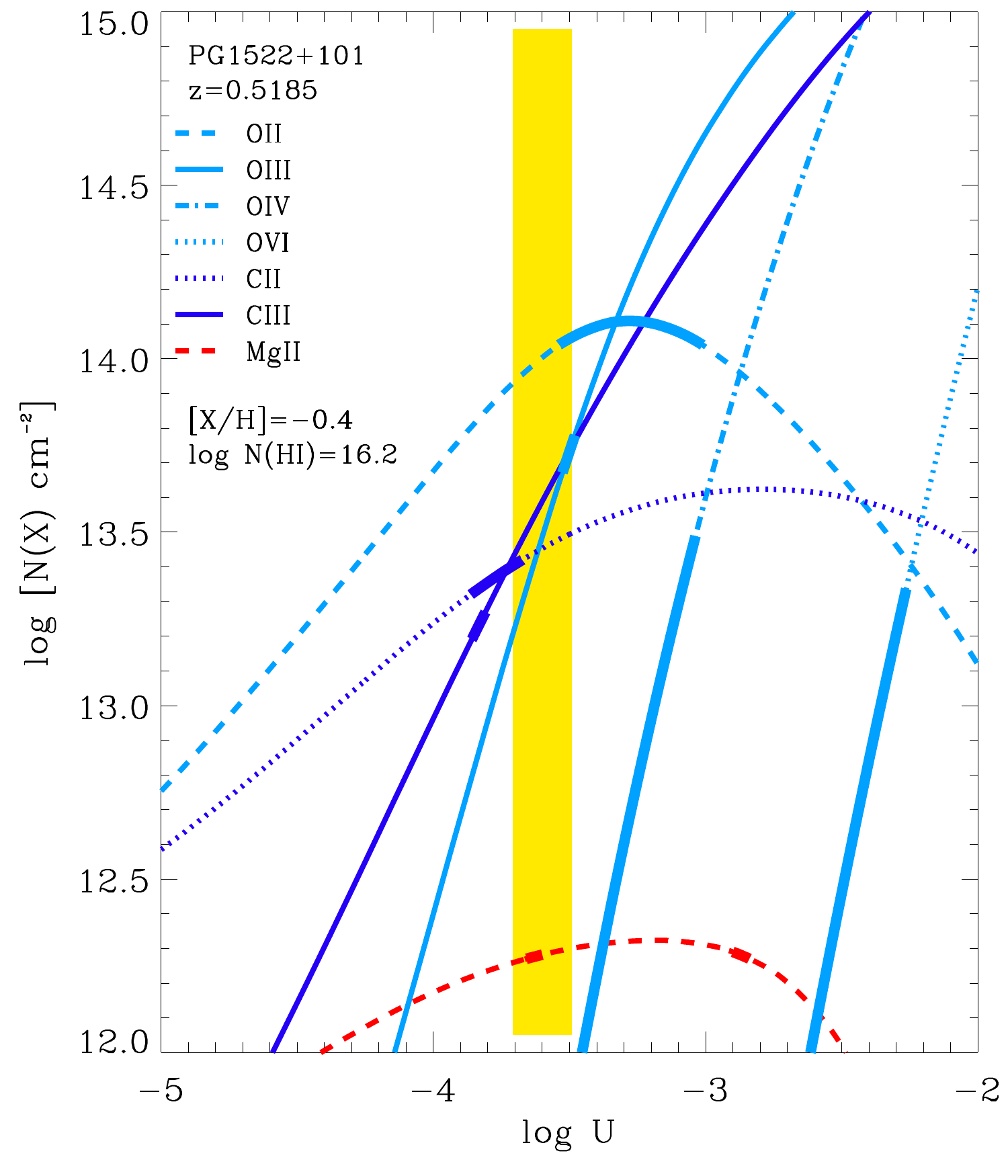}{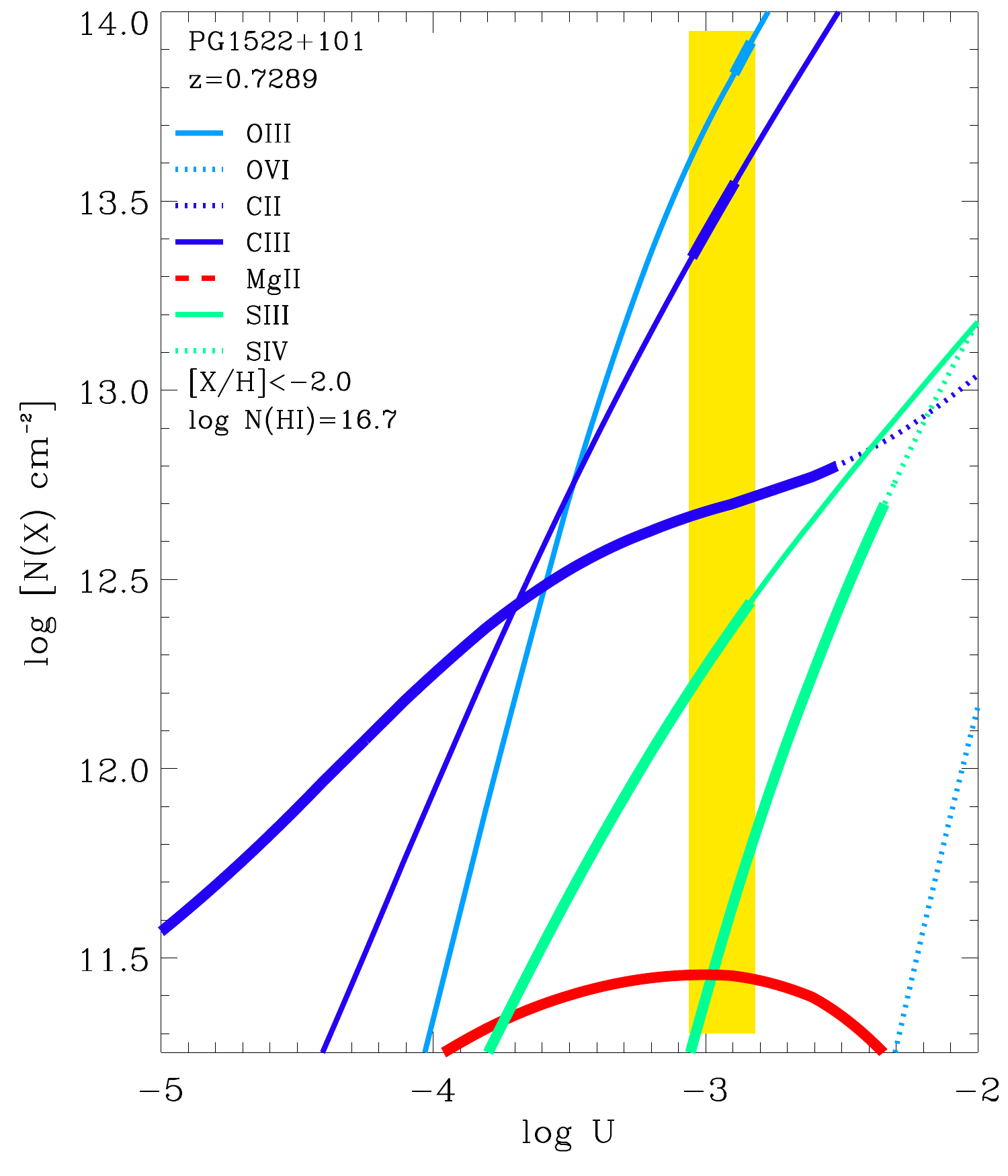}  
  \caption{Examples of Cloudy-predicted column densities as a function of ionization parameter $U$  toward the same sightline PG1522+101 for two different absorbers at $z=0.5185$ ({\it left panel}) and $z=0.7289$ ({\it right panel}). The gas is assumed to be photoionized by the background radiation field from quasars and galaxies, modeled as a uniform slab in thermal and ionization equilibrium. In both panels, the column density models are shown in the thinner lines, with the allowed values by the observations indicated by the thicker lines. The yellow region indicates the range of $\log U$ that fits best the observational constraints. Note that for the absorber on the right panel: i) a higher metallicity would predict too much \mgiit\ for any $U$; ii) $U$ is strictly a lower limit (higher $U$ are allowed with lower metallicities); and iii) \ovit\ is detected, but cannot be reproduced by photoionization. 
\label{f-cloudy}}
\end{figure*}

\smallskip
It is useful to first describe the Cloudy models for two illustrative systems, the $z=0.5185$ and $z=0.7289$ absorbers toward PG1522+101. The results of Cloudy modeling of these figures are shown in Fig.~\ref{f-cloudy}. These plots show the predictions of Cloudy models for an adopted metallicity as a function of ionization parameter $U$. The Cloudy results for these two and other absorbers are summarized in Tables~\ref{t-sum} and \ref{t-cloudy} for the new sample of LLS described here. We also list in these tables the results from the existing sample of LLS at $z\la 1$ prior to this work (the references are provided in Table~\ref{t-sum}). For each absorber, the column densities determined from the observations and the Cloudy simulations are matched by varying the ionization parameter ($U$) and metallicity. The errors that we provide on $U$ and the metallicity are not formally 1$\sigma$ error, but are ranges of values allowed by the models and errors on the column densities.  From the Cloudy solution, several other useful physical parameters can be determined (including, the total H column density ($N_{\rm H}$), the temperature of the gas -- $T$, the density of the gas -- $n_{\rm H}$, the pressure of the gas -- $P/k$, and the linear scale of the cloud -- $l \equiv N_{\rm H}/n_{\rm H}$). The values corresponding to the best solution are listed in Table~\ref{t-cloudy}. A description of these two absorbers follows. We comment below on all of the absorbers in our sample, but only show the models graphically for these first two, which were chosen because they are characteristics of the high and low metallicity branches of the metallicity distributions. Other models can also be found in, e.g., \citet{lehner09}, \citet{ribaudo11b}, \citet{tripp11}.

\begin{figure*}[tbp]
\epsscale{1} 
\plotone{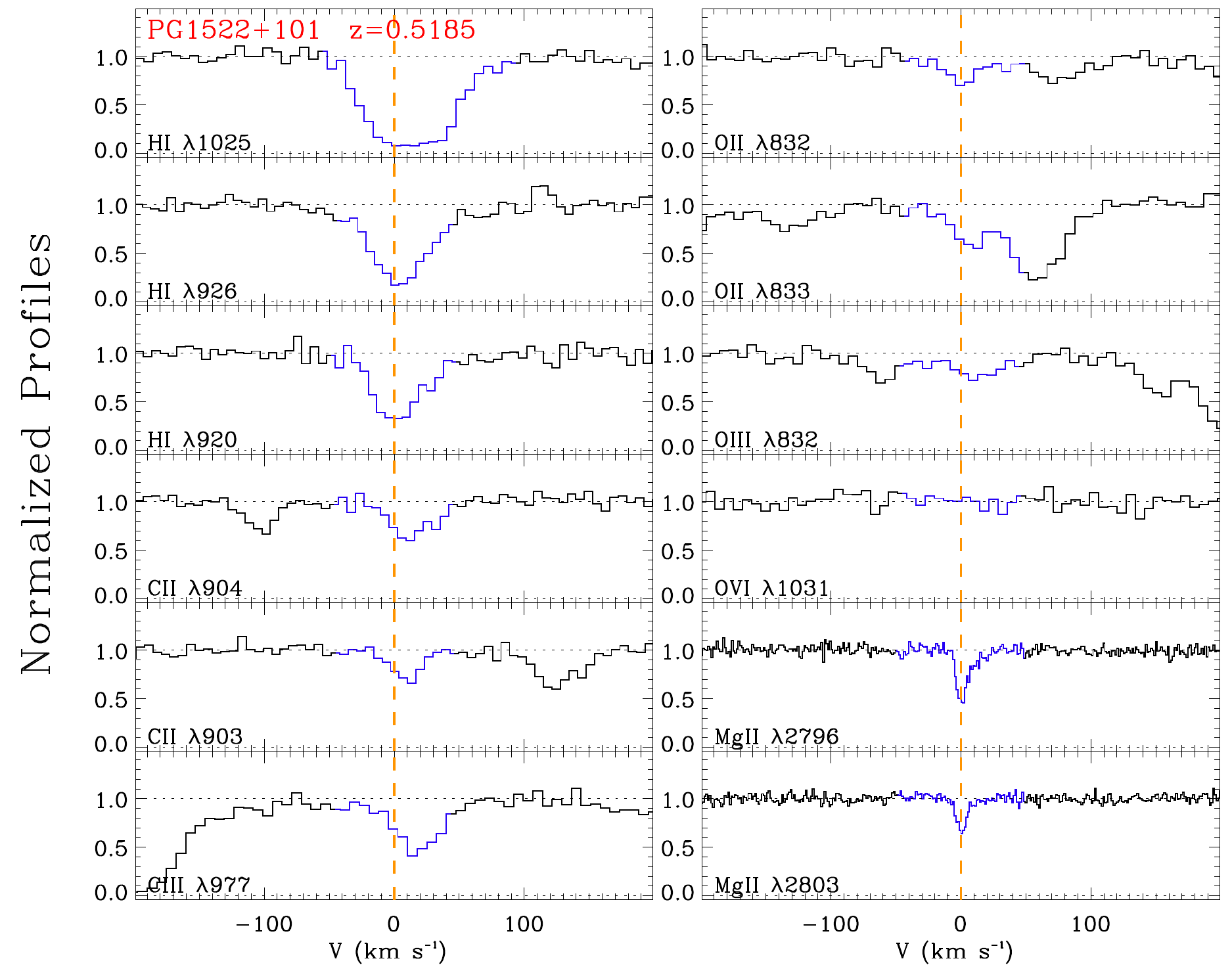}  
  \caption{Normalized absorption lines as a function of velocity centered on the absorber at $z=0.5185$ toward PG1522+101. The UV transitions are from COS G130M and G160M and \mgiit\ $\lambda$$\lambda$2796, 2803 are Keck HIRES spectra. The blue portion in each profile shows the approximate velocity range of the absorption (for \lya, the entire velocity interval is highlighted, while for the other species and other weak \hit\ transitions, generally on the main \hit\ component is highlighted). The reader should refer to Table~\ref{t-data} and the Appendix for the exact integration velocity intervals. The vertical dashed lines mark the zero velocity. \label{f-pg1522a}}
\end{figure*}

--  \underline{PG1522+101, $z = 0.5185$:} Fig.~\ref{f-pg1522a} shows that the \mgii\ absorption has more than one component, but the component at 0 \km\ dominates the absorption. This is also the case for \hi\ where most of the absorption in the weakest \hi\ transitions is observed at 0 \km. For \oii, \oiii, \cii, and \ciii, there is no evidence of multiple components at the COS resolution. The strongest \hi\ component aligns well with the strong \mgii\ component as well as the metal ions observed with COS accounting for a COS velocity calibration error of $\sim$5--$15$ \km. A single component COG fit to the \hi\ lines and the AOD measurements of the weakest \hi\ lines are consistent within $1\sigma$. We adopt a mean from the results of these two methods (in this case, the continuum break is too weak to estimate \nhi\ reliably).

Based on the comparison of the velocity profiles of the different ions and \hi, a single-phase photoionization model is justified. Fig.~\ref{f-cloudy} shows that a Cloudy photoionization model with $\log U = -3.6 \pm 0.1$ and   $[{\rm X/H}] =  -0.40 \pm 0.05$ can reproduce well the \oii/\mgii, \cii/\oii, \oii/\oiii\ ionic column density ratios. The well determined \ciii\ column density suggests that [C/$\alpha] \simeq -0.15$, consistent with Type II supernovae enrichment. As there is no evidence for highly ionized gas, this absorber is single phase. 
 
\begin{figure*}[tbp]
\epsscale{1} 
\plotone{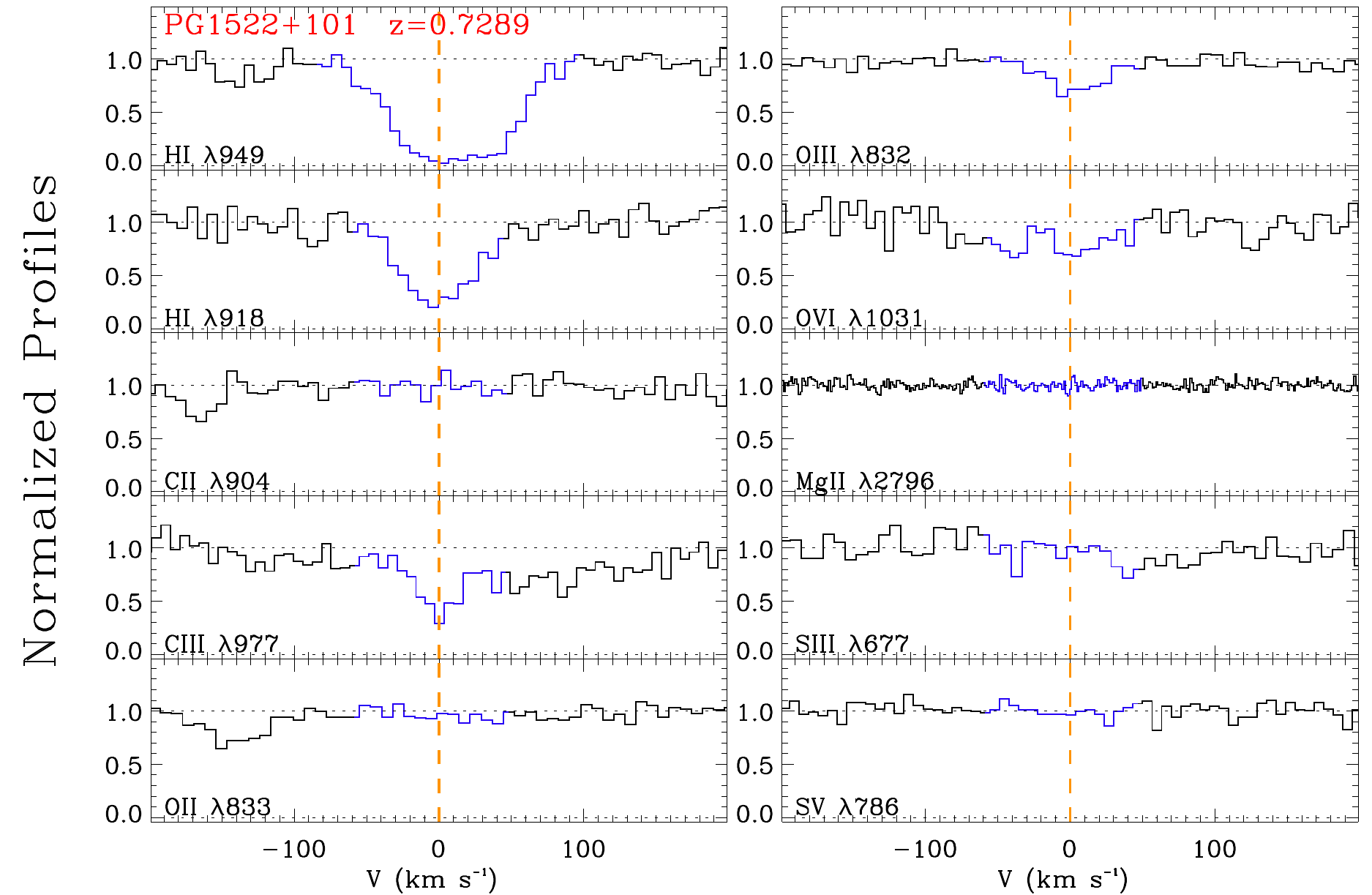}  
  \caption{Same as Fig.~\ref{f-pg1522a}, but for the absorber at $z=0.7289$ toward PG1522+101.
\label{f-pg1522b}}
\end{figure*}

--  \underline{PG1522+101, $z = 0.7289$:} Fig.~\ref{f-pg1522b} shows that there is no strong evidence for multiple components in the \hi, \ciii, or \oiii\ absorption profiles.  There is no detection of any singly ionized species (including the strong \mgii\ $\lambda$2796 line) and \Siii. \ciii\ $\lambda$977 is partially contaminated at higher velocities, but that does not affect much the component at 0 \km. Besides \ciii, \oiii\ $\lambda$832 and \ovi\ $\lambda$1031 are the only reliable metal line detections. All the useable \hi\ lines (i.e, lines that are uncontaminated and for which the continua can be reliably determined) remain quite strong, and therefore the AOD method only provides a lower limit on \nhi. We determine  \nhi\ from the continuum break and the COG. Both methods give consistent \nhi\ values within $1\sigma$ (and the AOD limit is consistent with these estimates), and we therefore adopt a mean from these two methods.

 Based on the comparison of the velocity profiles of the different ions, we used a single phase Cloudy photoionization model to predict the column densities or limits of all the observed species (see Fig.~\ref{f-cloudy}). A model with  $[{\rm X/H}] < -2.0$ and $\log U > -3.20$ can reproduce the observations. Higher metallicity would overproduce \mgii\ and \Siii. Note that even though we detect metals (\ciii\ and \oiii), the lower limit for $U$ leaves the total H poorly constrained, giving only an upper limit on the metallicity. In this case, ${\rm C/\alpha}$ is consistent with a solar despite the gas being extremely sub-solar. The model cannot produce enough \ovi\ ($\log N_{\rm O\,VI}= 13.92 \pm 0.07$, but we emphasize that this is only based on the strong transition of the doublet since \ovi\ $\lambda$1037 is contaminated), and hence this absorber is multiphase. As we discuss in the main text, for the LLS, photoionization alone fails to reproduce the columns of the high ions when they are detected, implying that other processes must be at play (e.g., collisional ionization).

\smallskip
For the remaining 14 LLS, similar Cloudy simulations and figures were produced, but we only discuss the results below. The order of the absorbers follow that of Table~\ref{t-data}.

\smallskip

\begin{figure*}[tbp]
\epsscale{1} 
\plotone{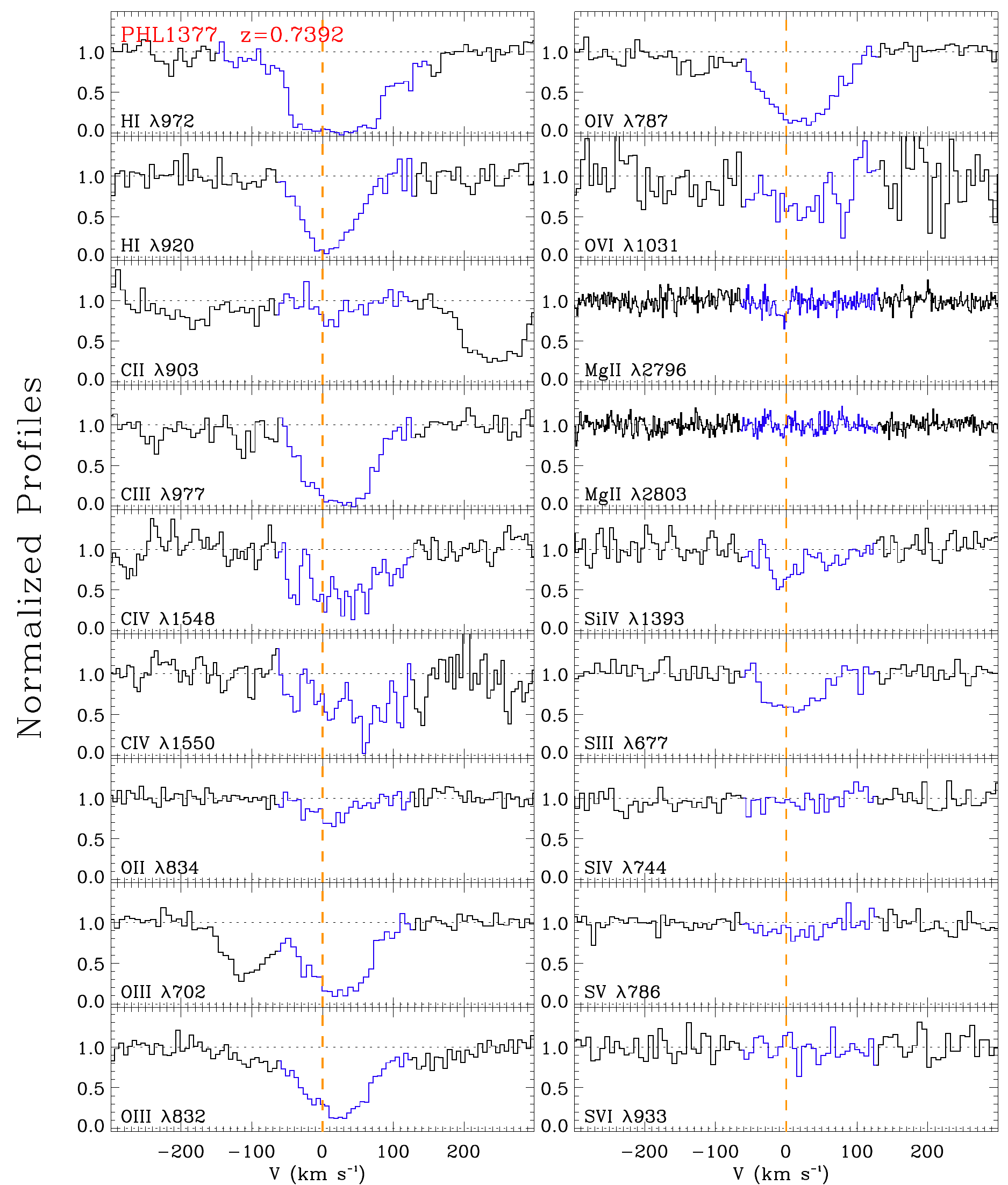}  
  \caption{Same as Fig.~\ref{f-pg1522a}, but for the absorber at $z=0.7390$ toward PHL1377; \civt\ $\lambda$$\lambda$1548, 1550 are STIS E230M spectra.  
\label{f-phl1377}}
\end{figure*}

--  \underline{PHL1377, $z = 0.7930 $:} Fig.~\ref{f-phl1377} shows that \mgii\ $\lambda$2796 is barely detected at 0 \km\ (the weaker transition at 2803 \AA\ is not detected at $3\sigma$). Most of the \hi\ column density is in the 0 \km\ component, but there is evidence of at least another component around $+40$ \km\ in the absorption profiles of \ciii, \oiii, \oiv, and \ovi\ (\siiv\ is, however, mostly present at 0 \km). The COG, AOD, and continuum break methods all give consistent results for \nhi\ and our adopted \nhi\ value is a mean of these results. \oiii, \oiv, \ciii\ have all large central optical depth, $\tau_0$, suggesting that the saturation might be important, and hence we treat the column densities for these ions as lower limits. \civ\ $\lambda$1550 appears slightly contaminated at $v>70$ \km, and we therefore used only \civ\ $\lambda$1548 to estimate $N_{C\,IV}$ (note that the \civ\ data -- as well as \siiv\ and \alii --  were obtained with STIS E230M). \ovi\ $\lambda$1037 is not covered at this redshift, and therefore only \ovi\ $\lambda$1031 is used to estimate $N_{\rm O\,VI}$. 

As the strong doubly ionized species (\ciii, \oiii) show at least two strong components, we focused mostly on the singly ionized species and weak \Siii\ transitions for the ionization model. The total column densities of \cii, \oii, \mgii, and \Siii\ can be satisfied  simultaneously within $1\sigma$ error in the Cloudy photoionization model $[{\rm X/H}] = -1.45 \pm 0.05$ and $\log U = -2.90 \pm 0.10$. Using, \cii, \oii, and \mgii, we derive  [C/$\alpha] \simeq -0.05$ despite the low metallicity of the gas. This model produces enough \ciii\ to match the observed lower limit, about 50--60\% of the observed \oiii\ and \siiv, and much too small amounts of \oiv, \ovi, \civ, and \sv\ relative to the columns determined from the observations. This absorber is therefore multiphase. 

\begin{figure*}[tbp]
\epsscale{1} 
\plotone{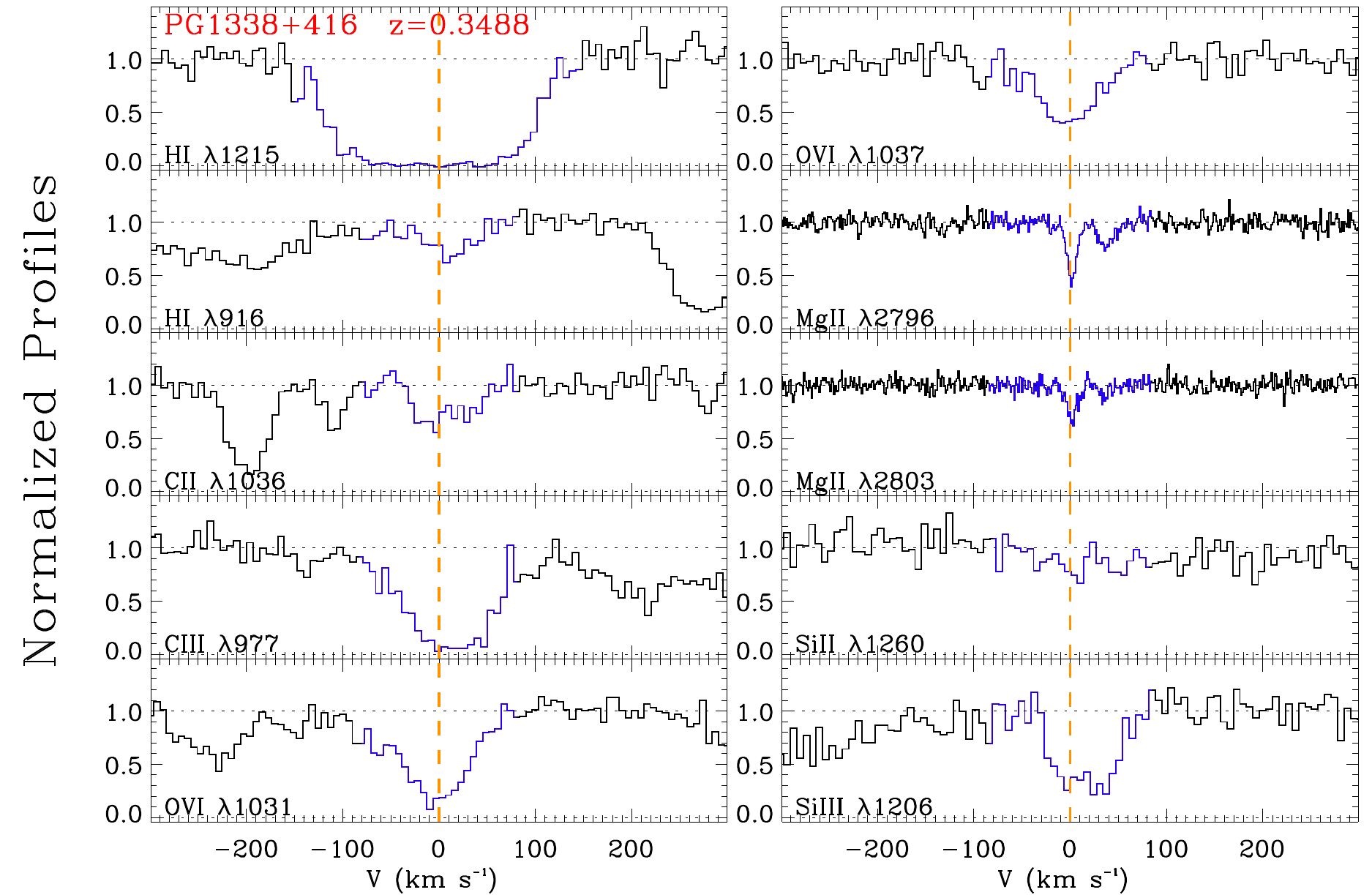}  
  \caption{Same as Fig.~\ref{f-pg1522a}, but for the absorber at at $z=0.3488$ toward PG1338+416. 
 \label{f-pg1338a}}
\end{figure*}

--  \underline{PG1338+416, $z =0.3488$:} Fig.~\ref{f-pg1338a} shows that the \mgii\ absorption profiles reveal two main components at 0 and $+35$ \km, but the column density is dominated by the 0 \km\ component. The \hi\ has at least two components at 0 and about $-80$ \km, but 90\% of the column density is in the 0 \km\ component. The one component COG fit to the \hi\ lines gives $\log N_{\rm H\,I} = 16.40 \pm 0.05$, while the AOD results on the weakest measurable \hi\ line ($\lambda$916) yields $\log N_{\rm H\,I} = 16.20 \pm 0.07$. The difference could be due to the presence of the $+35$ \km\ component in the strong \hi\ transitions that the COS data does not really resolve. We adopted a mean of the two results, where in this case the adopted error reflects the scatter between these two values (no reliable estimate could be made from the continuum break owing to the proximity of the Galactic \lya\ absorption component and the unreliable flux level near the edge of the detector). The \ciii\ $\lambda$977 and \siiii\ $\lambda$1206 profiles have strong absorption and may be saturated (although our Cloudy simulation implies the saturation should be mild, $<0.2$ dex). The \ovi\ $\lambda$$\lambda$1031, 1037 profiles are strong as well, but the lines are unsaturated based on the excellent $N_a$ agreement between the \ovi\ doublet lines. A similar agreement is found for the \mgii\ doublet. 

The total column densities of \mgii, \siii, and \siiii\ can be satisfied simultaneously in the Cloudy photoionization model with $[{\rm X/H}] = -0.75 \pm 0.15$ and $\log U = -3.2 \pm 0.10$. If \cii\ is not contaminated, then $[{\rm C/Si}] \simeq +0.15$, which is higher than typically observed at this metallicity.   The photoionization model cannot reproduce $N_{\rm O\,VI}$ by orders of magnitude, implying that the absorber is multiphase.  

\begin{figure*}[tbp]
\epsscale{1} 
\plotone{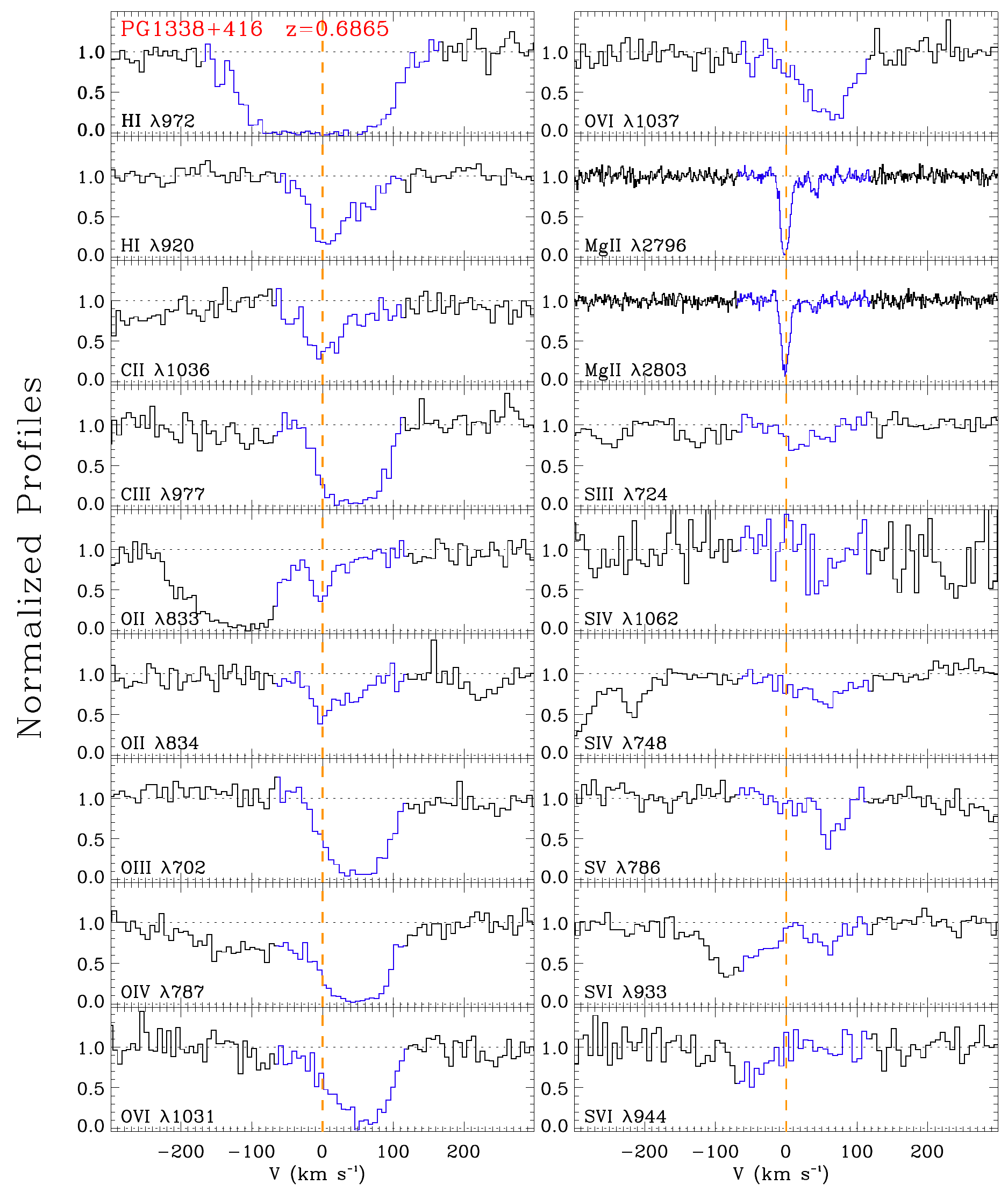}  
  \caption{Same as Fig.~\ref{f-pg1522a}, but for the absorber at $z=0.6865$ toward PG1338+416.
 \label{f-pg1338b}}
\end{figure*}

--  \underline{PG1338+416, $z = 0.6863 $:} Fig.~\ref{f-pg1338b} shows that the \mgii\ absorption profiles reveal two main components at 0 and 45 \km, but the column is dominated by the 0 \km\ component. This conclusion also applies to the other singly ionized species. However, for the doubly and higher ions, the absorption is dominated by higher velocity components (e.g., about 45 \km\ for \ciii\ and \oiii, and 70 \km\ for \ovi, \sv, \svi). The \hi\ has at least two components at 0 and about $45$ \km, but most of the column density is in the 0 \km\ component. The single component COG fit on the \hi\ lines gives $\log N_{\rm H\,I} = 16.39 \pm 0.02$. The AOD results on the weakest uncontaminated \hi\ lines ($\lambda$$\lambda$916, 917, 919, 920) yields $\log N_{\rm H\,I} = 16.45 \pm 0.04$. From the continuum break, we estimate  $\log N_{\rm H\,I} = 16.48 \pm 0.06$. We adopted a mean for \nhi\ from these 3 methods. For \ciii\ $\lambda$977,  \oiii\ $\lambda$702, and \oiv\ $\lambda$787, the absorption is strong and likely saturated. Based on the difference of 0.07 dex between the weak and strong lines of the \ovi\ and \mgii\ doublets, we applied a 0.07 dex correction on the weakest line of each doublet  \citep{savage91}. 

Based on the comparison of the velocity profiles, with the Cloudy simulations we only attempted to reproduce the column densities of \cii, \oii, and \mgii. The total column density of the ions can be satisfied simultaneously in the photoionization model with $[{\rm X/H}] = +0.10 \pm 0.15$ and $\log U = -3.90 \pm 0.15$. The metallicity cannot be much below $<-0.1$ dex because otherwise not enough \mgii\ would be produced for any $U$. If $U$ is higher, then \oii\ would be overproduced by $+0.5$ dex. For that absorber, $[{\rm C/Si}] \simeq +0.15$, which is typical for super-solar metallicity gas. We note that even if $U$ is small, the absorber is still nearly pre-dominantly ionized (more than 90\% ionized). The linear scale is, however, quite small, of the order of $3$ pc. Its small linear scale implies that it is very unlikely a ``clump'' or ``cloud'' of gas. Instead it might a somewhat dense ``shell'' within an extended ionized medium (possibly probed by the other velocity components).  This absorber has more than one component and is multiphase. 

\begin{figure*}[tbp]
\epsscale{1} 
\plotone{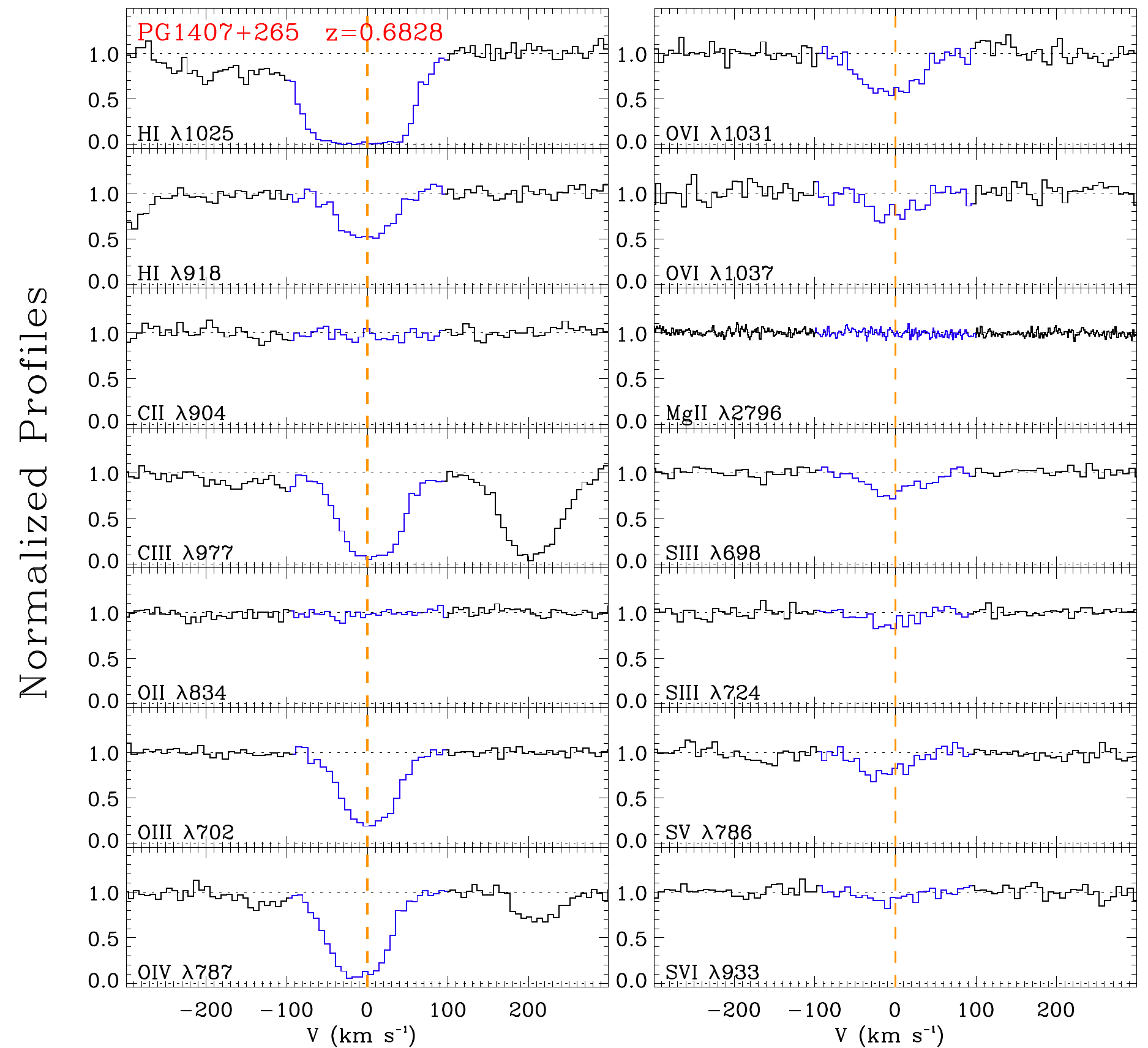}  
  \caption{Same as Fig.~\ref{f-pg1522a}, but for the absorber at $z=0.6828$ toward PG1407+265.
\label{f-pg1407}}
\end{figure*}

--  \underline{PG1407+265, $z = 0.6828$:}  Although the \hi\, \ciii, \oiii, and \oiv\ profiles are strong and suggest two components at 0 and about $-30$ \km\ (see Fig.~\ref{f-pg1407}), the profiles are dominated by the component at 0 \km\ as evidenced by the weaker transitions of \hi\ and \Siii\ $\lambda$$\lambda$698, 724 lines. The strongest transitions of \oii\ and \mgii\ are not detected, but \ovi\ and \sv\ are also observed at 0 \km. The \ciii\ absorption is quite strong and could be somewhat saturated.  The COG, AOD, and continuum break methods all give consistent results for \nhi\ and our adopted result is a mean of these results. All the ions with two transitions (\ovi, \Siii, \oiii) give consistent column density results. 

For this absorber, there is no evidence for very different kinematics between the different ions. We therefore considered all the limits and detections for the Cloudy model. The non-detection of \mgii\ and \cii\ and the detection of \Siii\ constrain this absorber to  $-2.0<[{\rm X/H}] < -1.5$ with $-2.3<\log U<-1.7$. In the case of the highest value of $U$ (lowest metallicity), all the observations except \ovi\  can be matched. For the lowest value of $U$, the model can only reproduce the singly and doubly ionized species.  Depending if \ciii\ is saturated or not, then  $-0.2 \la [{\rm C/alpha}]\la +0.2$.  \oiv, \ovi, \sv, and \svi, cannot be reproduced by photoionization, and therefore this absorber is also multiphase. 

\begin{figure*}[tbp]
\epsscale{1} 
\plotone{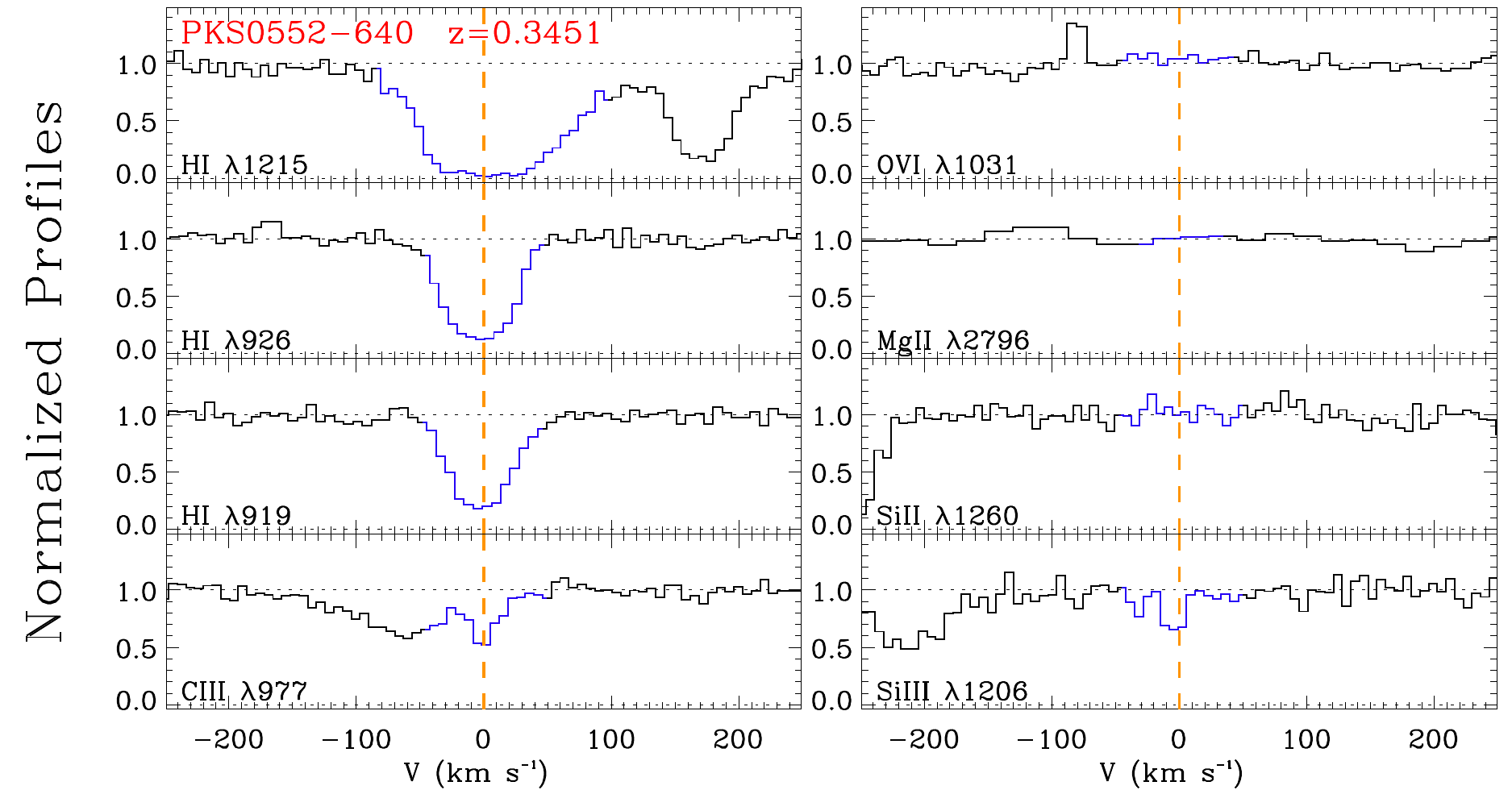}  
  \caption{ Same as Fig.~\ref{f-pg1522a}, but for the absorber at $z=0.3451$ toward PKS0552-640.
\label{f-pks0552}}
\end{figure*}

--  \underline{PKS0552-640, $z = 0.3451$:} Fig.~\ref{f-pks0552} shows that only the metal lines \siiii\ and \ciii\ are detected in this absorber, both with only one component at the COS resolution. \ciii\ is blended with another feature at  $v<-30$ \km, but based on \siiii\ we do not expect the absorption to extend much below $v<-30$ \km.  The single component COG fit of the \hi\ lines gives $\log N_{\rm H\,I} = 16.83 \pm 0.04$, overlapping within $2\sigma$ with the results from the continuum break, $\log N_{\rm H\,I} = 16.95 \pm 0.05$. The combination of the lines being strong (the weakest lines being on the knee of the COG) and the break occurring in the wing of the damped Galactic \lya\ likely explain these differences. We adopted a mean for \nhi\ between these 2 methods, where the adopted error reflects the scatter between these two values. 

Based on the comparison of the velocity between the different ions, we used a single-phase Cloudy photoionization simulation to model all the column densities of the observed species and limits for the non-detected species. A model with $[{\rm X/H}] < -1.60$ and $\log U > -3.70$ can reproduce the observations.  In this case, $[{\rm C/\alpha}] \simeq -0.15$. We emphasize that this absorber could be an extremely low metallicity absorber, and a better limit on \mgii\ would  be extremely beneficial to better constrain the metallicity.   As there is no \ovi\ detected or other ions that cannot be produced by the photoionization model, it is a single-phase absorber. 

\begin{figure*}[tbp]
\epsscale{1} 
\plotone{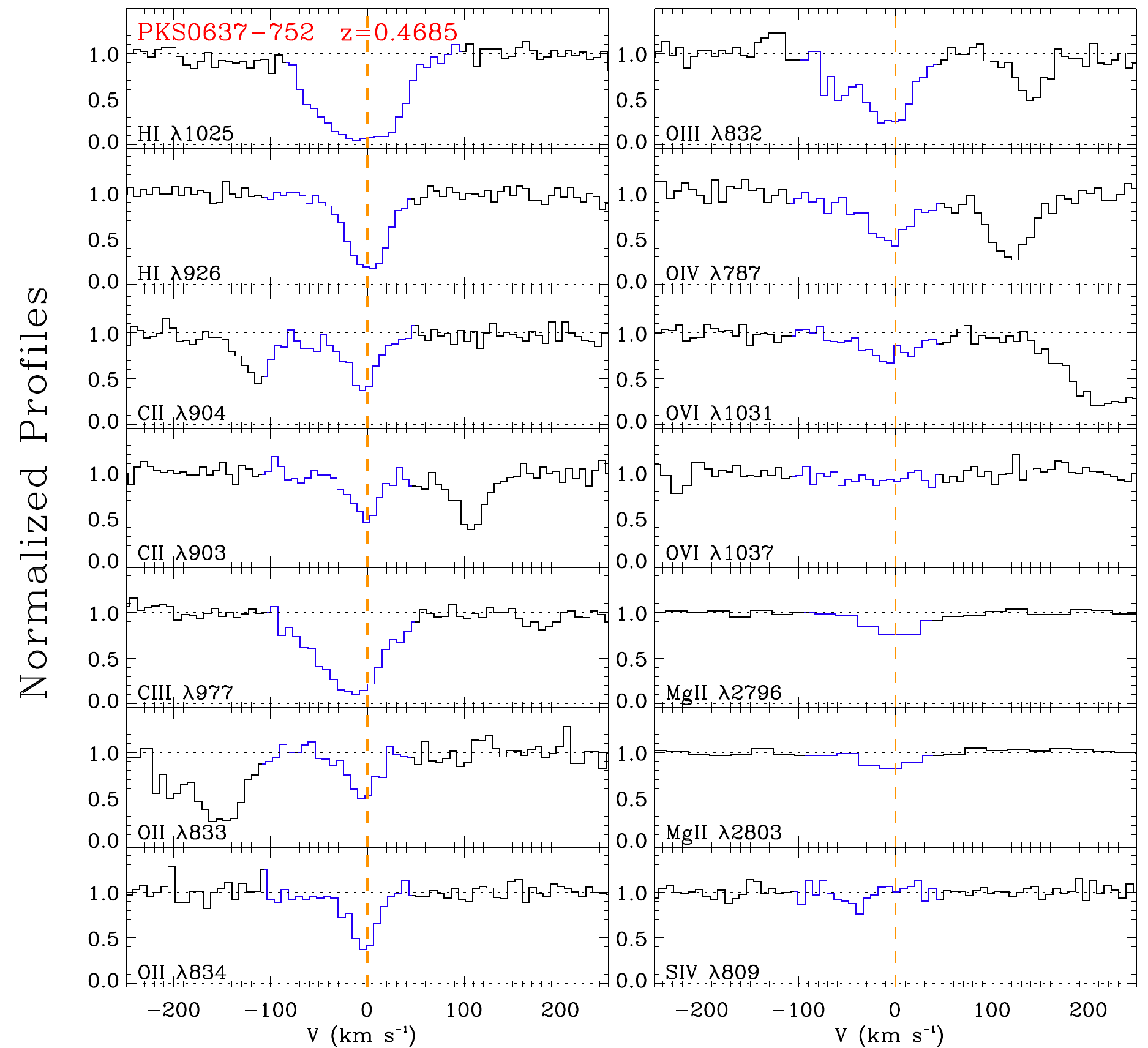}  
  \caption{Same as Fig.~\ref{f-pg1522a}, but for the absorber at $z=0.4685$ toward PKS0637-752.
 \label{f-pks0637}}
\end{figure*}

--  \underline{PKS0637-752, $z = 0.4685$:} The profiles of \oii\ and weak \hi\ transitions suggest that they are dominated by a single component (see Fig.~\ref{f-pks0637}), while the \ciii\ and \oiii\ absorption profiles show an additional but comparatively weak component at about $-50$ \km. The two detected unblended \oii\ lines and the \ovi\ doublet lines produce consistent column densities for these ions, respectively. The \ciii\ and \oiii\ lines are relatively strong and may be saturated. The \mgii\ doublet was only observed at $R\simeq 5,500$, and  there is evidence for weak saturation comparing the results from the weak and strong transitions. We corrected for the saturation ($0.05$ dex) following \citet{savage91}.  The single component COG fit on the \hi\ lines gives $\log N_{\rm H\,I} = 16.44 \pm 0.06$, overlapping within $1\sigma$ with the result from the continuum break, $\log N_{\rm H\,I} = 16.51 \pm 0.05$. We adopted a mean for \nhi\ between these 2 methods.

For this absorber, we can estimate a strong lower limit on the metallicity, $[{\rm Mg/H}]>-0.5$, based on \mgii. A model with  $[{\rm X/H}] = -0.4 \pm 0.1$ and $\log U = -3.8 \pm 0.1$ provides a good agreement between the observed and modeled column densities of \mgii, \oii, \cii, and \ciii. $N_{\rm O\,III}$ is, however, underproduced by a factor 6, possibly because it is partially contaminated or because the gas is multiphase as evidenced by the absorption of \oiv\ and \ovi\ (which are underproduced in the photoionization model by several orders of magnitude). A higher metallicity (and smaller $U$) would be consistent with the singly ionized species, but not \ciii, and we therefore favor the above solution. 

\begin{figure*}[tbp]
\epsscale{1} 
\plotone{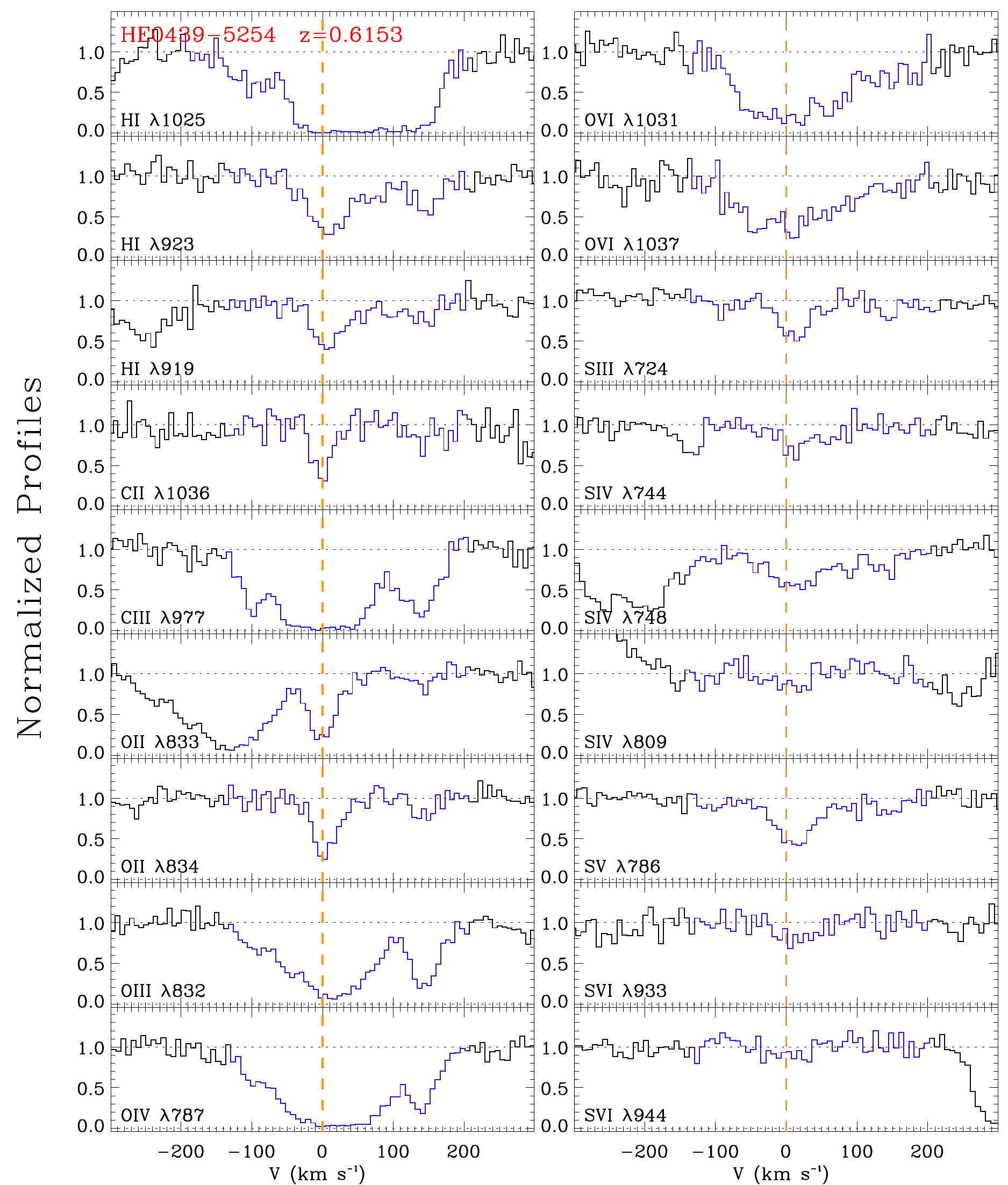}  
  \caption{Same as Fig.~\ref{f-pg1522a}, but for the absorber at $z=0.6153$ toward HE0439-5254. 
\label{f-he0439}}
\end{figure*}

--  \underline{HE0439-5254, $z = 0.6151$:} Fig.~\ref{f-he0439} shows that the profiles of this absorber have a complicated velocity structure, with the absorption spanning from about $-160$ to $+200$ \km. However, most of the \hi\ column density is located in the component at 0 \km\ between $-75$ and $+70$ \km\ as evidenced by the weakest \hi\ transitions. The weak \hi\ transitions also show a weaker component at $+145$ \km, which is cleanly separated from the low velocity component. The singly ionized species and \Siii\ profiles have a similar velocity structure than the weak \hi\ transition. The higher ions with strong transitions, including \ciii, \oiii, \oiv, and \ovi, show absorption over velocities comparable to that in the \lya\ absorption profile. The profiles of the high ions with weaker absorption (\siv, \sv, \svi) is dominated by the component at 0 \km. Therefore, except for  \ciii, \oiii, \oiv, and \ovi, the column densities of the metals and weak \hi\ transitions can be readily compared in the  0 \km\ component between $-75$ and $+70$ \km. The one component COG fit to the \hi\ lines (from $\lambda$916 to $\lambda$949; at higher wavelengths all the components are entirely blended) and AOD estimates using the weakest \hi\ lines ($\lambda<923$) give a similar $N_{\rm H\,I}$. We adopted a mean of the results.

We only attempted in our Cloudy simulations  to predict the column densities of \cii, \oii, and \Siii\ where the absorption is mostly confined to the strongest \hi\ component at 0 \km, where the \hi\ column density is estimated.  A model with  $[{\rm X/H}] = -0.30 \pm 0.05$, $\log U = -2.70 \pm 0.10$, and $[{\rm C/\alpha}]=+0.05 \pm 0.10$ reproduces very well the observations. In that case, the model predicts too much \oiii, which is consistent with the lower limit determined by the observations. This model, however, only produces a small amount of \oiv\ and \siv\ and negligible amount of \sv. It is therefore a multiphase absorber. 

\begin{figure*}[tbp]
\epsscale{1} 
\plotone{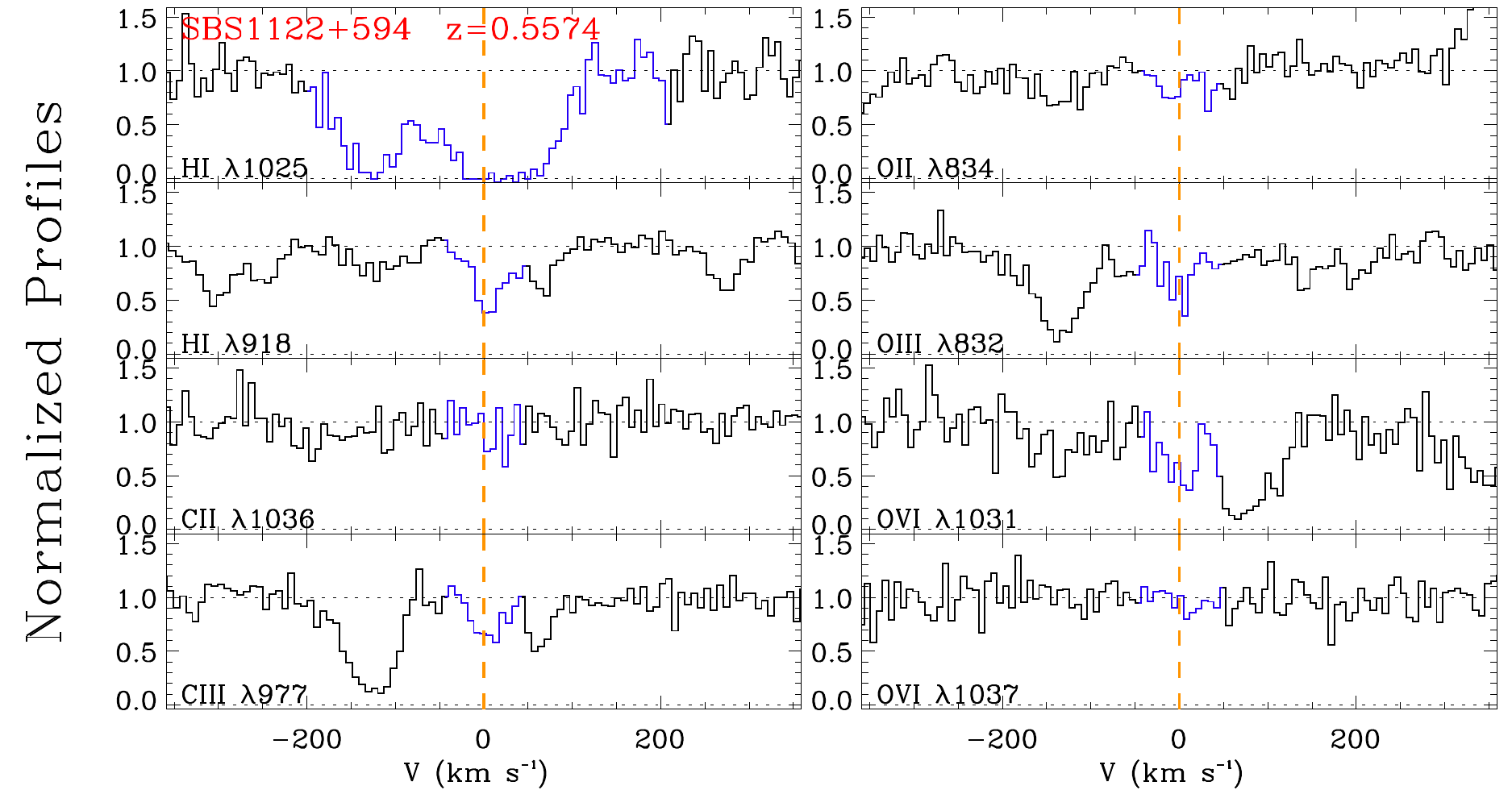}  
  \caption{Same as Fig.~\ref{f-pg1522a}, but for the absorber at $z=0.5574$ toward SBS1122+594
 \label{f-sbs1122}}
\end{figure*}

--  \underline{SBS1122+594, $z_{\rm abs}= 0.5574$:} The velocity profiles  for this absorber have a complicated structure (see Fig.~\ref{f-sbs1122}), with the absorption spanning from about $-200$ to $+120$ \km. However, most of the \hi\ column density is located in the component at 0 \km\ based on the weak \hi\ transitions. Within this component, there is only a weak absorption present in the profiles of the metal lines. The ionization and/or metallicity conditions must be different between the various components since a very strong \ciii\ and \oiii\ absorption (a factor $>6$ times larger in column than in the 0 \km\ component) is observed at about $-125$ \km\ with a smaller \nhi\ ($\log N_{\rm H\,I}\simeq 15.85$ or factor 2.8 smaller than that of the 0 \km\ component). In this case, we can estimate the column densities in the individual components and we only consider the component at 0 \km, i.e., the absorption between about $\pm 50$ \km. \cii\ $\lambda$904 is a tentative detection (\cii\ $\lambda$1036 is not detected, but the limit is consistent with $N_{\rm C\,II}$ estimated at 904 \AA). The one component COG fit to the \hi\ lines (from $\lambda$916 to $\lambda$937) and AOD estimates using the weakest \hi\ lines ($\lambda<920$ \AA) give the same $N_{\rm H\,I}$. We adopted a mean from the two results.

This absorber is complicated to model because C/O may not be solar, so we have to rely solely on the \oii/\oiii\ ratio (estimated between $-50$ and $+50$ \km) to estimate the metallicity. While the \oii\ absorption is weak, the two available transitions at 833 and 834 \AA\ provide a consistent $N_{\rm O\,II}$. The detection of \oii\ provides a strong lower limit on the metallicity $[{\rm X/H}] > -1.1$. Assuming that all the \oiii\ and \oii\ are in the same gas phase, then both the metallicity and U are tightly constrained with $[{\rm X/H}] = -1.05 \pm 0.05$ and $\log U = -3.2 \pm 0.10$. For that solution, we need to have $[{\rm C/O}] \simeq -0.6$. \ovi\ may be present at 0 \km, and therefore this absorber is multiphase, as the observed $N_{\rm O\,VI}$ cannot be reproduced by the photoionization model.

\begin{figure*}[tbp]
\epsscale{1} 
\plotone{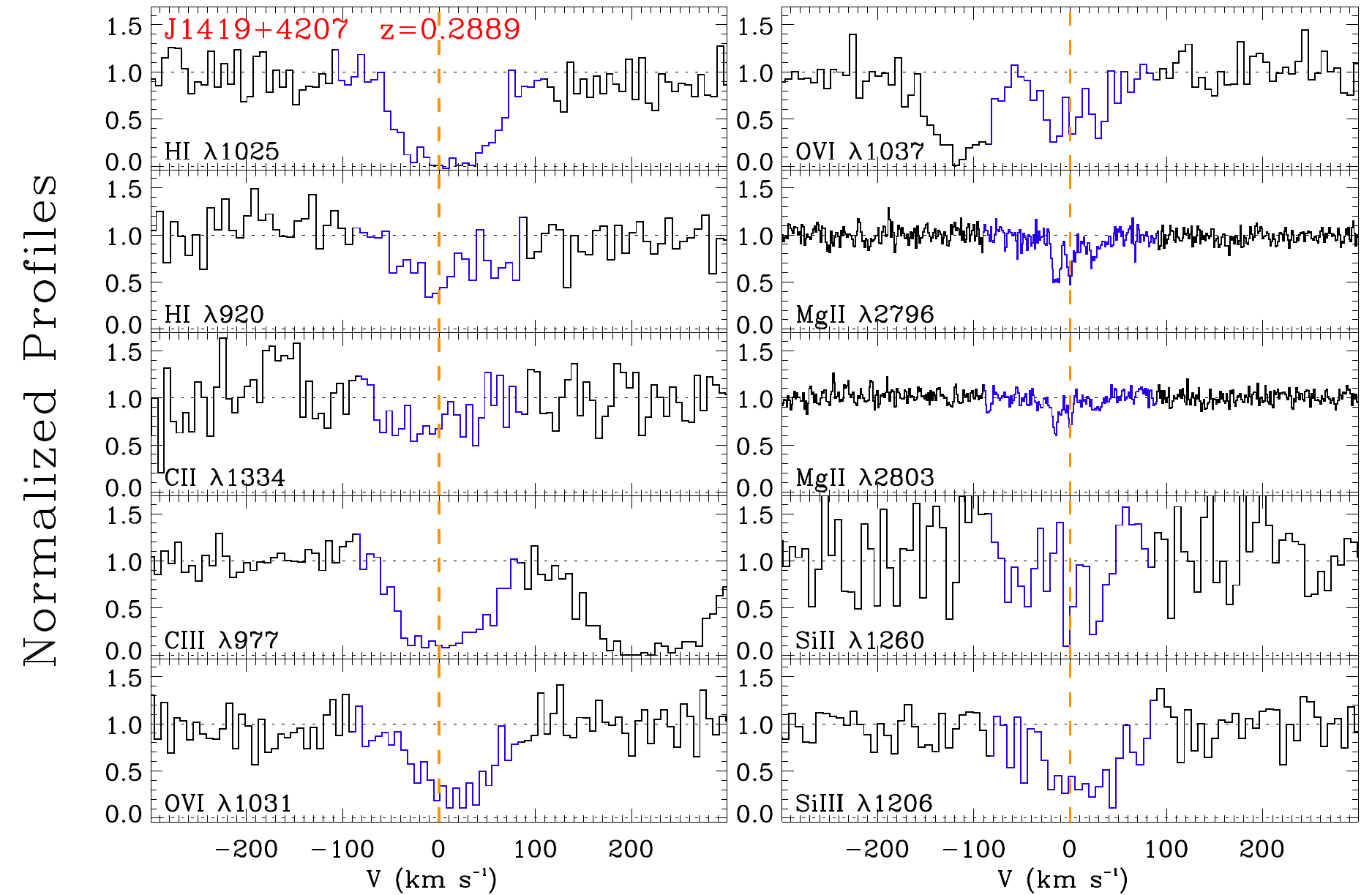}  
  \caption{Same as Fig.~\ref{f-pg1522a}, but for the absorber at $z=0.2889$ toward J1419+4207.
\label{f-j1419a}}
\end{figure*}

-- \underline{J1419+4207, $z_{\rm abs}= 0.2889$:} While the high resolution \mgii\ $\lambda$$\lambda$2796, 2803 spectra shown in Fig.~\ref{f-j1419a} reveal three components at $-14,0,24$ \km\ (the latter being the weakest), that velocity structure cannot be discerned in the lower resolution COS data. We therefore report the total column densities, i.e., $N_a(v)$ is integrated over the entire velocity profiles for that absorber. \ciii\ is quite strong and could be saturated. The one component COG fit to the \hi\ lines overlaps within $1\sigma$ with the AOD result. We adopted a mean from the two results.

A Cloudy photoionization model with  $[{\rm X/H}] = -0.65 \pm 0.10$, $\log U = -3.10 \pm 0.10$, and $[{\rm C/\alpha}]\simeq -0.15$ reproduces simultaneously the column densities of \cii, \ciii, \mgii, and \siiii\ within the estimated $1\sigma$ observational errors (allowing for about 0.2 dex saturation for \ciii), and the limit on \siii. A higher metallicity would over-predict the amount of \siii. A lower metallicity would under-predict the amount of \mgii. These two constraints set the $U$ in a narrow range of allowed values.  \ovi\ cannot be produced at all by this photoionization model, so it is a multiphase absorber.

\begin{figure*}[tbp]
\epsscale{1} 
\plotone{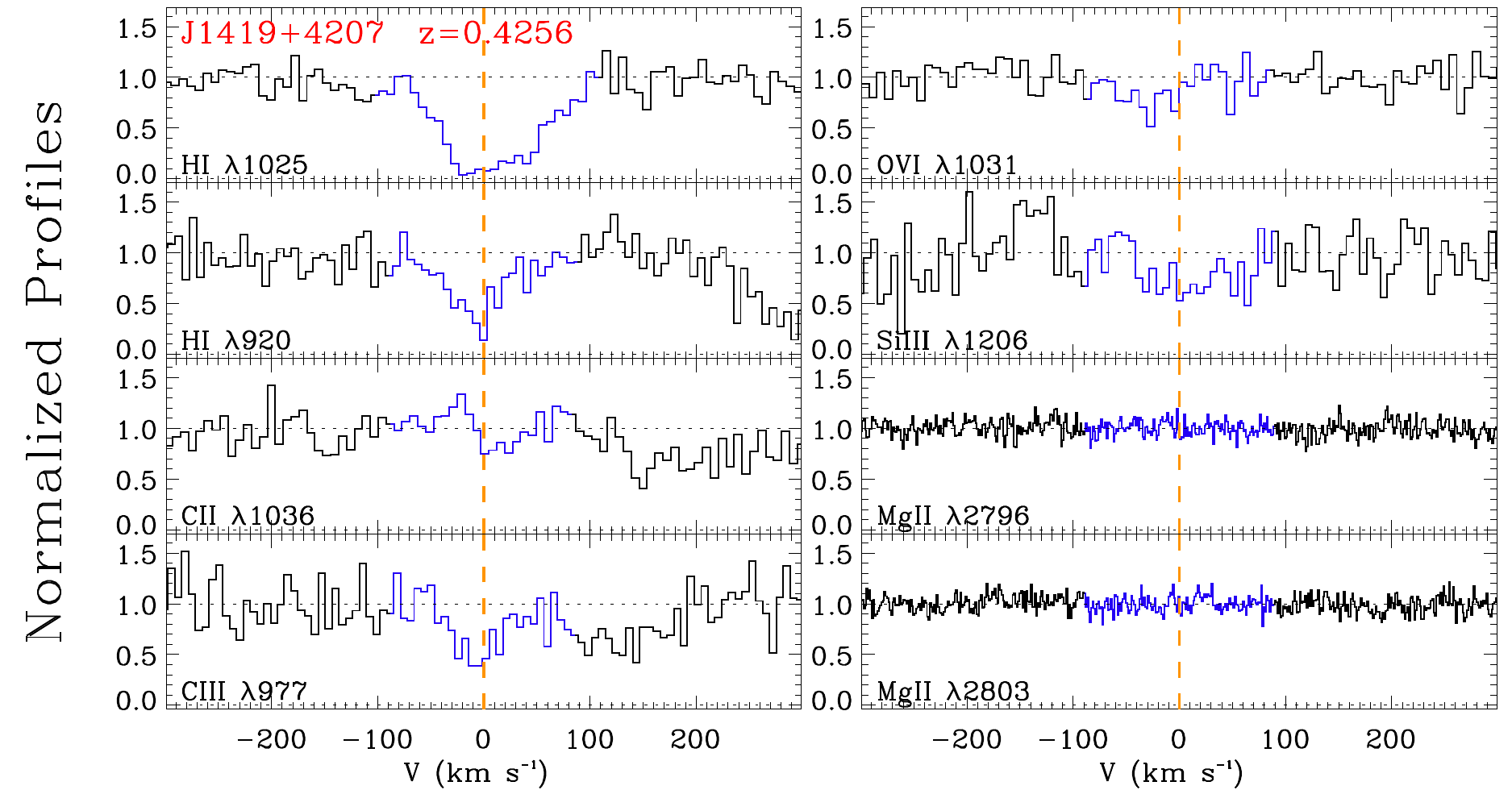}  
  \caption{Same as Fig.~\ref{f-pg1522a}, but for the absorber at $z=0.4256$ toward J1419+4207.
 \label{f-j1419b}}
\end{figure*}

-- \underline{J1419+4207,$z_{\rm abs}= 0.4256$:} No \mgii\ is detected in this absorber (see Fig.~\ref{f-j1419b}), and the COS absorption profiles do not reveal more than one component. All the metal lines are relatively weak, suggesting no saturation issue.  The one component COG fit to the \hi\ lines and AOD estimates using the weakest \hi\ lines ($\lambda<920$ \AA) give the same $N_{\rm H\,I}$. We adopted a mean of the two results.

A Cloudy photoionization model with  $[{\rm X/H}] = -1.40 \pm 0.20$ and $\log U = -2.90 \pm 0.20$ reproduces simultaneously the observed column densities of \siiii, \ciii, and agrees with the limit on \mgii.  A higher metallicity would over-predict the amount of \mgii. A lower metallicity would under-predict the amount of \siiii.  \ovi\ cannot be produced at all by this absorber, so it is a multiphase absorber (however, we note that \ovi\ is only detected in the stronger line, is weak, and its peak optical depth is shifted relative to the low ions; all this leading to an uncertain detection of \ovi). 

\begin{figure*}[tbp]
\epsscale{1} 
\plotone{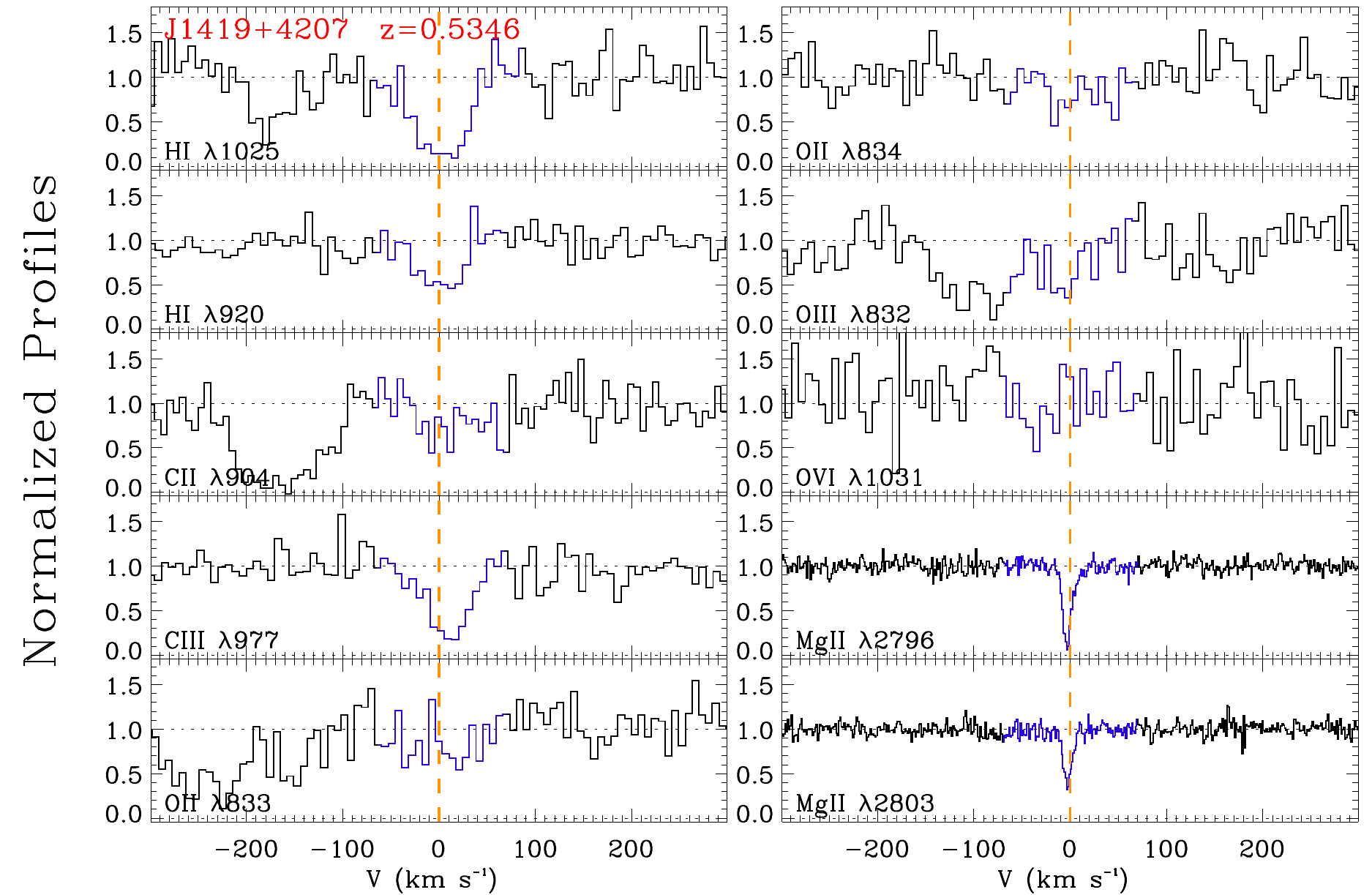}  
  \caption{Same as Fig.~\ref{f-pg1522a}, but for the absorber at $z=0.5346$ toward J1419+4207.
 \label{f-j1419c}}
\end{figure*}

-- \underline{J1419+4207, $z_{\rm abs}= 0.5346$:} Fig.~\ref{f-j1419c} shows that a strong single \mgii\ component is observed in this absorber. The only other significant detection of a metal line is \ciii. The \oii\ $\lambda$$\lambda$832, 834 absorption lines are weak, but both transitions give consistent results. \cii\ $\lambda$904 and \oiii\ $\lambda$832 are barely 3$\sigma$ detections. We estimated the same apparent column density for the weak and strong \mgii\ lines, implying that the lines are fully resolved. Unfortunately, for this absorber the \hi\ column density remains not very well determined (relative to the other absorbers) because the S/N is too low to provide reliable estimates of $N_a$ or $W_\lambda$ below 920 \AA: so all the available lines are near or on the flat part of the COG. The S/N is also too low to determine \nhi\ from the continuum break method. The adopted result is from the one component COG fit to the \hi\ lines. 

A Cloudy photoionization model with  $[{\rm X/H}] = -0.20 \pm 0.20$ and $\log U = -3.9 \pm 0.20$ reproduces simultaneously the column densities of \oii, \ciii, and \mgii. If \oiii\ is real, then the modeled column density is about a factor 5 too small. This is similar to the absorber at $z = 0.4685$ toward PKS0637-752. No \ovi\ is detected (although only down to a level of $13.90$ dex owing to the lower S/N spectrum of this QSO), but if \oiii\ is real, the gas could be multiphase. 

\begin{figure*}[tbp]
\epsscale{1} 
\plotone{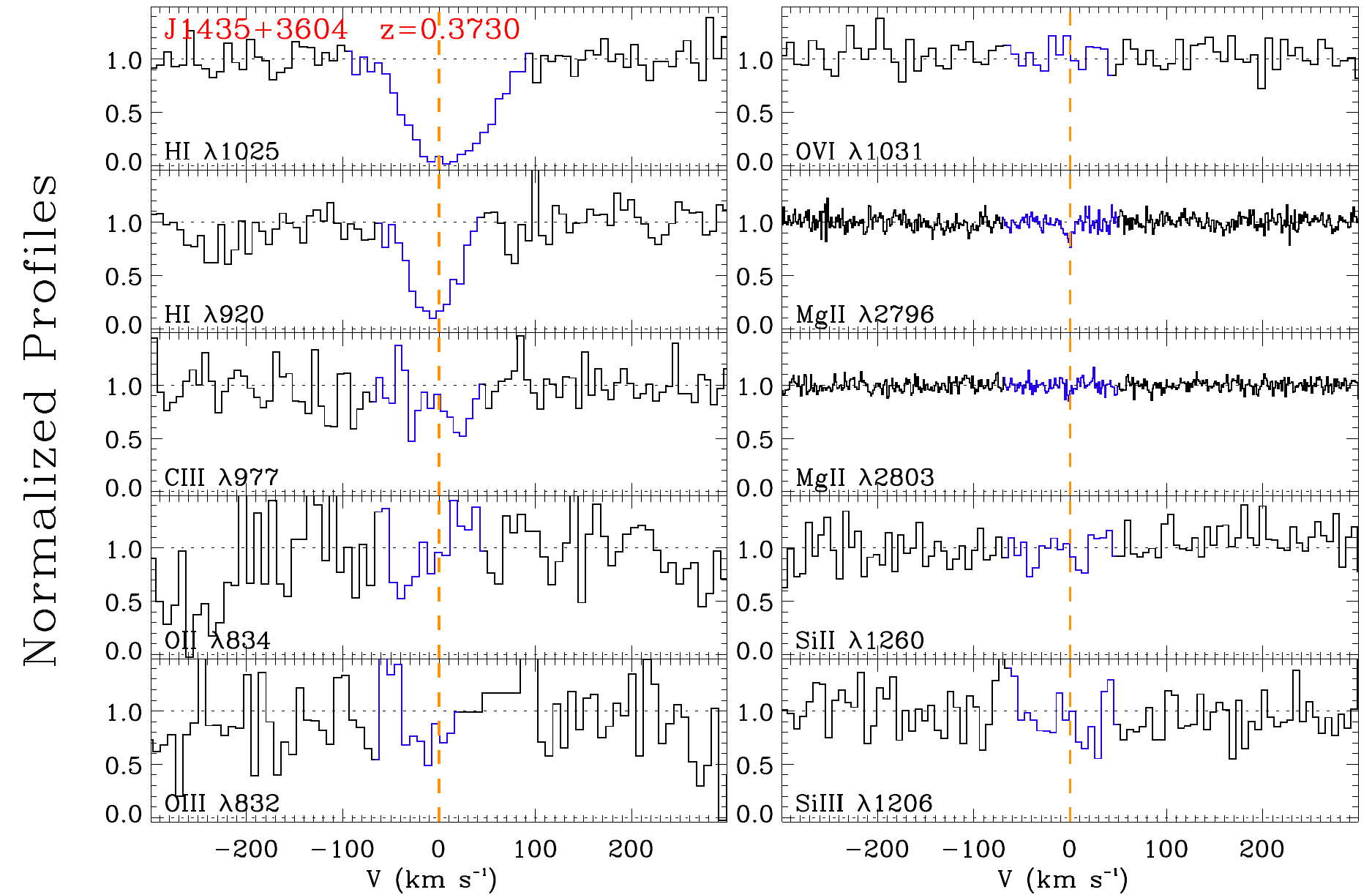}  
  \caption{Same as Fig.~\ref{f-pg1522a}, but for the absorber at $z=0.3730$ toward J1435+3604.
 \label{f-j1435a}}
\end{figure*}

-- \underline{J1435+3604, $z_{\rm abs}= 0.3730$:} Fig.~\ref{f-j1435a} shows that a very weak single \mgii\ component is observed. The only other detection of a metal line is \ciii. The AOD estimate of the weakest measurable \hi\ transition ($\lambda$916) and the COG fit to all the uncontaminated \hi\ lines give the same $N_{\rm H\,I}$. The estimate of the column density break is consistent with the cumulative effect of this absorber with the absorber at $z= 0.3878$.  

Based on the Cloudy photoionization model, the detection of \mgii\ implies that $[{\rm X/H}] > -2.0 $ as below that value, not enough \mgii\ is produced for any $U$. If $[{\rm X/H}] > -1.7 $, then $\log U < -4.1$ and  $[{\rm C/Mg}] > +1.2$, the latter being very unlikely. Therefore there is a small range of allowed metallicity that can reproduce the column densities of \ciii\ and \mgii\ (and limits on other ions),  $[{\rm X/H}] = -1.85 \pm 0.10$ and $\log U = -2.9 \pm 0.20$. No \ovi\ ($\log N_{\rm O\,VI}<13.51$) is observed in this absorber, so  there is no evidence for a multiphase absorber. 

\begin{figure*}[tbp]
\epsscale{1} 
\plotone{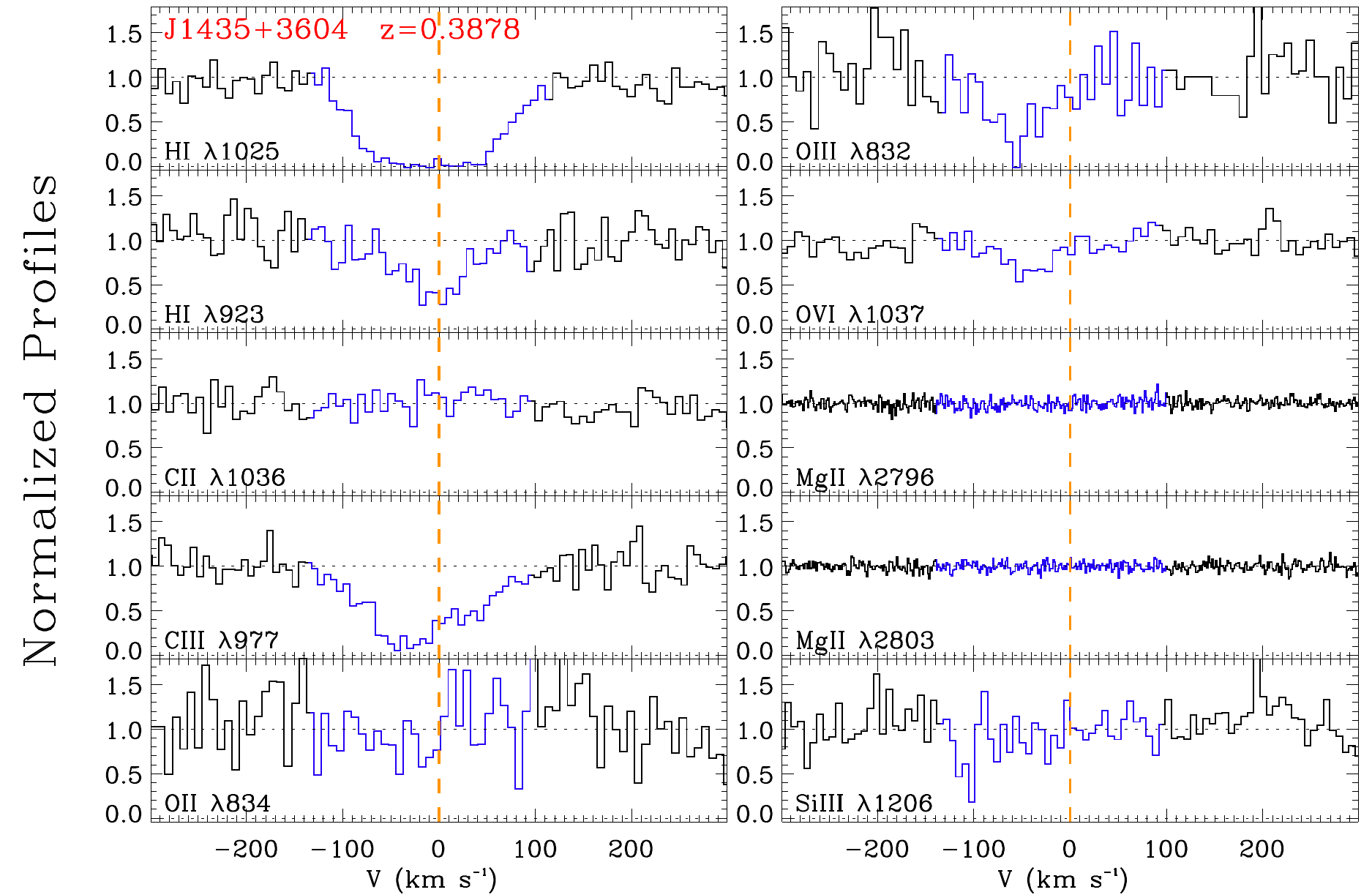}  
  \caption{Same as Fig.~\ref{f-pg1522a}, but for the absorber at $z=0.3878$ toward J1435+3604.
 \label{f-j1435b}}
\end{figure*}

-- \underline{J1435+3604, $z_{\rm abs}= 0.3878$:}  As shown in Fig.~\ref{f-j1435b}, no \mgii\ absorption is detected. The only detected metal lines are \ciii\ and \oiii. Both have more than one component, with the optical depth peaking in the second component that is shifted by about $-40$ \km. Similarly \ovi\ $\lambda$1037 is also shifted by the same amount from the strongest \hi\ component.   The one component COG fit to the \hi\ lines and AOD estimates using the weakest \hi\ lines ($\lambda<930$ \AA) yield similar $N_{\rm H\,I}$. We adopted a mean of the two results.

Based on the Cloudy photoionization model, the non-detections of \mgii\ and \siiii\ place a strong upper limit on the metallicity: for any $U$, $[{\rm X/H}] < -1.4$. This also consistent with the $3\sigma$ upper limits on \oii, although that limit is not very constraining. We are unable to better constrain the metallicity because, first, $N_{\rm O\,III}$ and $N_{\rm C\,III}$ are somewhat uncertain over the velocities where \nhi\ is estimated, and, second, the combination of increasing $U$ and decreasing the metallicity can fit all the observables. In order to be able to produce $N_{\rm O\,III}$ and $N_{\rm C\,III}$, $\log U$ must be $\ga -3$, but we emphasize that the velocities of these ions are shifted relative to the strongest \hi\ component. This absorber is also multiphase. 

\begin{figure*}[tbp]
\epsscale{1} 
\plotone{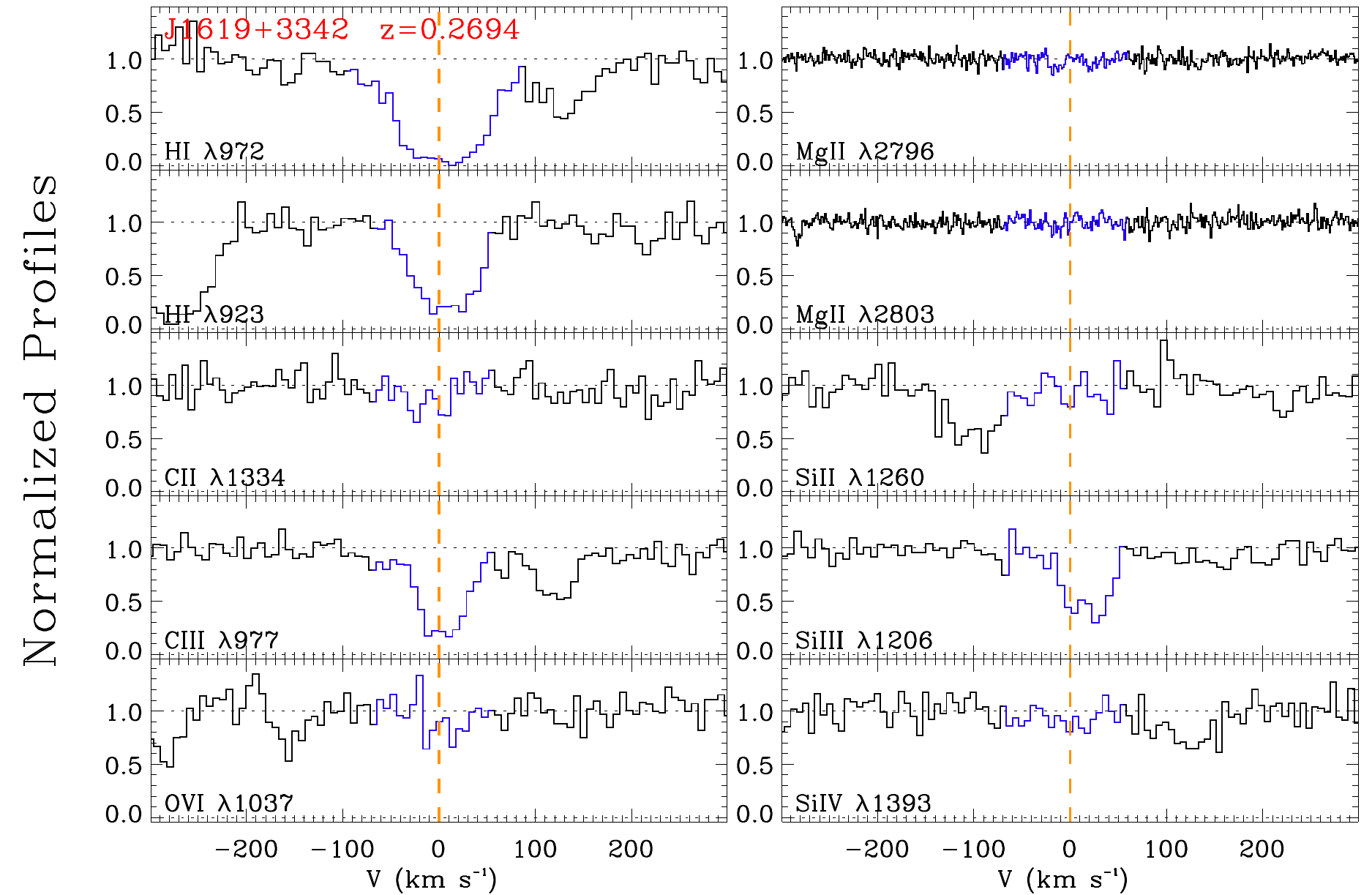}  
  \caption{Same as Fig.~\ref{f-pg1522a}, but for the absorber at $z=0.2694$ toward J1619+3342.
 \label{f-j1619}}
\end{figure*}

-- \underline{J1619+3342, $z_{\rm abs}= 0.2694$:} Fig.~\ref{f-j1619} shows that the \mgii\ absorption is very weak but detected in the stronger line of the doublet. There are possibly two components. The absorption of \cii\ $\lambda$1334 is barely detected at $3\sigma$, and \siii\ $\lambda$1260 and \siiv\ $\lambda$1393 are not detected. In contrast, \ciii\ and \siiii\ are both well detected and line up well with the \hi\ absorption within the COS velocity calibration error. A single component COG fit to the \hi\ lines, the AOD measurement of \hi\ $\lambda$$\lambda$917, 920, and the continuum break method all give overlapping results for \nhi\ within $1\sigma$. We adopted a mean for \nhi\ of the results from these 3 methods.

A Cloudy photoionization model with  $[{\rm X/H}] = -1.60 \pm 0.10$, $\log U = -2.9 \pm 0.15$, and $[{\rm C/\alpha}]\simeq -0.10$  reproduces simultaneously the column densities of \siiii, \mgii, and \ciii\ and limits on \cii. The metallicity cannot be higher as otherwise for any $U$, too much \mgii\ would be produced. There might be a 2--3\,$\sigma$ detection of \ovi\ $\lambda$1037, but unfortunately \ovi\ $\lambda$1031 is contaminated, therefore it is unclear if this absorber is multiphase. 

\begin{figure*}[tbp]
\epsscale{1} 
\plotone{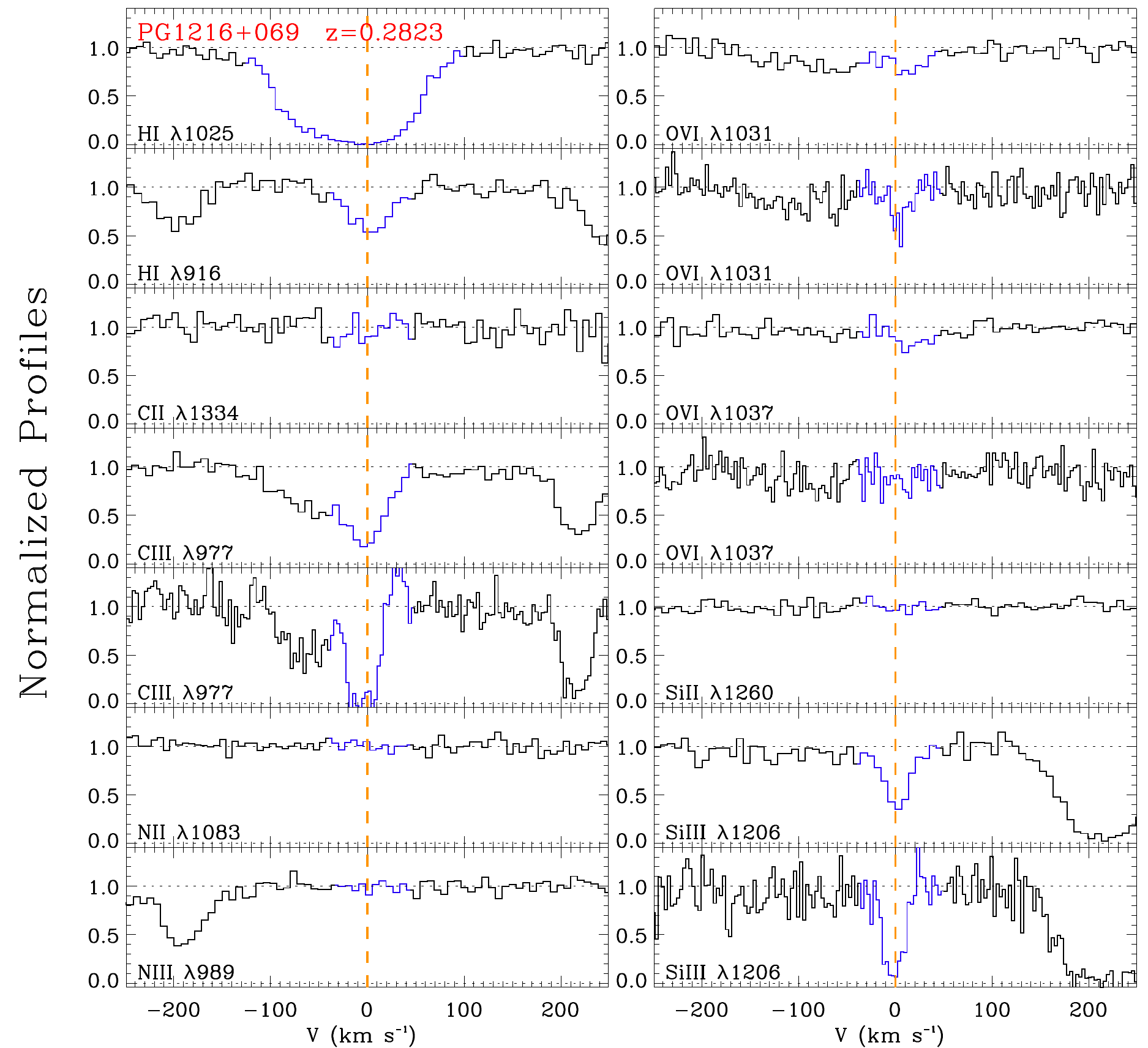}  
  \caption{Same as Fig.~\ref{f-pg1522a}, but for the absorber at $z=0.2823$ toward PG1216+069. For \ciiit, \ovit, and \siiiit, we also show the normalized STIS E140M profiles. 
 \label{f-pg1216}}
\end{figure*}

-- \underline{PG1216+069, $z_{\rm abs}= 0.2823$:} For that absorber, we use both the COS and STIS E140M observations (for a detailed description of the STIS data reduction, see \citealt{tripp05}).  In Fig.~\ref{f-pg1216}, we show the COS and STIS normalized profiles. While the STIS observations have a lower signal-to-noise level (typically S/N\,$\sim 10$ per resolution element), the 6.5 \km\ FWHM resolution (similar to Keck HIRES \mgii\ observations) helps to characterize the velocity structure and saturation level. Unfortunately, the STIS cannot be used to determine the \nhi\ because the weak Lyman series transitions lie in a region of the spectrum where  S/N\,$\la 3$. Only \ciii, \siiii, \ovi, and \hi\ are detected. The STIS \siiii\ profile show thats a single component is present, which lines up very well with the weak \hi\ transitions. The stronger \hi\ transitions and \ciii\ and \ovi\ show, however, additional components between about $-150$ and $-25$ \km. For \ciii\ and \ovi, these components are easily separated from the 0 \km\ component. For that absober, we were able to estimate the column densities in the narrow component at $0$ \km\ for all the metal lines and weak \hi\ transitions. The AOD estimates imply from COS and STIS \siiii\ spectra yield 12.80 and 12.99 dex, implying that saturation effects are not negligible. Our adopted result for \siiii\ is from a single Voigt component fit to the STIS profile, which yields $\log N = 13.10 \pm 0.09$ and $b = 9.9 \pm 1.4$ \km. For \ciii, we can only derive a lower limit as the STIS profile is black. A single component COG fit to the weak \hi\ lines ($\lambda$916 to $\lambda$923), the AOD measurement of  same weak \hi\ lines, and the continuum break method all give overlapping results for \nhi\ within $1\sigma$. We adopted a mean for \nhi\ of the results from these 3 methods.

A Cloudy photoionization model with  $[{\rm X/H}] < -1.65$, $\log U < -2.6$  reproduces simultaneously the column density of \siiii\ and limits on \siii\ and \siiv. If \siiii\ has a higher column density that derived from our profile fit, this would imply a higher $U$, but a lower metallicity (owing to the upper limit on \siii): therefore the adopted result implies a strict upper limit on the metallicity, but not on $U$. This model produces too much \ciii\ and \niii. As \ciii\ is saturated, the discrepancy can be explained in terms of both saturation or nucleosynthesis effects. However, \niii\ is not detected, which implies that $[{\rm N/Si}]<-0.6$, consistent with expected nucleosynthesis effect at low metallicity \citep[e.g.][]{henry00}. Although the \ovi\ directly associated with the LLS is weak and appears relatively narrow, the observed column density is several orders of magnitude predicted by the photoionization model; this absorber is therefore multiphase.

\smallskip

\begin{figure}[tbp]
\epsscale{0.6} 
\plotone{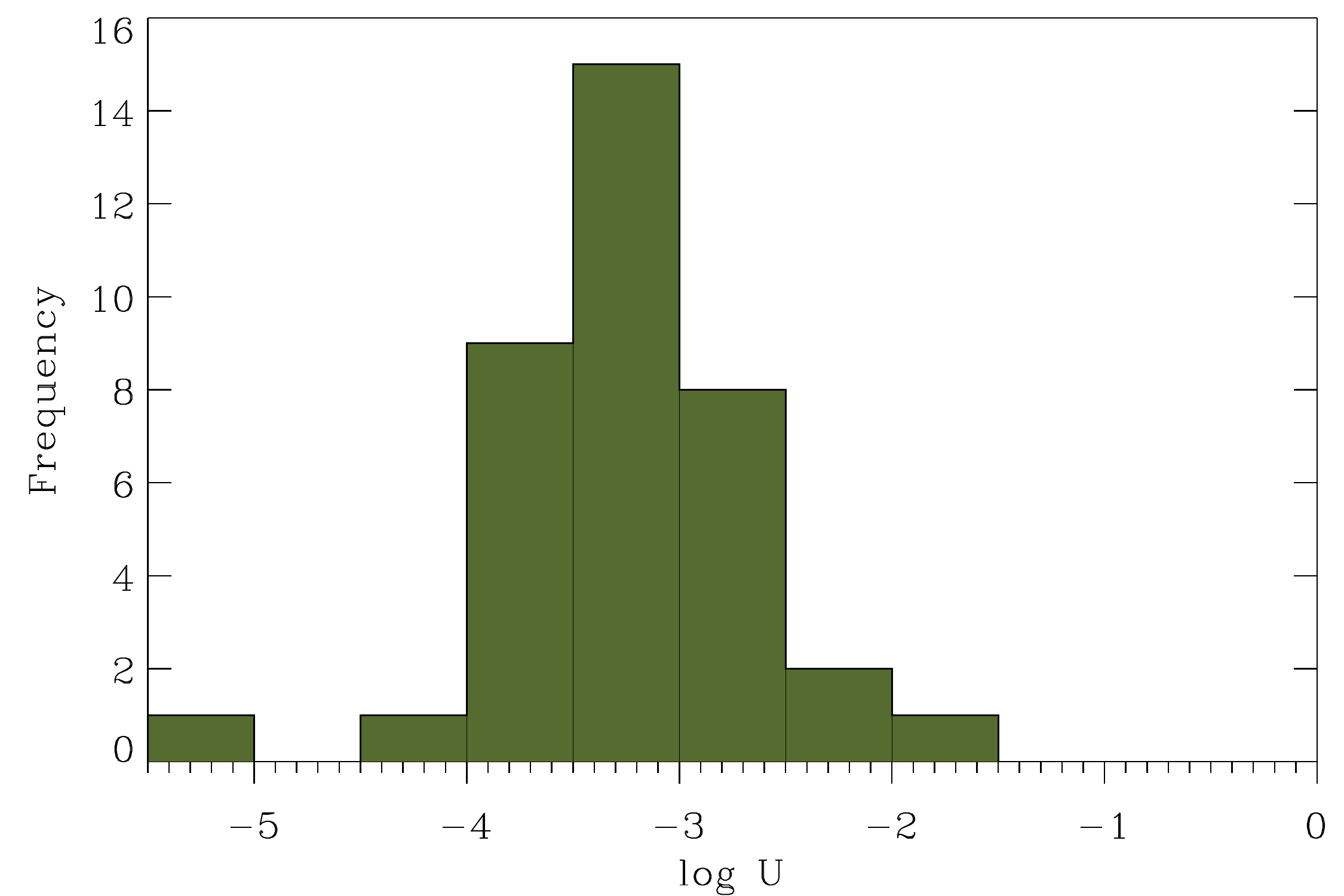}  
  \caption{Distribution of the central value of the ionization parameter, $\log U$, derived for the LLS from the Cloudy photoionization models. 
 \label{f-udist}}
\end{figure}

From the results above and other studies at $z\la 1$ (see Table~\ref{t-sum}), we show in Fig.~\ref{f-udist} the distribution of the central value of $\log U$. This indicates a relatively tight distribution in U, with $\langle \log U \rangle = -3.3 \pm 0.6$ (median $-3.2$). There is an evolution of $\log U$ with $z$ since \citet{fumagalli11a} reported that at $z>1.5$, $\log U> -3$. 

\begin{figure}[tbp]
\epsscale{0.55} 
\plotone{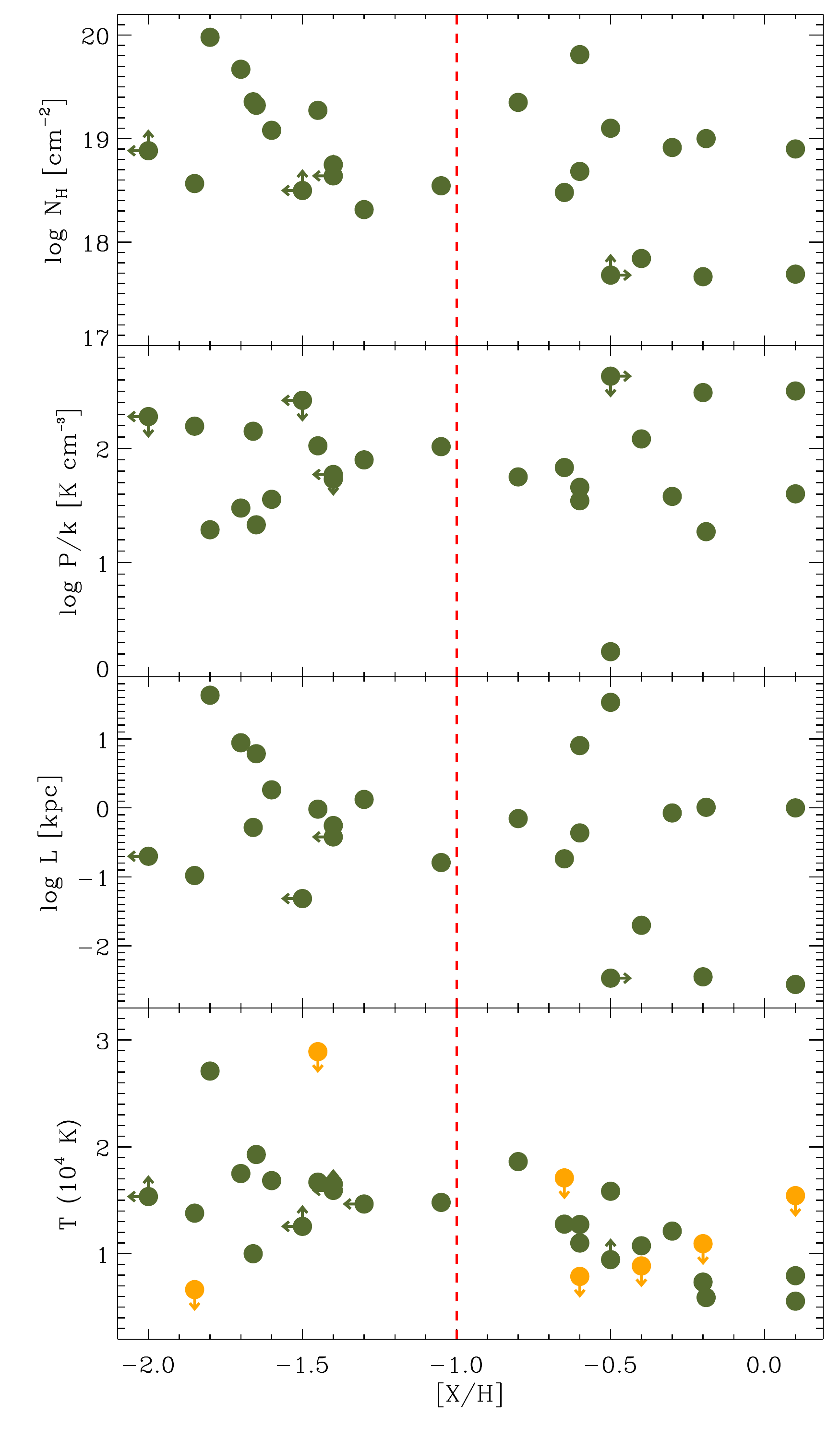}  
  \caption{\hit\ column densities determined from the observations and physical quantities determined from the Cloudy photoionization simulations as a function of the metallicity derived with the same models. In the bottom panel, the orange symbols show the result from the analysis of the temperature derived from the $b$-value of the individual components of \mgiit\ observed at high resolution with Keck HIRES. \label{f-docloudy}}
\end{figure}

Finally, in Fig.~\ref{f-docloudy}, we provide a visual summary of Table~\ref{t-cloudy} for the $N_{\rm H\,I}$, $N_{\rm H}$, $P/k$, $l$, and $T$ quantities as a function of the metallicity. There is no obvious trend between these quantities and the metallicity, except that the lowest values of $N_{\rm H}$ are observed in the high metallicity sample, which is expected (since the metallicity correction is smaller). Importantly, the low temperatures derived from the Cloudy simulations are confirmed by the observations (bottom panel of Fig.~\ref{f-docloudy} and \S\ref{s-llsprop}). 
\clearpage 

\begin{deluxetable}{lccccccccc}
\tabcolsep=3pt
\tablecolumns{10}
\tablewidth{0pc}
\tablecaption{Cloudy Summary for LLS with $16.2 \le \log N_{\rm HI} < 19$ \label{t-cloudy}}
\tabletypesize{\footnotesize}
\tablehead{\colhead{Name} & \colhead{$z$}& \colhead{$\log N_{\rm H\,I}$} & \colhead{$\log U$} & \colhead{$[{\rm X/H}]$}& \colhead{$\log N_{\rm H}$}& \colhead{$T$}& \colhead{$\log n_{\rm H}$}& \colhead{$P/k$}& \colhead{$l$}\\ %
\colhead{} &\colhead{} & \colhead{[cm$^{-2}$]} &\colhead{}  &\colhead{} &\colhead{[cm$^{-2}$]}&\colhead{($10^4$\,K)}&\colhead{[cm$^{-3}$]}&\colhead{}(K\,cm$^{-3}$)&\colhead{(kpc)} }
\startdata
	  J0943+0531   & $   0.3544 $ & $   16.11 $ & $	 \le -3.2 $ & $	 -1.3 $ & $\le 18.3 $ & $ \le 1.5 $ & $ \ge -2.3 $ & $\ge 79 $ & $    1.3    $ \\
          J1419+4207   & $   0.4256 $ & $   16.17 $ & $    -2.9 $ & $	 -1.4 $ & $    18.7 $ & $     1.7 $ & $    -2.5 $ & $	 54 $ & $     0.5    $ \\
          J1435+3604   & $   0.3878 $ & $   16.18 $ & $ \ge-3.0 $ & $	<-1.4 $ & $   >18.6 $ & $   > 1.6 $ & $   <-2.4 $ & $    <59$ & $     0.3   $ \\
          PG1116+215   & $   0.1385 $ & $   16.20 $ & $    -2.5 $ & $	 -0.5 $ & $    19.1 $ & $     1.6 $ & $    -4.0 $ & $	  2 $ & $ >    41   $ \\
          PG1522+101   & $   0.5185 $ & $   16.22 $ & $    -3.6 $ & $	 -0.4 $ & $    17.8 $ & $     1.1 $ & $    -1.9 $ & $	121 $ & $    0.02    $ \\
         SBS1122+594   & $   0.5574 $ & $   16.24 $ & $    -3.1 $ & $	 -1.0 $ & $    18.5 $ & $     1.5 $ & $    -2.2 $ & $	104 $ & $     0.2   $ \\
         HE0439-5254   & $   0.6153 $ & $   16.28 $ & $    -2.7 $ & $	 -0.3 $ & $    18.9 $ & $     1.2 $ & $    -2.5 $ & $	 38 $ & $     0.8   $ \\
	PG1338+416     & $   0.3488 $ & $   16.30 $ & $    -3.2 $ & $	 -0.6 $ & $    18.5 $ & $     1.3 $ & $    -2.3 $ & $	 68 $ & $     0.2   $ \\
         J1419+4207    & $   0.5346 $ & $   16.34 $ & $    -3.9 $ & $	 -0.2 $ & $    17.7 $ & $     0.7 $ & $    -1.4 $ & $	308 $ & $   0.004   $ \\		 
 	  PG1407+265   & $   0.6828 $ & $   16.38 $ & $    -2.0 $ & $	 -1.8 $ & $    20.0 $ & $     2.7 $ & $    -3.1 $ & $	 19 $ & $      41   $ \\
           PG1216+069  & $   0.2823 $ & $   16.40 $ & $\le -2.6 $ & $	<-1.7 $ &$  \le 19.3$ & $ \le 1.9 $ & $\ge -3.0 $ & $ \ge 21 $ & $     6.1   	 $ \\
	J1419+4207     & $   0.2889 $ & $   16.40 $ & $    -3.1 $ & $	 -0.6 $ & $    18.7 $ & $     1.3 $ & $    -2.4 $ & $	 46 $ & $     0.4   $ \\ 
          PG1338+416   & $   0.6865 $ & $   16.45 $ & $    -3.9 $ & $	  0.1 $ & $    17.7 $ & $     0.6 $ & $    -1.2 $ & $	318 $ & $   0.003    $ \\
         PKS0405-123   & $   0.1672 $ & $   16.45 $ & $    -3.1 $ & $	  0.1 $ & $    18.9 $ & $     0.8 $ & $    -2.7 $ & $	 40 $ & $      1.3   $ \\
         J1619+3342    & $   0.2694 $ & $   16.48 $ & $    -2.9 $ & $	 -1.6 $ & $    19.1 $ & $     1.7 $ & $    -2.7 $ & $	 36 $ & $      2.0 $ \\
         PKS0637-752   & $   0.4685 $ & $   16.48 $ & $    -3.8 $ & $	 -0.4 $ & $    18.0 $ & $     0.9 $ & $    -1.5 $ & $    261 $ & $  0.006    $ \\
	HE0153-4520    & $   0.2261 $ & $   16.61 $ & $    -3.8 $ & $	 -0.8 $ & $    19.4 $ & $     1.9 $ & $    -2.9 $ & $	 56 $ & $      6.4  $ \\
         J1435+3604    & $   0.3730 $ & $   16.65 $ & $    -3.5 $ & $	 -1.9 $ & $    18.6 $ & $     1.4 $ & $    -1.9 $ & $	156 $ & $     0.1   $ \\
          PG1522+101   & $   0.7292 $ & $   16.66 $ & $   >-3.2 $ & $	<-2.0 $ & $   >18.9 $ & $    >1.5 $ & $   <-1.9 $ & $  <190 $ & $     0.2    $ \\
             PHL1377   & $   0.7390 $ & $   16.72 $ & $    -2.9 $ & $	 -1.5 $ & $    19.3 $ & $     1.7 $ & $    -2.2 $ & $	105 $ & $  >   1.0  $ \\
         PKS0552-640   & $   0.3451 $ & $   16.90 $ & $  > -3.8 $ & $	<-1.5 $ & $   >18.5 $ & $    >1.3 $ & $   <-1.7 $ & $  <263 $ & $    0.05  $ \\
          PG1630+377   & $   0.2740 $ & $   16.98 $ & $    -2.8 $ & $	 -1.7 $ & $    19.7 $ & $     1.8 $ & $    -2.8 $ & $	 30 $ & $  >   10    $ \\
	     PHL1811   & $   0.0810 $ & $   17.98 $ & $    -4.0 $ & $	 -0.2 $ & $    19.0 $ & $     0.6 $ & $    -2.5 $ & $	 19 $ & $      1.0   $ \\
          PKS0312-77   & $   0.2026 $ & $   18.22 $ & $    -3.2 $ & $	 -0.6 $ & $    19.8 $ & $     1.1 $ & $    -2.5 $ & $	 35 $ & $      6.4  $ \\
           TON153      & $   0.6610 $ & $   18.30 $ & $    -3.2 $ & $	 -1.7 $ & $    19.4 $ & $     1.0 $ & $    -1.9 $ & $	141 $ & $     0.6   	 $ 
\enddata
\tablecomments{$z$ and $N_{\rm HI}$ are derived directly from the observations. $U$, $[{\rm X/H}]$, $N_{\rm H}$, $T$, $n_{\rm H}$, $P/k = n_{\rm H} T$, and 
$l =  N_{\rm H}/n_{\rm H}$ are derived from Cloudy photoionization simulations constrained by the column densities of metal ions and \hit\ determined from the observations. For the sightlines that are not in the new sample, the results were retrieved from the original papers (see references in Table~\ref{t-sum}).}
\end{deluxetable}

\clearpage

\end{appendix}

\end{document}